\documentclass[fleqn,usenatbib]{mnras}

\usepackage{newtxtext,newtxmath}
\usepackage[T1]{fontenc}
\usepackage{subfig}
\usepackage{caption} 
\usepackage{cleveref}
\usepackage{graphicx}
\usepackage{scrextend}
\usepackage{booktabs}
\usepackage{xcolor}
\usepackage{multicol}
\usepackage{multirow}

\crefname{figure}{Fig.}{Figs.}
\crefname{table}{Table}{Tables}
\crefname{chapter}{Chapter}{Chapters}
\crefname{section}{Section}{Sections}

\newcommand{\percSym}[1]{$#1\%$}
\newcommand{\perc}[1]{$#1$ per cent}
\newcommand{\units}[1]{\, \mathrm{#1}}

\newcommand{\diff}{\mathop{}\!\mathrm{d}}

\newcommand{\subrfig}[1]{\protect\subref{fig:#1}}

\newcommand{\orcid}[1]{\href{https://orcid.org/#1}{\includegraphics[scale=0.08]{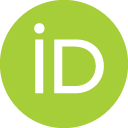}}}

\newcommand{\rgal}{r_\mathrm{gal}}

\newcommand{\rhStar}{r_{1/2,\ast}}

\newcommand{\tenMass}[1]{10^{#1}\msun}

\newcommand{\zeq}[1]{\mbox{$z=#1$}}

\newcommand{\msun}{\units{M_\odot}}
\newcommand{\Rv}{R_{\mathrm{200,c}}}   
\newcommand{\Mv}{M_{\mathrm{200,c}}}   

\newcommand{\cvir}{c_{\mathrm{vir}}}
\newcommand{\jness}{j_\mathrm{p}}
\graphicspath{ {figs/} }
\DeclareGraphicsExtensions{.pdf,.png,.svg}
\DeclareCaptionFormat{cont}{#1 (cont.)#2#3\par}

\title[Cosmological Jellyfish]{Jellyfish galaxies with the IllustrisTNG simulations -- Citizen-science results towards large distances, low-mass hosts, and high redshifts}

\author[Zinger et al.]{Elad Zinger$^{1,2}$\thanks{E-mail:\href{mailto:elad.zinger@mail.huji.ac.il}{elad.zinger@mail.huji.ac.il}}\orcid{0000-0002-6316-3996}, Gandhali Joshi$^{2,3}$, Annalisa Pillepich$^2$\orcid{0000-0003-1065-9274}, Eric Rohr$^{2}$\orcid{0000-0002-9183-5593}, and Dylan Nelson$^{4}$\orcid{0000-0001-8421-5890}\\\\
$^{1}$\,Centre for Astrophysics and Planetary Science, Racah Institute of Physics, The Hebrew University, Jerusalem 91904, Israel \\
$^{2}$\,Max-Planck-Institut f\"ur Astronomie, K\"onigstuhl 17, D-69117 Heidelberg, Germany\\
$^{3}$\,Department of Physics and Astronomy, University College London, London WC1E 6BT, UK\\
$^{4}$\,Universit\"{a}t Heidelberg, Zentrum f\"{u}r Astronomie, Institut f\"{u}r Theoretische Astrophysik, Albert-Ueberle-Str. 2, 69120 Heidelberg, Germany
}

\date{}
\pubyear{2023}

\begin{document}
\label{firstpage}
\pagerange{\pageref{firstpage}--\pageref{lastpage}}
\maketitle

\begin{abstract}
We present the ``Cosmological Jellyfish'' project - a citizen-science classification program to identify jellyfish galaxies within the IllustrisTNG cosmological simulations. Jellyfish (JF) are satellite galaxies that exhibit long trailing gas features -- `tails' -- extending from their stellar body. Their distinctive morphology arises due to ram-pressure stripping (RPS) as they move through the background gaseous medium. Using the TNG50 and TNG100 simulations, we construct a sample of $\sim 80,000$ satellite galaxies spanning an unprecedented range of stellar masses, $10^{8.3\--12.3}\,\mathrm{M_\odot}$, and host masses of $M_\mathrm{200,c}=10^{10.4\--14.6}\,\mathrm{M_\odot}$ back to $z=2$ \citep[extending the work of][]{yun_jellyfish_2019}. Based on this sample, $\sim 90,000$ galaxy images were presented to volunteers in a citizen-science project on the Zooniverse platform who were asked to determine if each galaxy image resembles a jellyfish. Based on volunteer votes, each galaxy was assigned a score determining if it is a JF or not. This paper describes the project, the inspected satellite sample, the methodology, and the classification process that resulted in a dataset of $5,307$ visually-identified jellyfish galaxies. We find that JF galaxies are common in nearly all group- and cluster-sized systems, with the JF fraction increasing with host mass and decreasing with satellite stellar mass. We highlight JF galaxies in three relatively unexplored regimes: low-mass hosts of $M_\mathrm{200,c}\sim10^{11.5-13}\,\mathrm{M_\odot}$, radial positions within hosts exceeding the virial radius $R_\mathrm{200,c}$, and at high redshift up to $z=2$. The full dataset of our jellyfish scores is publicly available and can be used to select and study JF galaxies in the IllustrisTNG simulations.
\end{abstract}



\section{Introduction}\label{sec:intro} 

One of the main determinants of the evolutionary pathway of a galaxy is the environment it resides in. Observations have long shown stark differences in a range of properties between galaxies that reside in the centers of their host halos versus galaxies that are satellites at larger distances \citep{dressler_galaxy_1980}. In particular, satellite galaxies are more likely to be quenched compared to field and central galaxies of the same stellar mass \citep{lewis_2df_2002,gomez_galaxy_2003,peng_mass_2010}, and have lower gas fractions, including both atomic and molecular components \citep{gavazzi_completing_2005,catinella_galex_2013,fumagalli_molecular_2009,boselli_cold_2014}.

A set of physical processes, that are environmental as opposed to secular, lead to these differences, primarily through the removal of gas from the galaxy body \citep{cortese_dawes_2021}. Some of these processes affect all components of the galaxy via gravitational interactions -- such as tidal stripping \citep{gnedin_tidal_2003,villalobos_simulating_2014} or galaxy harassment \citep{moore_galaxy_1996,gnedin_dynamical_2003}. Others arise from hydrodynamical effects, including ram-pressure stripping \citet{gunn_infall_1972} and viscous stripping, \citet{nulsen_transport_1982}, or thermo-dynamical channels \citep[e.g. thermal evaporation,][]{cowie_thermal_1977}, and so affect only the gaseous component, leaving the stellar structure of the galaxy unchanged. 

Ram pressure stripping (RPS) occurs when the ram pressure is generated by the motion of a satellite through the ambient medium. The ram-pressure force is $\propto\rho_\mathrm{medium}v_\mathrm{sat}^2$, where $\rho_\mathrm{medium}$ is the gas density of the medium and $v_\mathrm{sat}$ is the velocity of the satellite with respect to the medium. Stripping occurs when this force exceeds the gravitational restoring force of the galaxy \citep[see][for a recent review]{boselli_ram_2022}. Viscous (or turbulent) stripping removes gas from the outer layers of a galaxy through momentum transfer by viscosity. This process is thought to be less important for satellite galaxies than RPS \citep{roediger_ram-pressure_2008,roediger_ram_2009}. Both processes lead to extended gas structures that emanate from the stellar body and trail behind it, a configuration that resembles a jellyfish. Due to this similarity, such galaxies have come to be known as `jellyfish' (JF) galaxies and have by now been observed across a broad wavelength range \citep[e.g.][]{chung_vla_2009,bekki_ram-pressure_2009,smith_ultraviolet_2010,ebeling_jellyfish_2014, poggianti_gasp_2017}.

Due to its dependence on the medium properties, RPS is expected to be more important in more massive halos and in the inner regions of the halos since the ambient medium is denser and the infall velocities are higher, due to the deeper potential wells. In addition, smaller galaxies, with weaker gravitational binding are expected to be more affected by RPS. These trends are by and large consistent with observations \citep{boselli_ram_2022,roberts_lotss_2021-2}

For these reasons, observational efforts to find JF galaxies usually focus on the inner regions of galaxy cluster-sized systems, {\i.e.\@ in hosts with $\Mv\gtrsim \tenMass{14}$} \citep[even though recent surveys have begun to include group-sized hosts in their search for JF galaxies,][]{roberts_lotss_2021} and at low redshifts of $ z < 1$ \citep[e.g.][]{boselli_evidence_2019}, where such massive systems are more common. However, some theoretical studies have suggested that RPS can occur at and even beyond the virial radius \citep{bahe_why_2013,cen_origin_2014,jaffe_budhies_2015,zinger_quenching_2018,ayromlou_new_2019}, and JF galaxies have also been observed to occupy these regions \citep{chung_vla_2009,boselli_virgo_2018,roberts_lotss_2021-2}. 

Galaxies affected by ram pressure have been detected first in the radio \citep{miley_active_1972,gavazzi_determination_1978}, with subsequent multi-wavelength observations revealing that the tails of JF galaxies are multi-phase and can contain ionized gas \citep[$\mathrm{H\alpha:}$][]{boselli_spectacular_2016,gavazzi_ubiquitous_2018}, neutral Hydrogen \citep[HI:][]{shostak_ngc_1982,kenney_vla_2004,healy_h_2021}, molecular gas \citep{jachym_abundant_2014,jachym_molecular_2017,verdugo_ram_2015}, and hot ionized gas \citep[X-rays:][]{machacek_chandra_2006,sun_70_2006,wood_infall_2017}.

Many of the observational efforts have focused on single galaxies or the population of a single host cluster \citep[e.g.][]{boselli_virgo_2018}, but recent years have seen the advent of several surveys targeting larger samples of JF galaxies, such as the GAs Stripping Phenomena in galaxies project \citep[GASP][]{poggianti_gasp_2017}, LOFAR Two-Meter Sky Survey \citep[LoTSS][]{roberts_lotss_2021-2}, and OSIRIS Mapping of Emission-line Galaxies \citep[OMEGA][]{roman-oliveira_omegaosiris_2019}, each of which compiled JF samples of $50-70$ objects. These surveys have made statistical studies of $\sim 100$ observed JF galaxies possible \citep{smith_new_2022,peluso_exploring_2022}.

Numerical simulations enable us to build a theoretical picture of JF galaxies. Idealized `wind-tunnel' type simulations focus on a single galaxy in high spatial and temporal resolution and can explore the impact of different physical effects \citep[][]{tonnesen_gas_2009,roediger_ram_2007,roediger_ram_2009}. Large-scale cosmological simulations such as Illustris \citep{vogelsberger_introducing_2014, genel_introducing_2014, sijacki_illustris_2015}, EAGLE \citep{crain_eagle_2015,schaye_eagle_2015}, and IllustrisTNG \citep[][and references below]{pillepich_simulating_2018}, have achieved sufficiently high resolution, large-enough volumes, and physical sophistication to generate large samples of JF galaxies. These large-volume simulations enable statistical studies over a large range of satellite and host properties, and the ability to follow the history of each JF galaxy in a cosmological context \citep{yun_jellyfish_2019}. Zoom-in simulations of galaxy cluster sized systems, e.g.\@ RomulusC \citep{tremmel_introducing_2019}, offer additional opportunities to study the evolution JF galaxies in the high-resolution, cosmological setting \citep{ricarte_link_2020}. Semi-analytic models also commonly incorporate RPS effects in order to successfully model satellite galaxy evolution in dense environments \citep{somerville_semi-analytic_2008-1,lagos_shark_2018,zinger_quenching_2018,ayromlou_new_2019}  

Despite the advantages of numerical studies, it is not computationally possible to include all the relevant physical processes in simulations. For example, many simulations do not include magnetic fields although they have been suggested to be important in shaping the tails of JF galaxies \citep{tonnesen_ties_2014,ruszkowski_impact_2014,muller_highly_2021}. Large-scale cosmological simulations, such as IllustrisTNG, do not typically include physical viscosity and thus do not model gas stripping by viscous momentum transfer. For this reason, JF galaxies in cosmological galaxy simulations, which are the subject of this study, are shaped predominantly by RPS. 

To study JF galaxies within a large sample of systems, whether observed or simulated, one must first find them. The asymmetric tail pattern that is the hallmark of these objects is usually easy to identify visually. And indeed, in most observational endeavors \citep[e.g.][]{ebeling_jellyfish_2014,poggianti_gasp_2017,roman-oliveira_omegaosiris_2019,roberts_lotss_2021-2}, the classification of galaxies as JF relies on a visual inspection by the observer(s) and, even when automated identification methods are used, the results are usually visually confirmed \citep{mcpartland_jellyfish_2016,roberts_lotss_2021}. 

In an earlier project \citep[][hereafter Yun19]{yun_jellyfish_2019}, our team identified and studied a large sample of JF galaxies in the TNG100 simulation, one of the flagship runs of the IllustrisTNG project. To find JF galaxies among the many thousands of simulated satellites, five members of the research team carried out a visual classification of galaxy images depicting gas column density. The project included $\sim2,600$ inspected satellites selected from 4 snapshots of the TNG100 simulation box, ranging between \zeq{0.6} and \zeq{0}, of which about 800 were identified as JF galaxies. 

In Yun19 we used this sample to study the demographics of JF galaxies, finding that their frequency increases with host mass and decreases with satellite stellar mass. No strong dependence was detected across the redshift range. Roughly equal numbers of JF galaxies were found on both infalling and outgoing trajectories. We also found that JF galaxies exhibit higher velocities, higher Mach numbers and experience stronger ram-pressure than other satellites, showing that the physical driver of the tails in JF galaxies is indeed ram-pressure. 

In this work, we build upon and extend the study initiated in Yun19. Namely, we use the TNG50 and TNG100 simulations of the IllustrisTNG project to generate an unprecedentedly-large sample of about $90,000$ images (of more than $80,000$ galaxies) in which we search for JF galaxies. The sample spans satellite stellar masses of $10^{8.3}\--10^{11.5}\msun$, host masses of $10^{10.5}\--10^{14.5}\msun$ and extends back to \zeq{2}, exceeding currently existing observational and simulation-based studies. Furthermore, the IllustrisTNG simulations are fully cosmological, include magneto-hydrodynamics and hence magnetic fields, as well feedback models from stars as well as super massive black holes (SMBHs).
%
%
To identify JF within the gas-based images, we developed a citizen-science project: the Cosmological Jellyfish Zooniverse Project (hereafter CJF Zooniverse project). Its goal was to produce reliable identifications of JF galaxies for our $\sim 90,000$ images generated from IllustrisTNG. Our citizen-science project was launched on June 14\textsuperscript{th}, 2021 and was completed in two consecutive phases: in June 2021 (Phase 1) and between August and November 2021 (Phase 2), involving a total of more than $6,000$ volunteers. This paper collects and analyzes the classifications produced therein, presents the first scientific results based upon them, and serves as a reference guide for future scientific and outreach usages of the data i.e.\@ of the classifications, which we publicly release here.

Citizen science is a relatively recent development -- one of the first, large-scale citizen science projects, Galaxy Zoo \citep{lintott_galaxy_2008}, was founded to contend with the unprecedented large datasets from the Sloan Digital Sky Survey and the need to identify the morphology of nearly a million galaxies. Over 100,000 volunteers took part in the various classification tasks. Galaxy Zoo has been active since its launch in 2007 to this day. Volunteers have even discovered several unknown objects and novel phenomena \citep[e.g.\@][]{lintott_galaxy_2009,cardamone_galaxy_2009}. After these initial successes the Zooniverse platform\footnotemark for citizen-science projects was established  \citep{borne_zooniverse_2009,smith_zooniverse_2011,borne_zooniverse_2011}. The platform enables research scientists in a variety of fields to enlist the help of volunteers in carrying out scientific tasks.\footnotetext{\url{www.zooniverse.org}}.

For our work, the citizen-science approach to classify JF galaxies was a natural one. Their unique characteristics, i.e., asymmetrical gas tails, make them easy to identify visually, even for participants with no prior knowledge. Further, inspecting tens of thousands of galaxy images is a formidable challenge, since their sheer number (over 13 times more than in Yun19) exceeds our team's ability to inspect them all. While automated identification of tails and JF morphologies will be an important tool in the future, a citizen-science project provided a compelling way to complete the classification task. Overall, the existence of the Zooniverse framework, access to a pre-existing pool of interested volunteers, and a high level of engagement and dedication all made the process highly successful.

In this paper, we describe our Cosmological Jellyfish (CJF) Zooniverse project in detail, and present results on the demographics of JF galaxies according to the outcome of the IllustrisTNG simulations. \cref{sec:methods} describes the dataset we use, i.e.\@ galaxy images generated from the TNG50 and TNG100 simulations, and details the classification process carried out in the Zooniverse platform. In \cref{sec:scoreProcess} we describe how the volunteer classification data is synthesized to generate a galaxy `score' and compare these scores to expert classifications, validating the citizen-science results. We then set a threshold score for JF galaxy identification in \cref{sec:jf_guidlines} and suggest a statistical framework for evaluating the confidence level associated with a given threshold score. With a fully classified sample in hand, we present results about the demographics of the JF galaxy population, and their host halos in \cref{sec:results}, with an emphasis on the JF population found in low-mass hosts and at large radial distances from the host center. In \cref{sec:discuss} we discuss the validity of our classification method, and also the public-outreach value of or project. Finally we summarize our findings in \cref{sec:summary}.

 
\section{The Cosmological Jellyfish Zooniverse Project}\label{sec:methods} 

\subsection{The TNG50 and TNG100 cosmological galaxy simulations}
\label{sec:sims}

Throughout this paper and to construct the sample of satellite galaxies to be visually inspected, we make use of two simulations from the IllustrisTNG project\footnotemark, a suite of magneto-hydrodynamic cosmological simulations carried out in three volumes of varying size and resolution. Namely, we use the flagship runs called TNG100 \citep{marinacci_first_2018,naiman_first_2018,nelson_first_2018,pillepich_first_2018,springel_first_2018} and TNG50 \citep{pillepich_first_2019, nelson_first_2019}. All IllustrisTNG simulations were run with the AREPO code \citep{springel_e_2010,pakmor_simulations_2013} and are based on a physical model of galaxy formation \citep[the IllustrisTNG model,][]{weinberger_simulating_2017,pillepich_simulating_2018} that has been shown to reproduce, with a reasonable level of accuracy, a large number of observational properties of galaxies. \footnotetext{\url{www.tng-project.org}} 

TNG50 and TNG100 evolve cubic volumes of comoving side length of roughly $50\units{Mpc}$ and $100\units{Mpc}$, respectively. By combining both these simulations, we obtain a large statistical sample of satellite galaxies residing in a large variety of hosts -- from Milky-Way sized halos through groups and clusters of galaxies (up to $\tenMass{14.6}$ in total mass). The higher-resolution TNG50 box allows us to probe the lower-mass satellites $(\gtrsim \tenMass{8.3})$ regime, as well as providing a more detailed outcome. Galaxies are resolved with baryonic mass resolution of $8.5\times\tenMass{4}$ and $1.4\times\tenMass{6}$ in TNG50 and TNG100, respectively. The full simulation data of TNG50 and TNG100 are publicly available \citep[see][for details]{nelson_illustristng_2019}. 

The cosmological framework for these simulations is a $\Lambda$CDM cosmological model with the parameter values based on the \citet{planck_collaboration_planck_2016} data: cosmological constant \mbox{$\Omega_\mathrm{\Lambda}= 0.6911$}, matter density \mbox{$\Omega_\mathrm{m}=\Omega_\mathrm{dm}+\Omega_\mathrm{b}=0.3089$}, with a baryonic density of \mbox{$\Omega_\mathrm{b}= 0.0486$}, Hubble parameter $h = 0.6774$, normalisation \mbox{$\sigma_\mathrm{8} = 0.8159$}, and spectral index \mbox{$n_\mathrm{s} = 0.9667$}.

\subsection{Selection of TNG50 and TNG100 satellite galaxies inspected in the CJF project}\label{sec:sample} 

Among the many tens of thousands of satellite galaxies simulated within TNG50 and TNG100 across cosmic epochs, a sample (denoted from now on as ``inspected'') was selected for visual inspection from the \textsc{subfind} subhalo catalogues at various snapshots. The selection was based on the galaxy stellar mass and gas fraction, as well as central or satellite status within their host Friends-of-Friends (FoF) halo. Unless otherwise specified, all stellar and gas masses are measured within a radius $\rgal$, which is set to be twice the radius that encloses half of the stellar mass in a given subhalo ($ \rgal \equiv 2r_{1/2,\mathrm{stars}}$). Furthermore, we exclude subhalos that are likely to be clumps of matter and unlikely to be galaxies of cosmological origin, based on the SubhaloFlag defined in \cite{nelson_illustristng_2019}. In particular, we applied the following criteria to select our galaxy sample:
\begin{itemize}
    \item Only satellites, i.e.\@ excluding central galaxies.
    \item Satellite stellar mass: $\mathrm{M}_{\ast}>10^{8.3}\msun$ in the case of TNG50, and $\mathrm{M}_{\ast}>10^{9.5}\msun$ in the case of TNG100.
    \item Gas fraction: $f_{\mathrm{gas}}>0.01$, with $f_{\mathrm{gas}}$ defined either as $M_{\mathrm{gas, r<\rgal}}/M_{\mathrm{\ast, r<\rgal}}$ (Phase 1), or $M_{\mathrm{gas, all}}/M_{\mathrm{\ast, r<\rgal}}$ (Phase 2).
\end{itemize}

In Phase 1, galaxies were selected from the following snapshots:
\begin{enumerate}
  \item TNG50 \& TNG100: Full snapshots up to redshift $z=2$ i.e.\@ snapshots 99, 91, 84, 78, 72, 67, 59, 50, 40, and 33. The time interval between pairs of these snapshots is $\sim 1 \units{Gyr}$. 
\end{enumerate}

\noindent while in Phase 2 the snapshot selection was:

\begin{enumerate}
  \item TNG50 \& TNG100: Full snapshots up to redshift $z=2$, excluding galaxies already considered in Phase 1. These galaxies are those with $f_{\mathrm{gas,2\rhStar}}<0.01$ but $f_{\mathrm{gas,all}}>0.01$, making the samples complete down to the lower of the two gas fraction criteria.
  \item TNG50: All snapshots up to redshift 0.5 i.e.\@ snapshots 68-98 (inclusive), excluding those already considered in Phase 1. The time interval between the snapshots is $\sim 150 \units{Myr}$.
\end{enumerate}

Furthermore, in Phase 2 we also included, as a test, an additional $2,918$ images from TNG50 and $5,844$ images from TNG100 of the same galaxies selected as above solely from snapshots 99 and 67 (i.e.\@ at $z=0$ and $z=0.5$) and shown in a preferred, rather than random, orientation: these will be described and studied in \cref{sec:viewAngleComp}. 

Note that no selection was imposed a priori on the mass of the underlying dark-matter host halos. Furthermore, 20 galaxies were removed from the Phase 1 inspected sample, as their images were deemed to be too messy for classification. These galaxies were inspected by the expert team members and manually assigned a score of 0, as described in later Sections. Importantly, the galaxies selected for inspection are biased towards larger gas fractions than the randomly-selected satellite in groups and clusters at any given time and stellar mass: this is because they need to have at least some gas, otherwise they could not exhibit tails of stripped gas and could not be identified as jellyfish. 

\begin{figure}
  \centering
   \includegraphics[width=8.5cm,keepaspectratio]{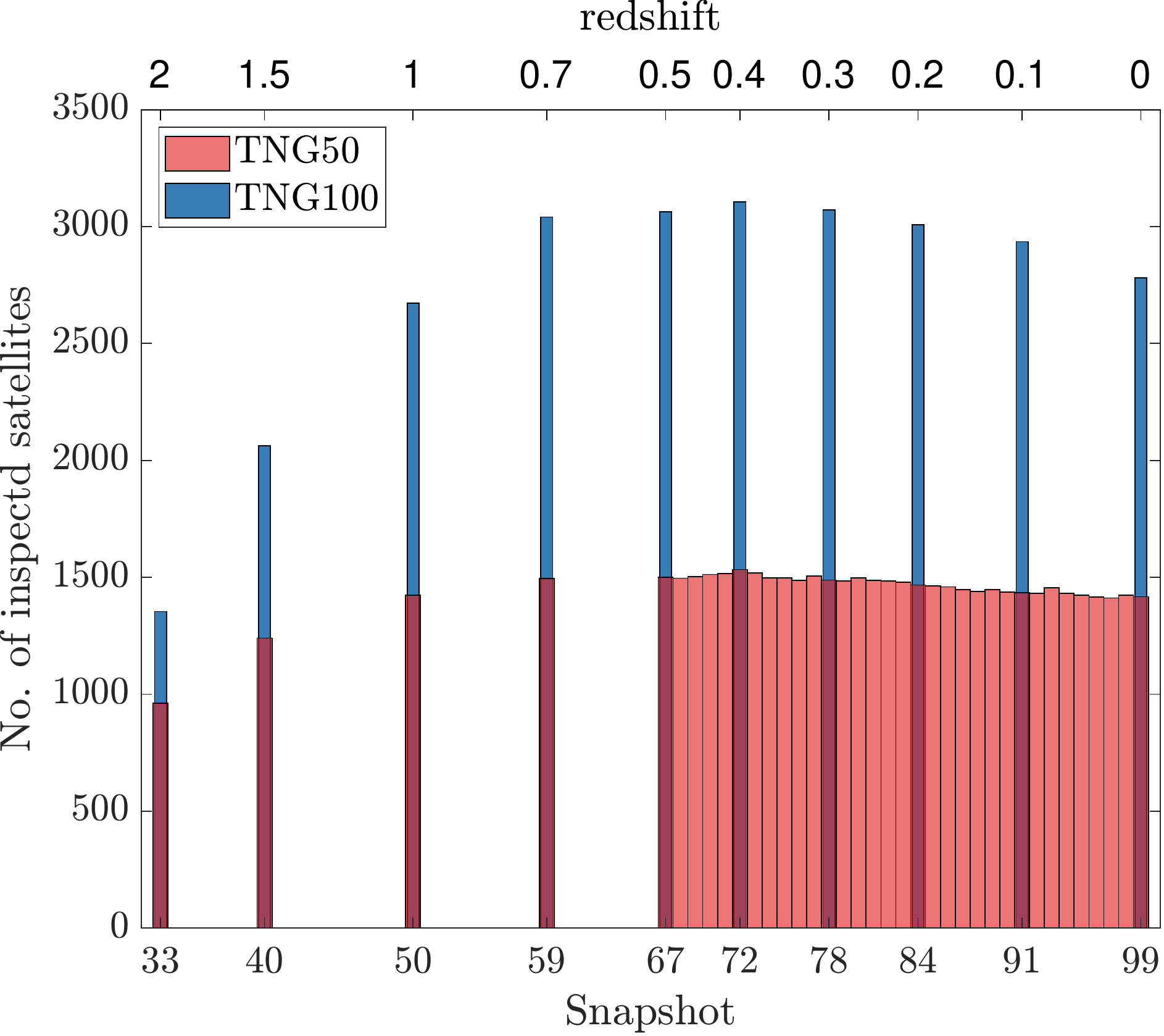}     
   \caption{Number of satellite galaxies selected for the visual inspection in our Cosmological Jellyfish (CJF) Zooniverse project from the TNG50 (red) and TNG100 (blue) simulations, at each selected snapshot. The corresponding redshift values are shown in the upper axis. We deliberately chose to inspect TNG50 satellites at $z\le0.5$ as frequently as allowed by the available output data, i.e.\@ approximately every 150 Myr.}
  \label{fig:sampleNzred}
\end{figure}

In total, we selected for the visual inspection on Zooniverse $13,592$ and $24,394$ galaxies from TNG50 and TNG100, respectively, in Phase 1 and a further $40,001$ and $2,697$ galaxies in Phase 2, respectively. The total number of galaxies inspected during the two phases of our CJF project is $80,704$ (including the 20 `messy' galaxies), and they reside within $30,169$ hosts. We compared these numbers to the total number of satellite galaxies in the TNG50 and TNG100 simulations, above the respective mass thresholds of $\tenMass{8.3}$ and $\tenMass{9.5}$, and their respective hosts. We find that the CJF inspected sample comprises \perc{\sim 90} of all relevant satellites at high redshifts (\zeq{2}), with this fraction dropping at low redshifts to \perc{\sim 60} for TNG100 and \perc{\sim 50} for TNG50. The host halos in the CJF sample comprise between \perc{96} (low redshifts) to \perc{99} at \zeq{2} of all relevant hosts. 

The total number of images put up for inspection was $89,446$, which includes the additional $8,762$ images of galaxies repeated from the preferred viewpoint. In all that follows, unless specified otherwise, the tables, figures and results, all describe the inspected sample of $80,704$ galaxies, with the identification of these galaxies as JF or not relying on images generated from a random viewing angle. 

In \cref{fig:sampleNzred} we show the number of objects in each snapshot/redshift from the TNG50 and TNG100 simulations. In \cref{tab:sampleNumbers} we list the number of inspected objects, in particular of hosts and satellites in the TNG50 and TNG100 samples divided into three relevant redshift ranges. Since our total sample of inspected galaxies includes all the snapshots of TNG50 at $z\le 0.5$, we can see that TNG50 objects from this timeframe account for \perc{\sim 58} of the satellites and  \perc{\sim 55} of the hosts, even though TNG50 simulates a much smaller volume than TNG100. 

\begin{figure*}
  \centering
  \subfloat[] {\label{fig:stellarMass_hist}
  \includegraphics[width=8cm,keepaspectratio]{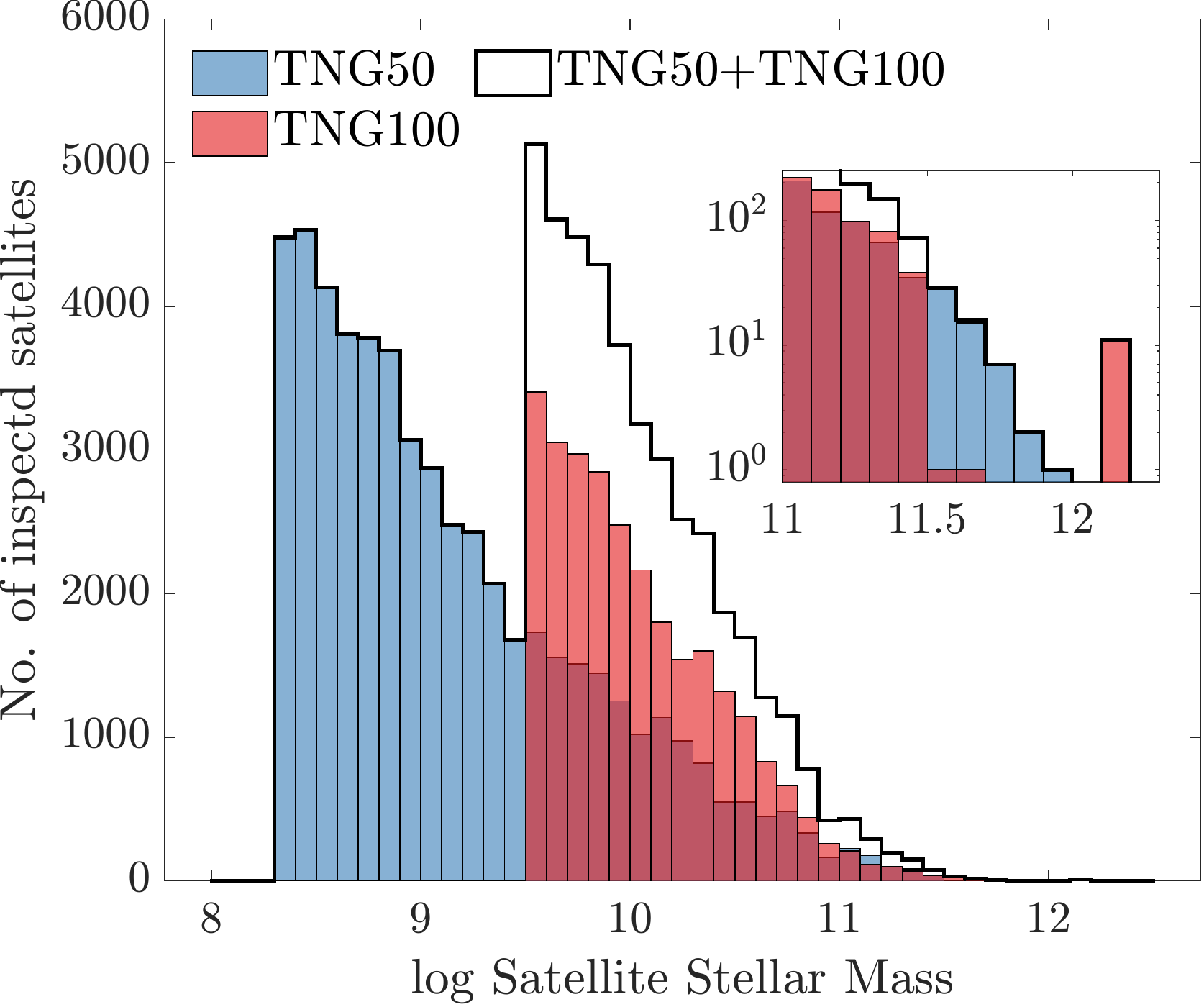}}
  \subfloat[] {\label{fig:hostMass_hist}
  \includegraphics[width=8cm,keepaspectratio]{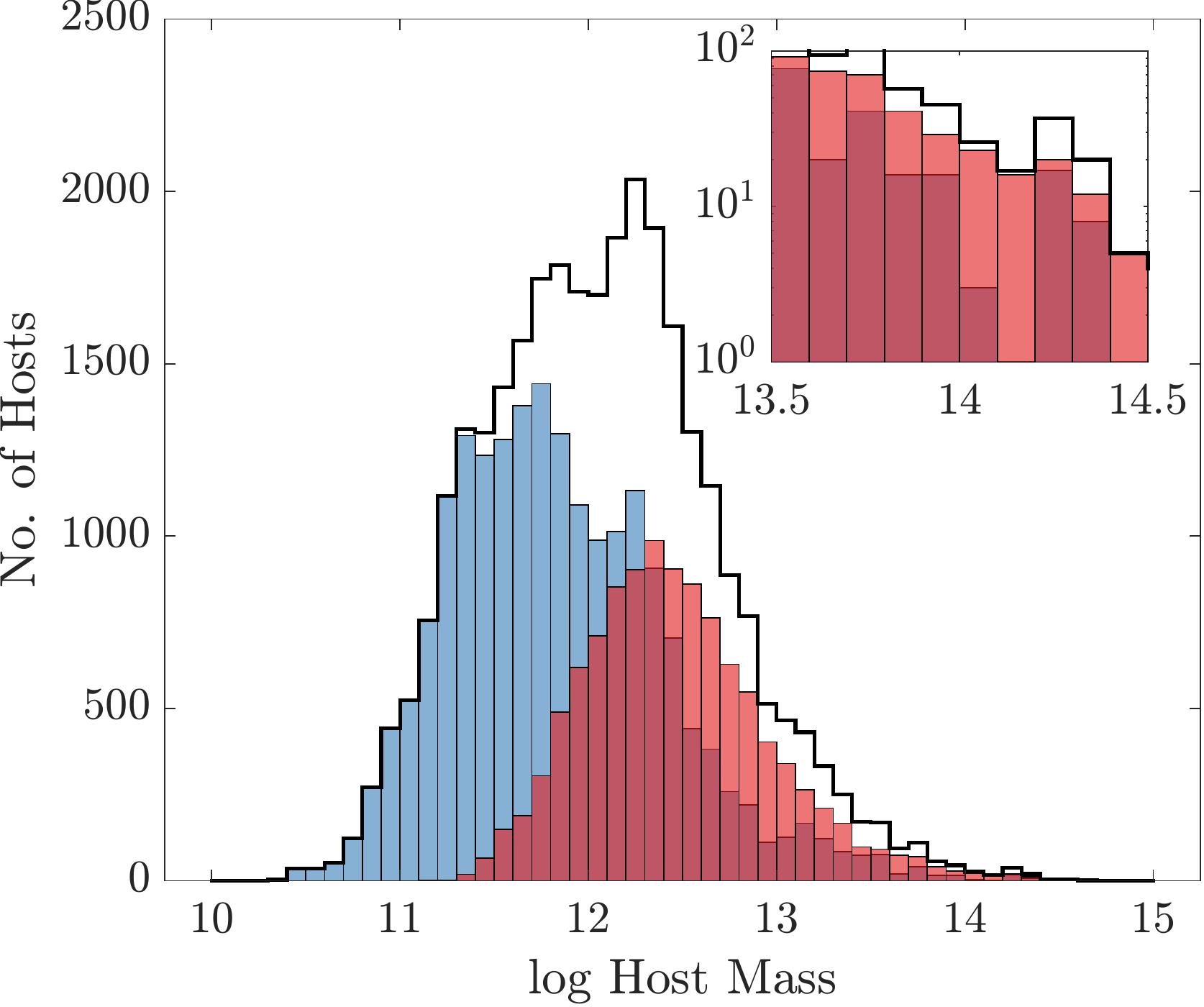}} \\
  \subfloat[] {\label{fig:massRatio_hist}
  \includegraphics[width=8cm,keepaspectratio]{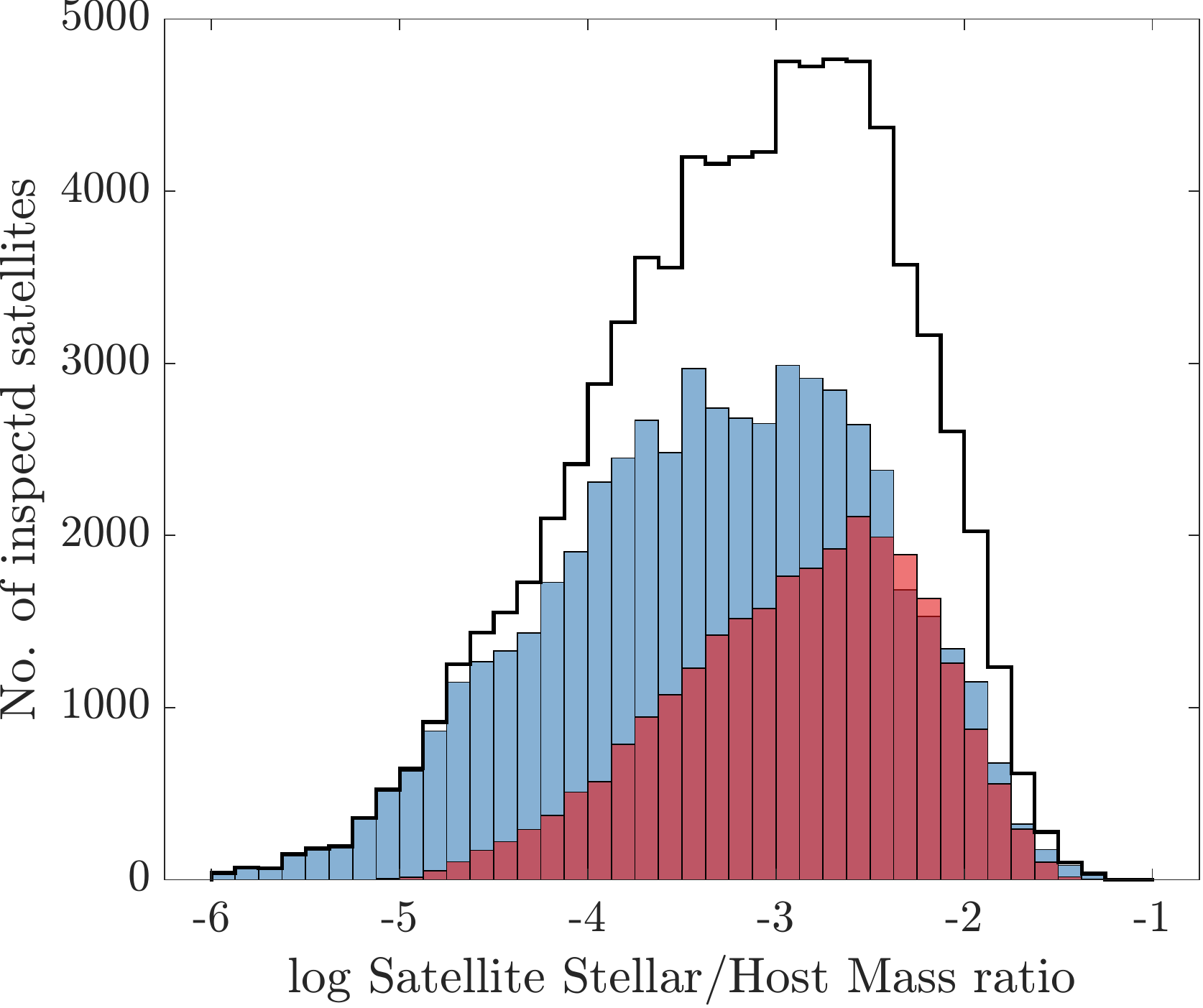}}
  \subfloat[] {\label{fig:satNum_hist}
  \includegraphics[width=8cm,keepaspectratio]{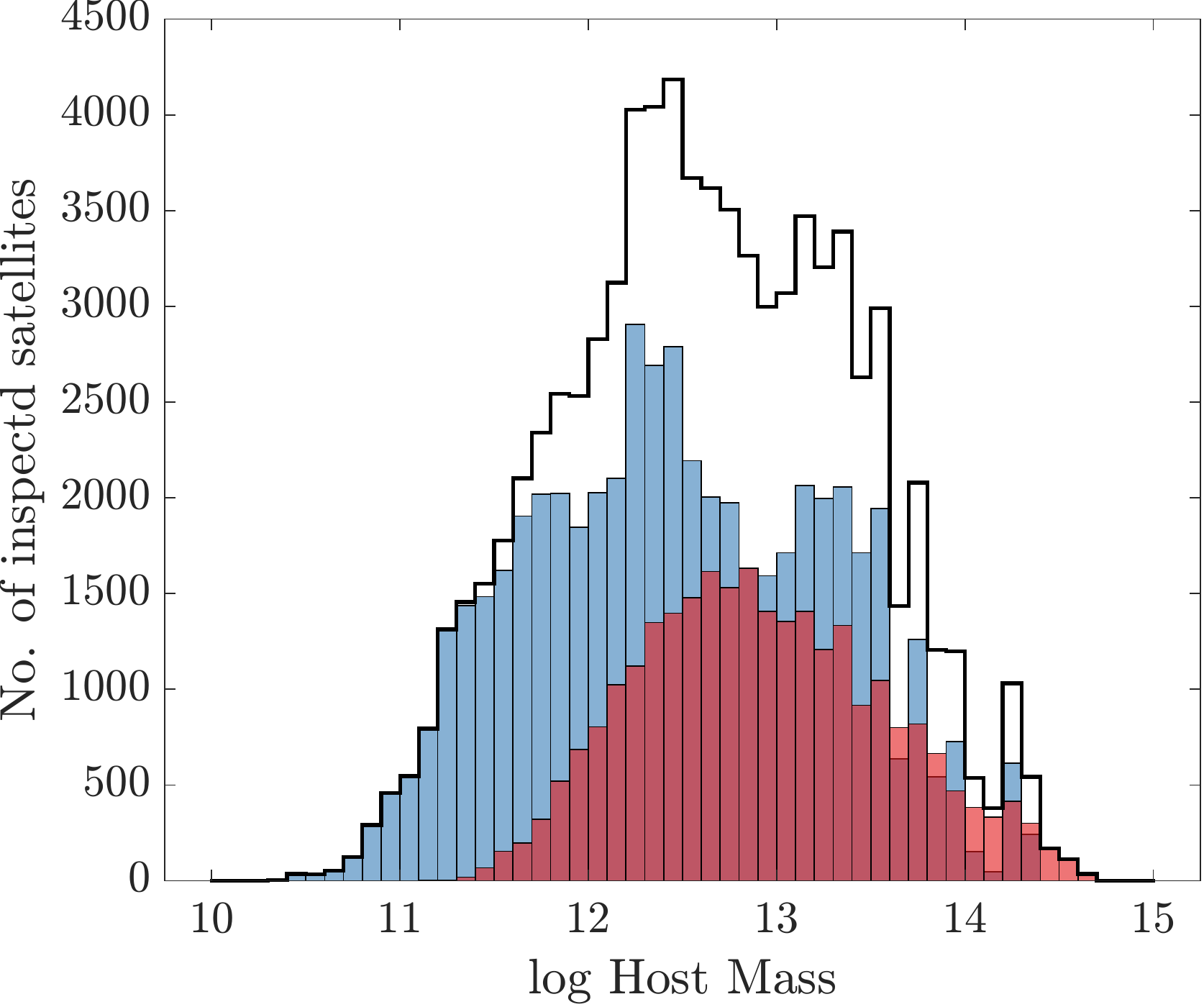}} 
  \caption{Demographics of the satellite galaxies selected for visual inspection in our CJF Zooniverse project. The panels show the distribution of satellite stellar mass \subrfig{stellarMass_hist}, host mass $\Mv$ \subrfig{hostMass_hist}, the satellite stellar to host mass ratio \subrfig{massRatio_hist} and the total number of satellites in hosts of a given mass \subrfig{satNum_hist} across all selected snapshots ($0\le z\le2$, see text for details). The distributions for the satellites selected from the TNG50 and TNG100 simulations, and thus visually inspected, are shown separately in blue and red histograms, respectively, with the black line showing the combined population from both simulation.  In panels \subrfig{stellarMass_hist} and \subrfig{hostMass_hist}, the inset shows the distribution at the high-mass end.} 
  \label{fig:sample_demograf}
\end{figure*}

\begin{figure*}
  \centering
  \subfloat[No. of Satellites in hosts - TNG50] {\label{fig:satNumByHost50}
  \includegraphics[width=8.5cm,keepaspectratio]{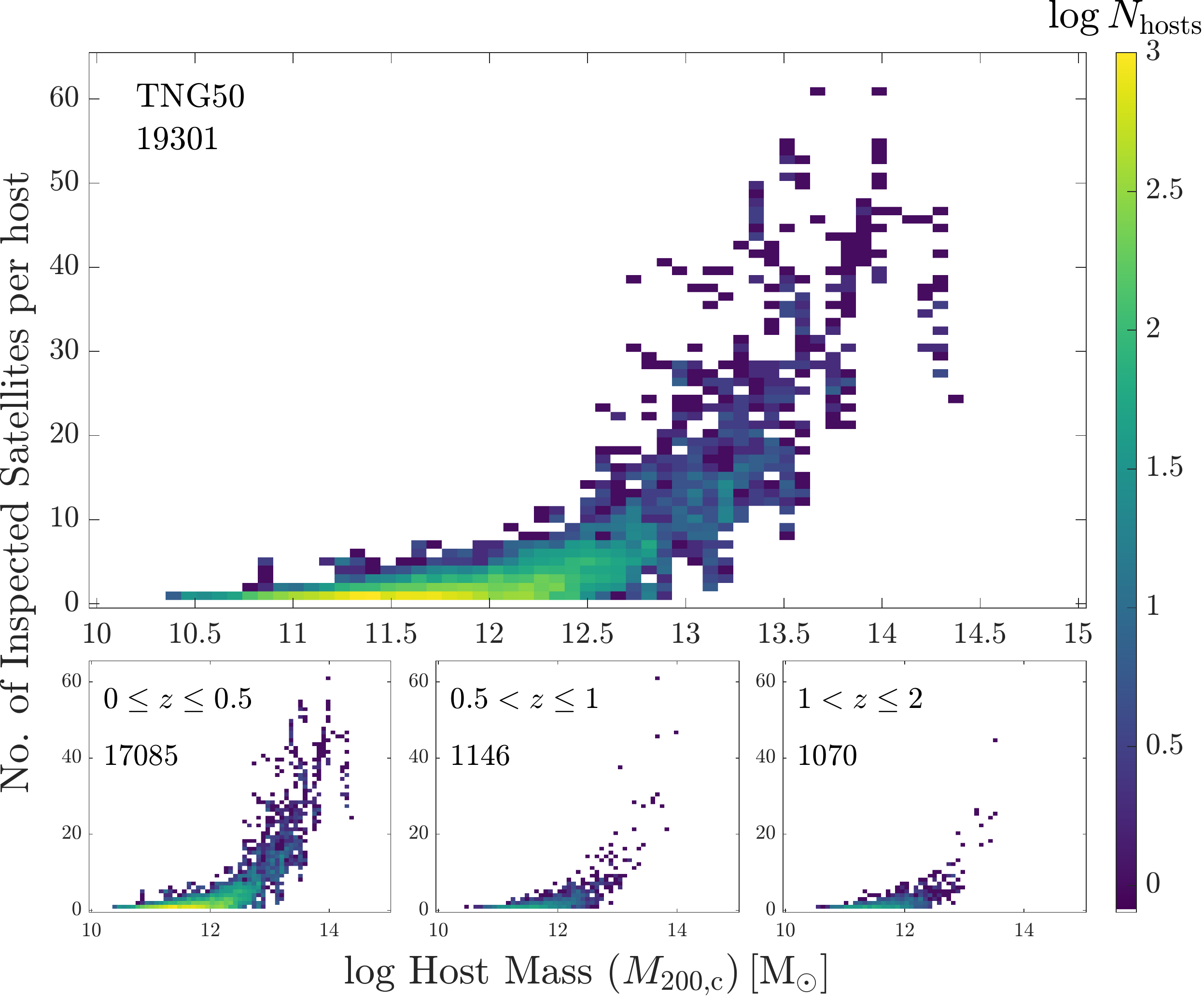}}
  \subfloat[No. of Satellites in hosts - TNG100] {\label{fig:satNumByHost100}
  \includegraphics[width=8.5cm,keepaspectratio]{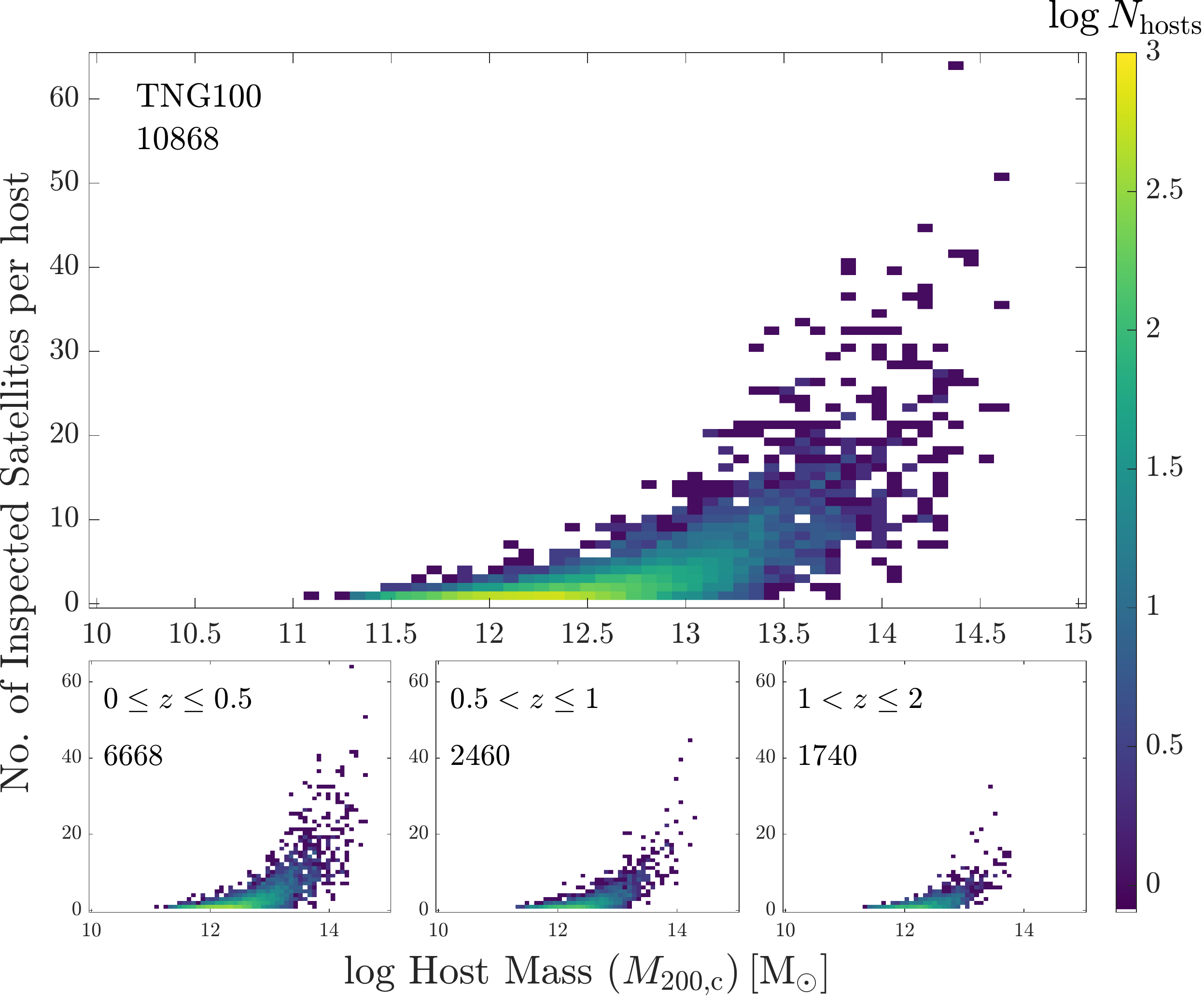} }
  \caption{The number of inspected satellites that reside in individual hosts, as a function of host mass. The color-map corresponds to the number of hosts in each bin. The host population of TNG50 (TNG100) is shown in panel \subrfig{satNumByHost50} (\subrfig{satNumByHost100}). The number of simulated hosts inspected within our CJF Zooniverse project is shown in the top left corner of each panel. The three smaller panels on the bottom show the inspected host sample in three redshift bins.}
  \label{fig:sample_hostDemograf}
\end{figure*}

In \cref{fig:sample_demograf} we show the basic demographics of the TNG50 and TNG100 selected and thus inspected objects, over all selected redshifts. In \cref{fig:stellarMass_hist} we show the distribution of the satellite stellar masses of the galaxies in the inspected sample, whereby the imposed lower mass limits are clearly evident. The high stellar mass end is shown in detail in the inset: we can study satellites up to a a few $\tenMass{11}$ in stars. Even though the TNG100 volume is roughly 8 times larger, there are more galaxies in the TNG50 inspected sample, mainly due to there being more snapshots included from the TNG50 simulation (see \cref{fig:sampleNzred}). 

\cref{fig:hostMass_hist} quantifies the mass distribution of the hosts in which the galaxies reside at the time of inspection. The host mass shown is $\Mv$ of the host), i.e., the mass enclosed within a radius $\Rv$, which in turn is defined as the radius where the mean density of the halo is equal to 200 times the critical density of the universe at that time. The TNG50 inspected sample allows us to probe low-mass satellite galaxies and, by extension, low-mass hosts. The TNG100 inspected sample supplies us with a very large number of group- and cluster-sized hosts. In \cref{fig:massRatio_hist} we show the distribution of the satellite to host mass ratio, where again we see than TNG50 allows us to explore a wide range of satellite-host interactions. 

Finally, the total number of satellites found and inspected within all hosts in a given mass range is shown in \cref{fig:satNum_hist}. Due to the volume of TNG100, there are many satellites in high-mass hosts. Overall, the $80,704$ satellite galaxies in the CJF Zooniverse encompass a large range of satellite stellar masses, host halo masses and redshift, namely: $\tenMass{8.3\--12.3}$, $\Mv=\tenMass{10.4\--14.6}$, and $z\le z \le 2$, respectively. As we expand upon in \cref{sec:branches} and in \textcolor{blue}{Rohr et al 2023}, it is important to note that not all satellites in the CJF project are unique, in that many of them represent different evolutionary stages of the same galaxy selected and inspected at different redshifts.

In \cref{fig:sample_hostDemograf} we show the number of inspected satellites \emph{that reside in individual hosts} of a given mass. The distribution of hosts is shown by a color-map that signifies the  number of inspected hosts within a given host mass vs.\@ satellite number bin. The  {\it total} number of hosts inspected throughout the project is given in the top-left corner (see also \cref{tab:sampleNumbers}). The distributions from TNG50 and TNG100 are shown separately in \cref{fig:satNumByHost50,fig:satNumByHost100}, with a breakdown of the host population into three redshift bins shown in the three smaller panels. As expected, the number of hosts, and the number of satellites within them, grows towards lower redshift.

Because the galaxy and halo populations from large-volume cosmological simulations like IllustrisTNG are volume limited by construction, the majority of hosts are found in the low mass range, with mass \mbox{$\Mv\lesssim \tenMass{13}$}(as already manifest in \cref{fig:sample_demograf}): in such hosts, the typical number of inspected satellites is of order of a few and very rarely above 10. The higher resolution and lower mass limit afforded by TNG50 result in slightly higher numbers of satellites per host. The larger volume of TNG100 results in many more hosts at masses of \mbox{$\Mv\gtrsim \tenMass{13}$}.

\begin{table}
\begin{center}
\begin{tabular}{@{}rrrrrr@{}}
   \multicolumn{1}{l}{}  & \multicolumn{2}{c}{TNG50}   &\multicolumn{2}{c}{TNG100}    \\ 
 \cmidrule(lr){2-3} \cmidrule(l){4-5} 
 \multicolumn{1}{c}{Redshift} & Hosts & Satellites & Hosts &Satellites  \\ 
 \cmidrule(l){1-1}  \cmidrule(l){2-3} \cmidrule(l){4-5} 
 $0\leq z \leq 0.5$     &   17085    & 48491    & 6668 & 17964          \\
 $0.5< z \leq1$         &   1146    & 2918     & 2460 & 5714          \\
 $1< z \leq2$           &   1070     & 2201     & 1740 & 3416          \\ 
\cmidrule(l){1-1}  \cmidrule(lr){2-3} \cmidrule(l){4-5} 
\multicolumn{1}{r}{Total}  & 19301 & 53610 &10868 &27094        \\ 
 \end{tabular}
\caption{Number of hosts and satellites from the TNG50 and TNG100 simulations proposed for visual inspection in our CJF Zooniverse project, in three redshift bins. In total, $80,704$ satellites were classified based on randomly oriented images.} 
\label{tab:sampleNumbers}
\end{center}
\end{table}

\subsection{Visual classification \& identification of jellyfish}\label{sec:visClass} 

As discussed above, the identification of JF galaxies was carried out through visual inspection on the Zooniverse platform for citizen science.\footnotemark The CJF project considered images of all the galaxies in the sample defined in \cref{sec:sample}. A total of $6,494$ volunteers participated in the classification effort, with a total of 1.8 million classifications. Here we describe the proposed tasks, the workflow, and the characteristics of the inspected images. \footnotetext{\url{www.zooniverse.org/projects/apillepich/cosmological-jellyfish}}

In our CJF Zooniverse project, after a short training process, volunteers were shown one galaxy image at a time and were asked to classify the galaxy in the center of the image by answering a simple yes/no question: "Do you think that the galaxy at the center looks like a jellyfish?". Once an image was classified by 20 different people, it was retired from the image pool. For context, the original Galaxy Zoo project relied on $\sim 38$ inspectors per image \citep{lintott_galaxy_2008}. As detailed below, the training included a number of examples and guidelines, under the form of a Tutorial and a Field Guide, available to the volunteers at any step of the classification.

\begin{figure*}
  \centering
       \includegraphics[width=\textwidth]{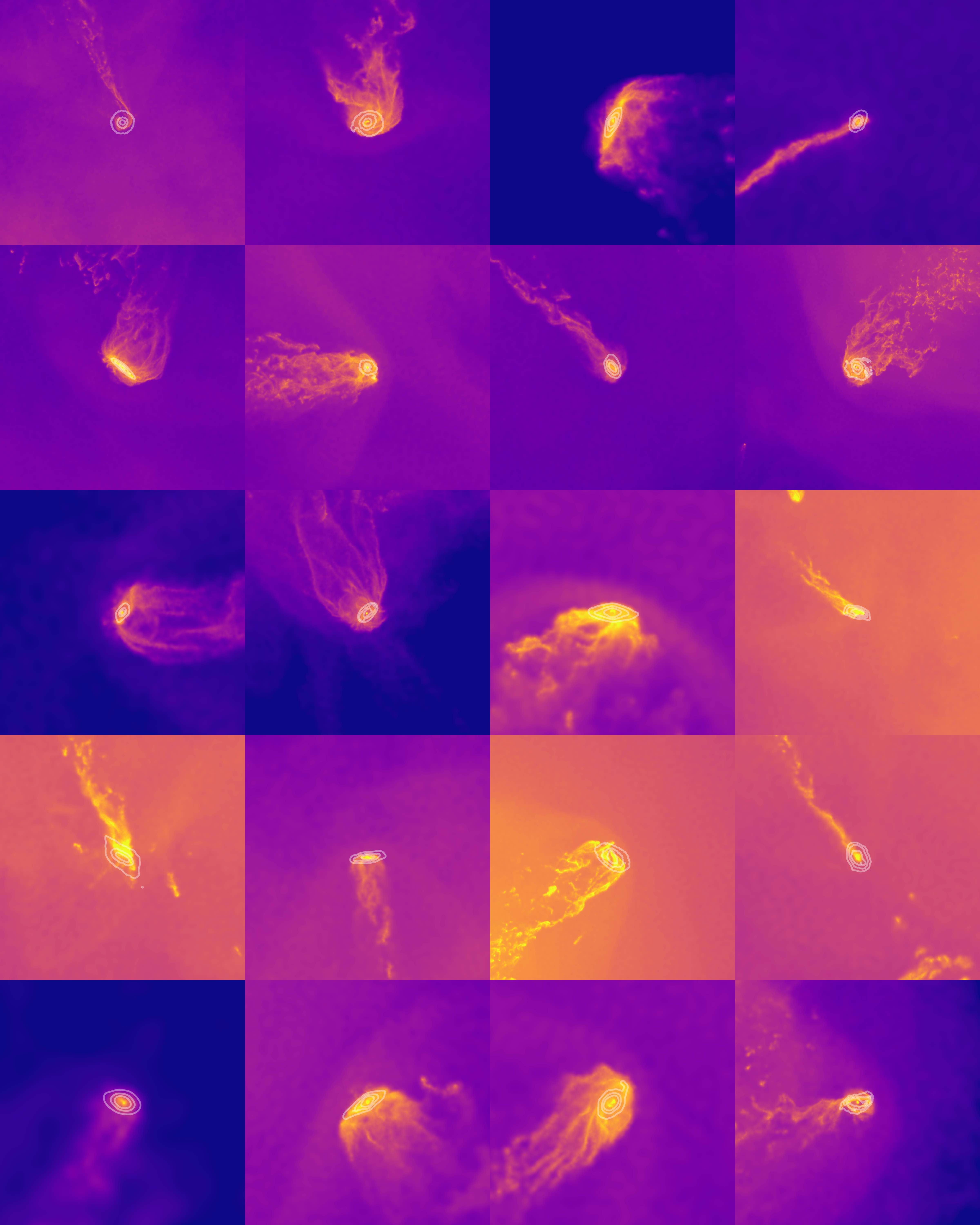}
       \caption{Examples of jellyfish galaxies identified via the ``Cosmological Jellyfish'' (CJF) Zooniverse project, including galaxies from both the TNG50 and TNG100 simulations at various redshifts. These are the exact stamps shown to the volunteer inspectors, depicting gas column density as color map and stellar mass density as white transparent contours. The gas density ranges between $10^5$ (blue) and $10^8$ (yellow) $\msun$ kpc$^{-2}$, the stamps have a side length of $40\times$ the stellar half-mass radius of the galaxy. All jellyfish galaxies identified in this paper can be seen at \url{www.tng-project.org/explore/gallery/zinger23/}. The procedure we use to identify JF galaxies based on the visual classifications is described in \cref{sec:scoreProcess,sec:jf_guidlines}.}
  \label{fig:jf_mosaic}
\end{figure*}

\subsubsection{Image specifications}
\label{sec:visClassImage} 

The images generated for the classifications consist of maps of two main matter components: i) gas mass surface density shown as a coloured 2D histogram and ii) stellar mass surface density shown as white contours overlaid on the gas map. For these, we followed the methodology and visualization technique of \citet{nelson_illustristng_2019}.

The map of the gas distribution was generated by measuring the (log of the) mass projected on an $800\times800$ pixel grid of side length $40\times\rhStar$, centred on the given galaxy. All of the gas belonging to the host halo was included, but restricted to a cube of side length $40\times\rhStar$. The color range corresponded to $10^{5}-10^{8}\msun/\mathrm{kpc}^{2}$ in column density (colorbars not shown to classifiers). 

Stellar contours were then overlaid on the gas maps to allow the classifier to determine the extent of the stellar mass in the galaxy under consideration as well as to indicate the presence of other galaxies in the image. In order to generate these contours, we selected all galaxies within the field of view down to a mass limit 0.5 dex lower than that used for the actual inspected sample i.e.\@ $10^{7.8}\msun$ for TNG50 and $10^{9}\msun$ for TNG100. This lower limit was used to ensure that we captured most galaxies that would be evident in the images. As with the gas maps, we first obtained a 2D histogram of the (log) stellar mass -- all gravitationally-bound stellar mass to each galaxy --  projected on the same grid as that used for the gas maps, but \emph{separately for each galaxy in the field of view}. Contours were then generated based on these 2D histograms, corresponding to 75, 80 and \perc{85} of the log of the peak mass surface density for the given subhalo. These levels were determined manually and after inspection of many systems in order to adequately capture the extent of the stellar mass distribution of each galaxy. Finally, the images were saved with a resolution of 300 dpi.\footnote{For a small number of cases where the image file size exceeded the platform limit the image was saved at a resolution of 100 dpi. We confirmed that this made no visual difference.} Examples of the images as they appeared on the website can be seen in \cref{fig:jf_mosaic,fig:score_mosaic} and in \cref{sec:app_imageSample}.

The simulated galaxies were projected along random orientations, i.e.\@ from an arbitrary view point irrespective of e.g.\@ the orientation of the stellar disk or the galaxy location within the host halo. To do so we projected the mass distribution of each galaxy along the z-axis of the simulated volume. A subset of galaxies (\zeq{0,0.5} snapshots from TNG50 and TNG100) were also projected along an orientation optimized for tail identification, for the purpose of studying the effect of viewing angle on classification. We discuss this study in \cref{sec:viewAngleComp}.

\begin{figure}
  \centering
   \includegraphics[width=8.0cm,keepaspectratio]{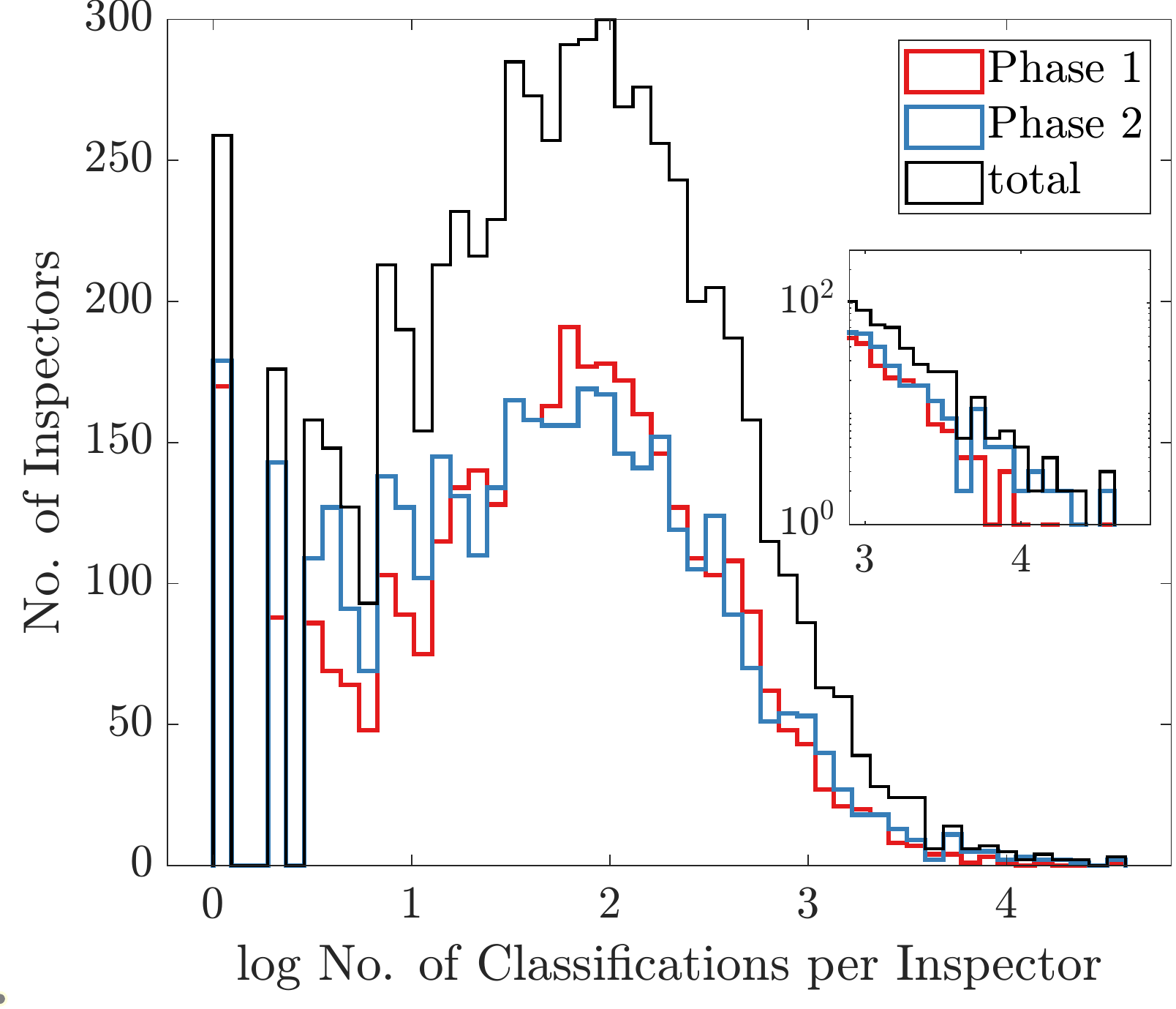}
   \caption{Distribution of the number of inspectors who classified a certain number of images in our CJF Zooniverse project, for the two phases (red and blue lines, respectively) and for the entire inspected sample (black line).}
  \label{fig:classificationHistogram}
\end{figure}

\subsubsection{Visual classification process}\label{sec:visClassProcess} 

The classification process on the Zooniverse platform was open to anyone, with no requirements on previous training or experience.  We provided ample background information, as well as an extensive training guide. New volunteers were shown this training guide upon their first entry to the site. In the training guide, a JF galaxy was described as a ``galaxy (that) exhibits one or more `tails' of gas that stem from the main galaxy body and stretch in one preferred direction. Such galaxies almost look like the jellyfish in the sea!''. The volunteers were asked to focus on the gas features, shown in a density color map (see \cref{sec:visClassImage}), in relation to the stellar body of the galaxy, shown by thick white contours (see e.g.\@ \cref{fig:jf_mosaic}).

Several examples of what we considered to be clear-cut JF galaxies were shown, but most of the training guide was to point out, via examples, what a JF galaxy is \emph{not}: galaxies with no gas tails, galaxies with very little gas, galaxies with gas tails pointing in many different i.e.\@ opposite directions, galaxies with tails that do not appear to connect to the galaxy and galaxies with very messy gaseous surroundings. In addition, volunteers were asked to treat galaxies with nearby neighbors as non-JF, \emph{even if they exhibited gas tails}, to avoid confusion with merger events.  

In addition, we added a training set of images that were previously classified by our research team, either as a part of the Yun19 project or as part of a pilot project that was carried out within the team prior to the public release of the official CJF Zooniverse project (see \cref{sec:visClassCompare} for more details). Volunteers were shown images from the training set sporadically, and asked to classify them without being aware that they are from the training set: upon completion of the associated task, they were then notified whether or not their yes/no choice matched the expert one, receiving an immediate feedback on their classification. 

Finally, volunteers could discuss specific cases on a public forum, and often sought (and received) assistance in the classification process from the members of the research team.

In \cref{fig:classificationHistogram} we show the distributions of classifications per inspector, quantifying how many volunteers classified a certain number of images, for the two project phases separately as well as for the entire inspected sample. As described above, the classification process was carried out in two phases: Phase 1 included $37,986$ galaxies, produced $759,720$ classifications (20 classifications per image), while Phase 2 included $51,460$ images and resulted in $1,029,200$ classifications. In summary:

\begin{itemize}
 \item $6,494$ inspectors performed a total of $1,795,292$ classifications -- a few objects actually have more than 20 classifications;
\item \perc{4} of the inspectors only classified a single object (of them, \perc{60} were not logged on);
 \item \perc{19} of the inspectors classified fewer than 10 objects;
 \item the two phases exhibit very similar distributions of participation;
 \item the average number of classifications per inspector is 276 overall; the median number of classifications per inspector is 45 overall;
\item \perc{19} of all classifications came from anonymous users who did not register on the Zooniverse website or registered volunteers who did not log-on before beginning to classify;
\item \perc{5} among the inspectors \emph{each} classified more than 1,000 images, being responsible for \perc{56} of all classifications;
\item of these, \perc{0.67} (44 persons (17), including a few of us, each classified more than $5,000$ ($10,000$) objects, being responsible for about \perc{27} of all classifications;
\end{itemize}

Whereas most inspectors classified several tens or even a few hundred objects each, a small number of very dedicated inspectors (about \perc{5}) are responsible for more than half of all classifications. In light of this, we assess the quality of individual inspectors and weigh the scores accordingly, as we detail in \cref{sec:InspectorWeighting}. 

\section{Assessment of the outcome of the CJF Zooniverse classification}\label{sec:scoreProcess}

\subsection{Raw jellyfish scores}\label{sec:rawScore}

At the end of the classification process, all the classifications of a given object were tallied and a final score between 0 and 20 was assigned to each galaxy. A small percentage of objects (\perc{\sim 2} in phase 1 and \perc{\sim 1.16} in phase 2) received more than 20 classifications due to technical issues. In order to generate a standardized score between 0 and 20 for each of these objects, we created 200 random sub-samples of 20 votes (out of the entire classification pool for that object) and assigned the median value over these 200 scores as the final score for that object. The scores are then normalized to values between 0 and 1.

\begin{figure}
  \centering
   \includegraphics[width=8.0cm,keepaspectratio]{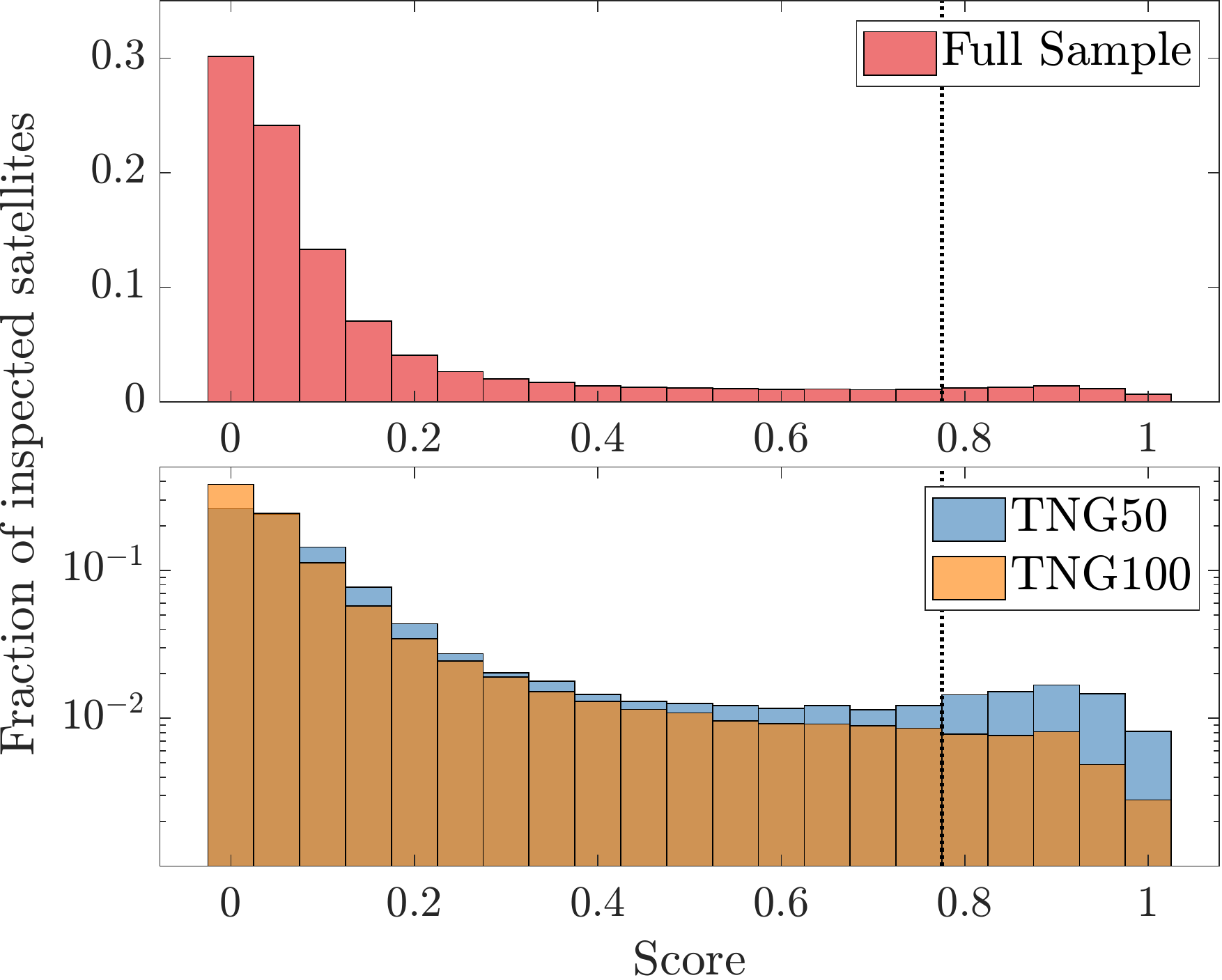}           
   \caption{Distribution of the raw jellyfish scores from our CJF Zooniverse project after about 1.8 million classifications. We show results for the entire inspected sample in the top panel, and separately for the TNG50 (blue) and TNG100 (orange) simulated inspected galaxies in the bottom panel. The \mbox{y-axis} denotes the fraction of the sample population -- note that the bottom panel is shown in log scale. The vertical dotted line denotes the threshold chosen in this work (0.8 in a range from 0 to 1) above which we define a galaxy to be a jellyfish.}
  \label{fig:scoreHistogram}
\end{figure}

In \cref{fig:scoreHistogram} we show the distribution of these raw scores of our inspected galaxy sample, both for the entire inspected sample and also for the TNG50 and TNG100 samples separately. The dashed vertical lines denote the score threshold we choose in this study to define jellyfish galaxies (see also \cref{sec:jfDef}): galaxies with a raw score of 0.8 or higher are dubbed JF or, in other words, 16 of 20 inspectors deemed a given galaxy to be a JF. 

By comparing the results from the TNG50 and TNG100 simulations, we find that the TNG50 sample is skewed towards higher scores. This may be due to the higher resolution of the TNG50 simulation run: the features that lead to a classification of JF are more pronounced at higher resolution. However, differences in the two populations exist and may be responsible for the different results without being directly connected to the underlying numerical resolution: the TNG50 sample is dominated by lower-mass and lower-redshift objects (\cref{tab:sampleNumbers}), which are more likely to be JF galaxies (see e.g.\@ Yun19 and the next Sections).

\cref{fig:jf_mosaic} shows 20 random examples of galaxy images identified as jellyfish galaxies, i.e., with a score of 0.8 and above. 
 
\subsection{Comparison to previous classification projects}\label{sec:visClassCompare} 

To assess the public classifications of the CJF project, we compared them with classifications completed by a team of experts. This allows us to asses the extent to which the classifications from the general public align with our understanding of what comprises a JF. We compare results for a subset of objects that have also been independently classified by members of our research team. In particular, we make two comparisons: 1) against a subset of TNG50 objects classified in a pilot project completed  \emph{prior} to the public opening of the CJF Zooniverse website and 2) against the galaxies inspected and studied in Yun19.

\begin{figure*}
  \centering
  \subfloat[Score comparison with the TNG50 pilot project] {\label{fig:tng50Comp_scores}
  \includegraphics[width=8.5cm,keepaspectratio]{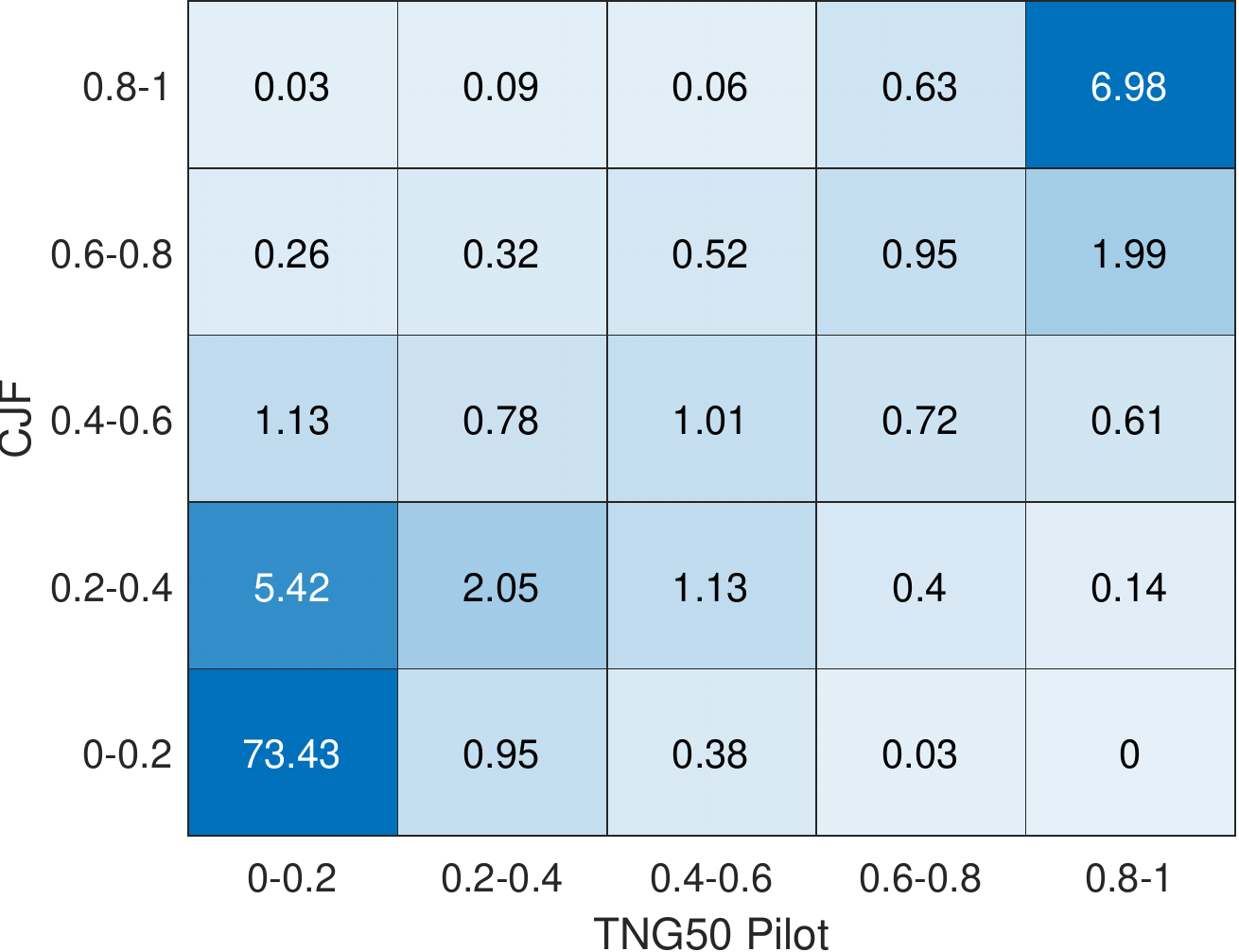}}  
  \subfloat[Score comparison with the Yun19 project] {\label{fig:tng100Comp_scores}
  \includegraphics[width=8.5cm,keepaspectratio]{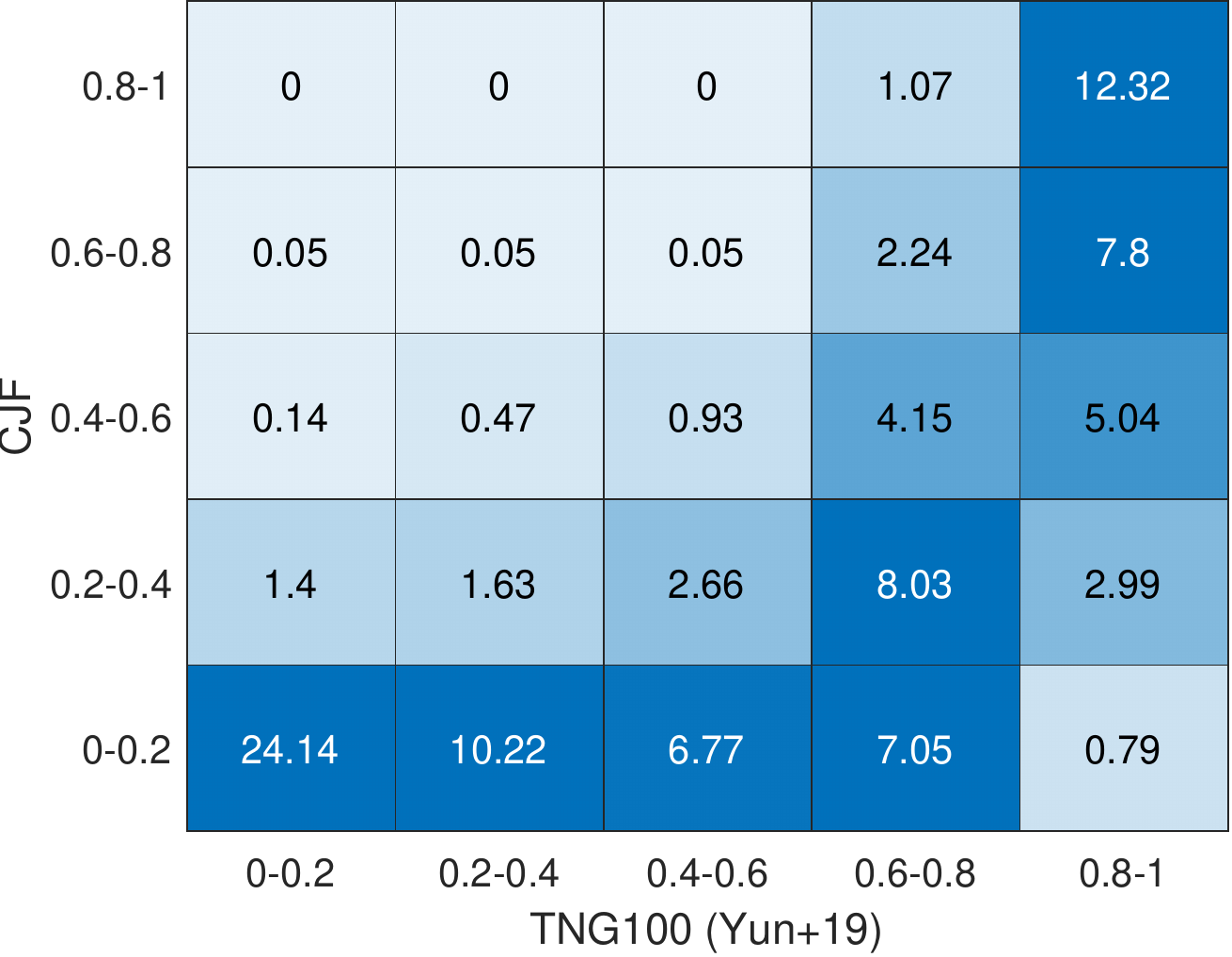}} \\
  \subfloat[JF comparison with TNG50 Pilot] {\label{fig:tng50Comp_jf}
   \includegraphics[width=8.5cm,keepaspectratio]{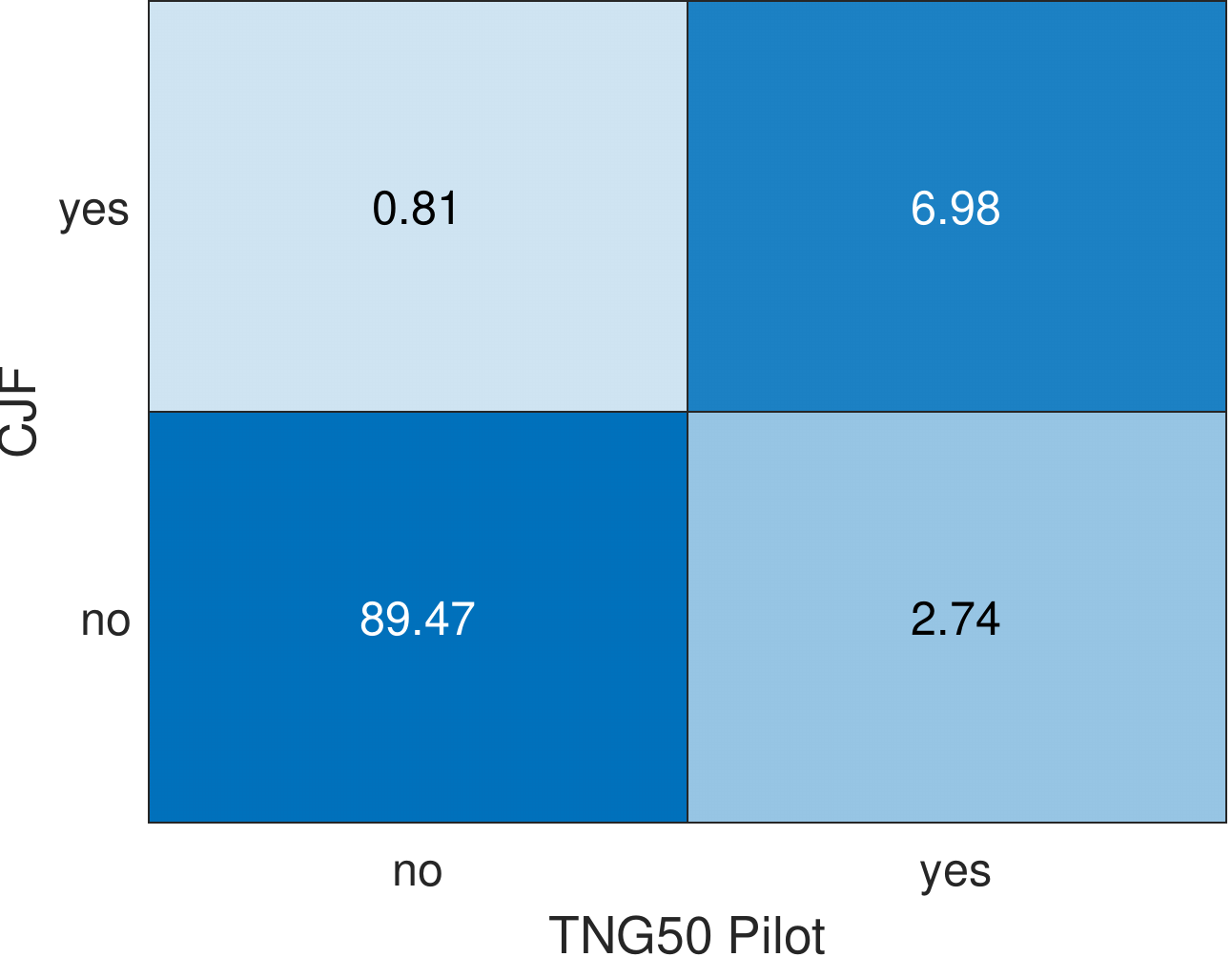}}
   \subfloat[JF comparison with Yun19 project] {\label{fig:tng100Comp_jf}
   \includegraphics[width=8.5cm,keepaspectratio]{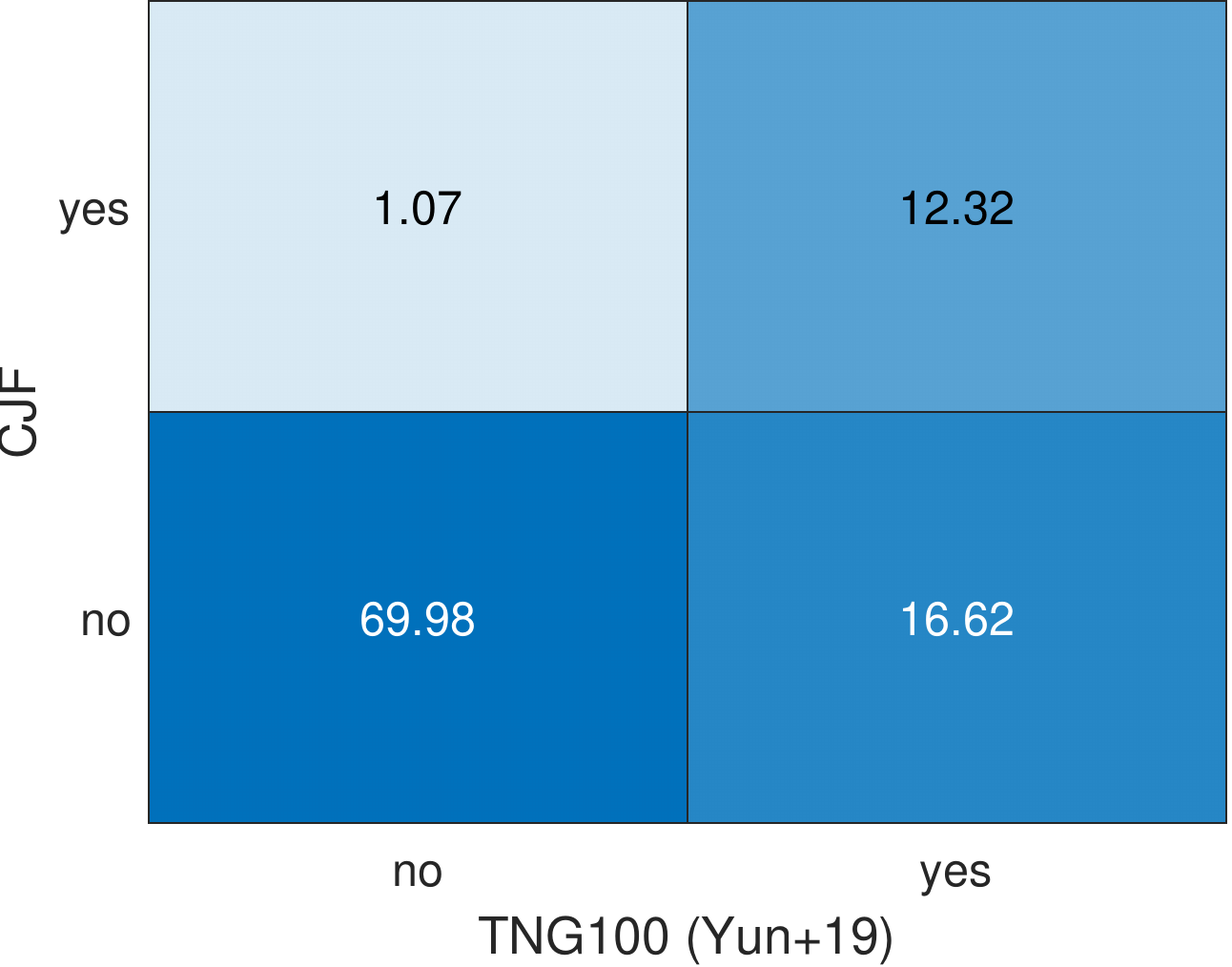}}
 
  \caption{Comparison of  the CJF Zooniverse results to previous studies carried out by expert inspectors from the research team on a subset of objects. We compare to a pilot project on TNG50 galaxies (\cref{fig:tng50Comp_scores,fig:tng50Comp_jf}) and to the Yun19 TNG100 sample and classifications (\cref{fig:tng100Comp_scores,fig:tng100Comp_jf}). The top rows show the raw score comparison in 5 score bins, with the bottom rows showing the `collapsed' comparison for the binary options jellyfish/non jellyfish (corresponding to raw scores above/below the value of 0.8). The TNG50 pilot, performed on identical images and using the same CJF Zooniverse platform, shows very high agreement. The agreement is not as strong for the comparison with the Yun19 visual inspections, most likely due to differences in the adopted images.}
  \label{fig:compare_with_past}
\end{figure*}

\subsubsection{Comparison with the TNG50 pilot project}\label{sec:visClassCompare50} 

Our in-house pilot study is functionally identical to the final CJF project. The sample includes $3,466$ objects from the TNG50 simulation, from the \zeq{0,0.2,0.5,1,1.5, 2} snapshots (541, 585, 690, 673, 552,and 425 objects, respectively): this is a subset of the CJF Zooniverse project of Figs.~\ref{fig:sampleNzred}, \ref{fig:sample_demograf} and Table~\ref{tab:sampleNumbers}. The classification team consists of six team members, five of which also classified for Yun19. 

In \cref{fig:compare_with_past}, left panels, we show a comparison of the results and the raw jellyfish scores for the CJF Zooniverse project and our pilot project (both normalized to values in the range $0\-- 1$). In \cref{fig:tng50Comp_scores}, the numbers in each score bin show the percentage of the sample. Values along the secondary diagonal (bottom-left to top-right) show the percentage of objects for which there is complete agreement (within the shown bins). Summing along the diagonal we find complete agreement for about \perc{84} of the objects. To answer the binary choice: `Is the galaxy a JF, yes or no?', we consider the collapsed matrix, with all scores below 0.8 considered to be not a JF: this is given in \cref{fig:tng50Comp_jf}. We find that for the question of whether or not a galaxy is a JF, there is an agreement of \perc{96.5} of all objects. Most of the objects with inconsistent outcome are considered JF by the experts but not by the general public. The degree of agreement remains the same for JF thresholds of either 0.66 or 0.9.

The high degree of agreement shows that the classification by non-experts, with a high enough number of classifications per object, is a viable alternative to expert classification for the purpose of identifying jellyfish galaxies in gas maps. 

\subsubsection{Comparison with the Yun19 Project}\label{sec:visClassCompare100} 

Of the sample used and studied in Yun19, $2,142$ galaxies are included in the CJF Zooniverse project: these are all from the TNG100 simulation. The Yun19 classifications use galaxy images that are similar but not identical to those in the subsequent TNG50 pilot and CJF projects, and without a dedicated common platform for classification. In both cases, images are based on a combination of a color map for gas column density and of stellar mass contours (see Section~\ref{sec:visClassImage}), with the same density limits. However, the smoothing procedure is not necessarily identical. Furthermore, a major difference between the images is the background subtraction in the Yun19 project. In that case, two side by side images of gas density for each galaxy, with one of the images subtracted by the mean gas density, enhances identification of features such as gas tails and bow shocks \citep[see Fig 1 of][]{yun_jellyfish_2019}.

\cref{fig:compare_with_past}, right panels, summarizes the comparison for the commonly-inspected images in the CJF Zooniverse and Yun19 projects. Summing along the diagonal of \cref{fig:tng100Comp_scores} we find complete agreement for  about \perc{41} of the objects. Most of the discordant cases are for galaxies for which the experts give a higher jellyfish score than the volunteers -- consistent with the advantages inherent in the background-subtracted images. The collapsed matrix for the binary states (JF vs.\@ non-JF) shown in \cref{fig:tng100Comp_jf} gives an agreement for \perc{82.3} of inspected and common galaxies.

We speculate on the source of this disagreement, and why the outcome is so different than for the comparison with the TNG50 pilot project. As noted above, the Yun19 project used different images for the classification, including images with background subtraction. In addition, the Yun19 project was the first classification campaign for the team, and it is possible that with increased experience, the subjective criteria for identifying JF might have refined and improved. Finally, the TNG50 pilot project uses a setting almost identical to that of the CJF campaign, with respect to the images, and the classification platform. In contrast, for the Yun19 visual inspection, the classifiers all use the same images, but view them in disparate ways. 

Despite the differences between the Yun19 and CJF classifications, the overall agreement is greater than 80 per cent. In both comparisons, we find that most objects for which there is disagreement are of the `false-negative' variety, i.e.\@ galaxies that experts see as JF but are not identified as such by the general public, suggesting that the JF population identified by the general public is pure but perhaps not complete. Overall, we find that we can trust volunteer classifications so long as a sufficiently large number per object are available.

\begin{figure}
  \centering
  \subfloat[Raw and adjusted Score distribution] {\label{fig:scoreHistogram_weighted}
  \includegraphics[width=8.5cm,keepaspectratio]{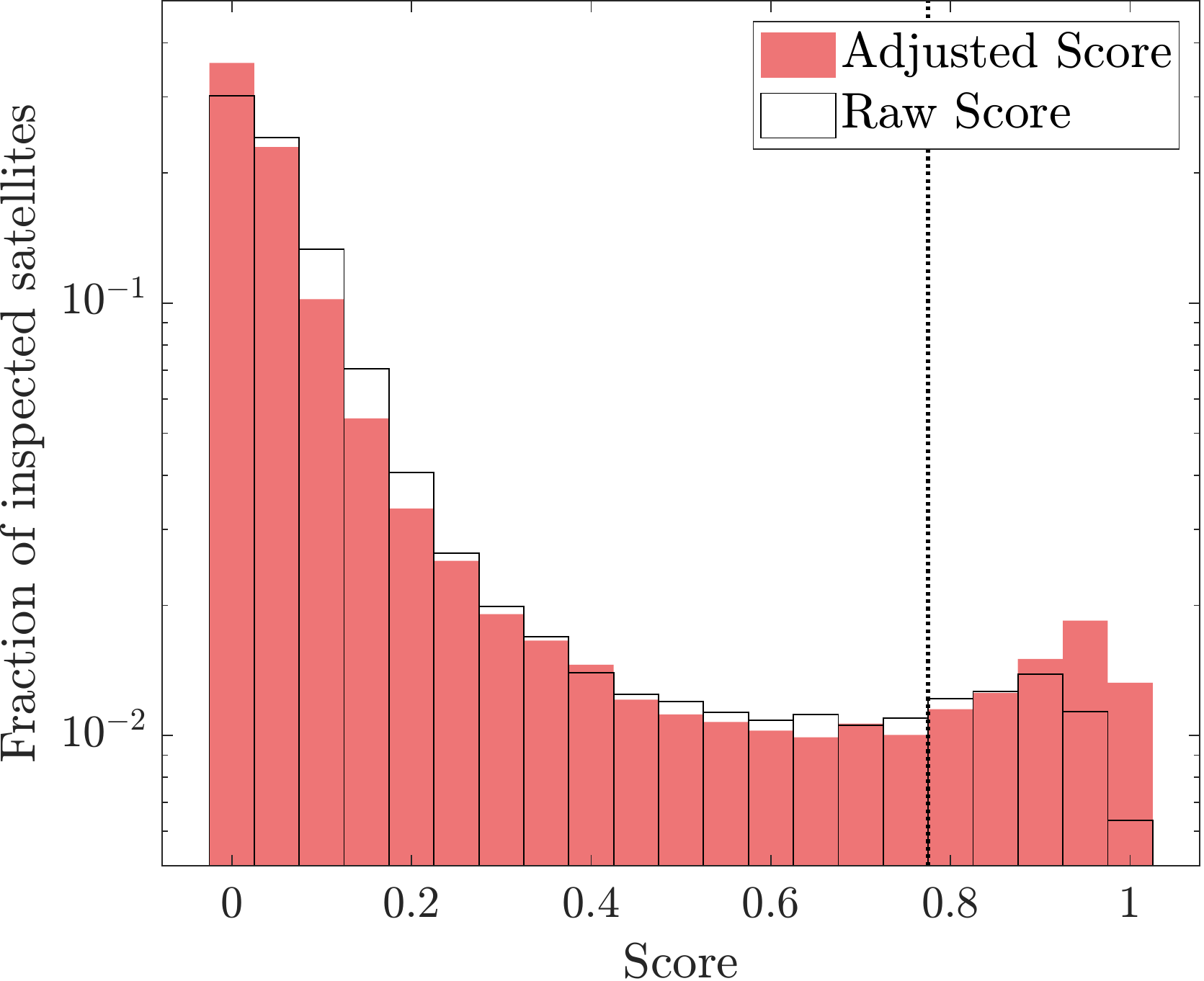}}\\
  \subfloat[Raw vs.\@ adjusted Scores] {\label{fig:scoreComp_raw_weight}
  \includegraphics[width=8.5cm,keepaspectratio]{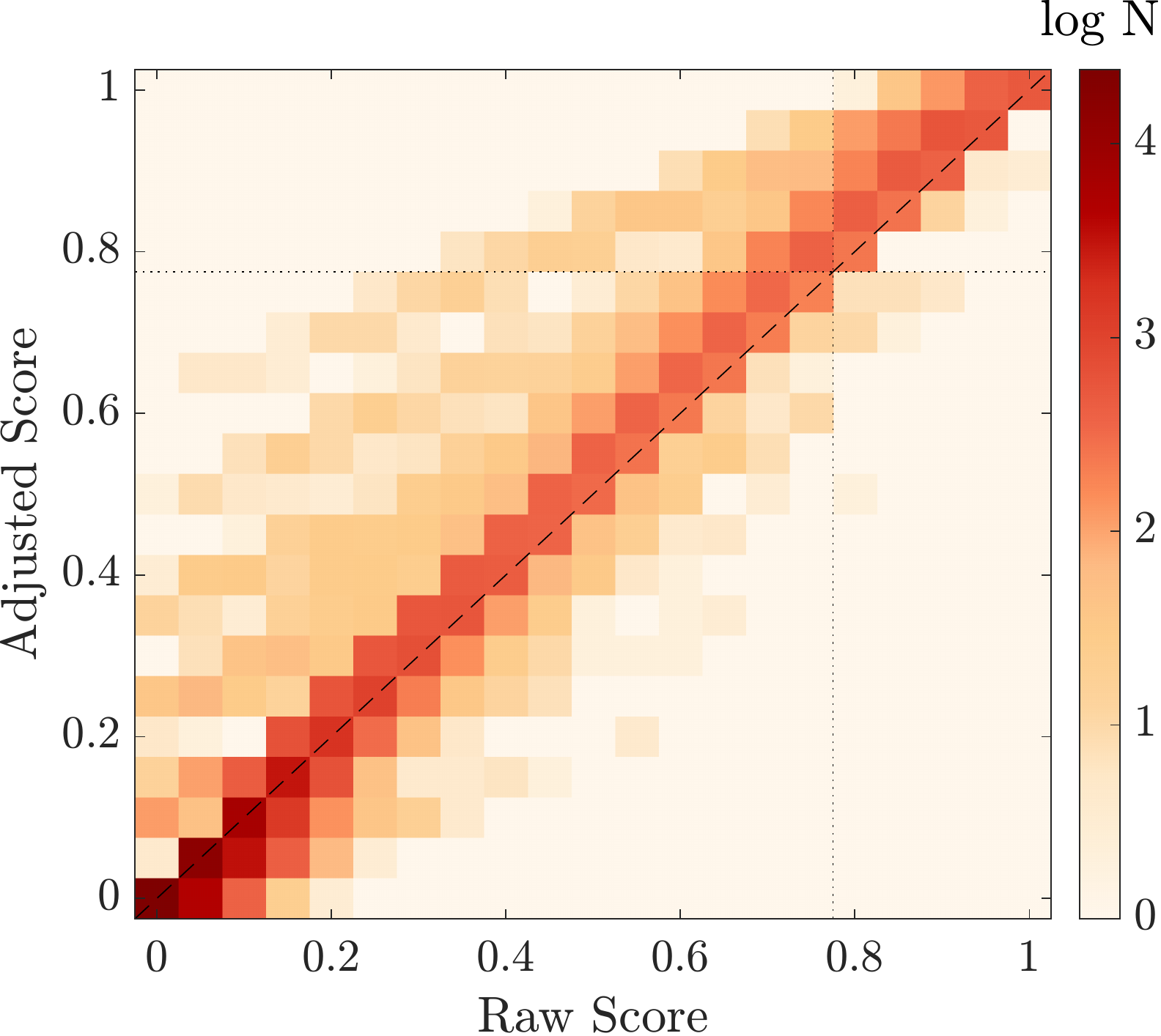}}
  \caption{Comparison between raw and adjusted jellyfish scores. The distributions of the raw and adjusted jellyfish scores for the entire inspected sample (TNG100+TNG50 galaxies) are shown in the top panel, with the numbers in each bin denoting the percentage of the population. A one-to-one comparison between the two scores for the entire inspected galaxy sample is shown in the bottom panel. The dotted horizontal and vertical lines mark the JF threshold: here we denote galaxies with score $\geq 0.8$ to be JF. Due to the score adjustment, an additional \perc{\sim 1} of the population is identified as a JF  bringing the total number of jellyfish from $4,553$
  to $5,307$. }
  \label{fig:weightScoreComp}
\end{figure}

\subsection{Adjusted jellyfish scores}\label{sec:InspectorWeighting}

Given the initial analysis of the visual inspections discussed above and the resultant galaxy scores, we propose to adopt a more nuanced interpretation of the CJF Zooniverse classifications. To do so, we assess the expertise of each inspector, and assign a weighting accordingly \citep[e.g.][]{lintott_galaxy_2008}.

As could be seen in \cref{fig:classificationHistogram}, roughly \perc{19} of inspectors classify 10 images or fewer. As with any learned task, increased experience (usually) leads to higher proficiency, and this should be reflected in assessing the classifications. Conversely, there are several hundred inspectors who have classified more than a thousand images, with some having classified more than $10^4$ images (see \cref{sec:visClassProcess}). These inspectors are responsible for more than half of the total classifications, and appraising the quality of their classifications is also important. 

One issue which may impair our inspector weighting scheme is the issue of classifications by unidentified inspectors. There is no way to generate and evaluate the voting record of an inspector if they are identified differently in each session. We suspect that in many cases, a non logged-in user is someone who only classified a few images before growing dis-interested and moving on. For these cases, the inspector-weighting scheme works as intended.  

Finally, we wish to take advantage of the classification of objects by experts, i.e.\@ members of the team with experience in image classification in previous projects, both for jellyfish identification or other galaxy-inspection tasks. 

To this end we assign weights to individual inspectors based on their experience and voting history. The revised score of an object is then set to be 
\begin{equation}
    \mathrm{score}=\frac{\sum w_\mathrm{i}v_\mathrm{i} }{\sum w_\mathrm{i}},
\end{equation}
where the summation is over all inspectors who classify the object, $w_\mathrm{i}$ is the inspector weight and $v_\mathrm{i}$ is the vote (0,1) given by the inspector. The inspector weight is set by the following scheme: 
\begin{itemize}
    \item {\bf Inexperienced inspectors} that have classified fewer than 10 objects are all given a weight of 0.5. 
    \item {\bf Expert inspectors} who are members of the research team are all given a uniform weight of $5$. The classifications from the TNG50 Pilot and the Yun19 projects are incorporated into the final score as additional expert classifications\footnotemark. 
    \item {\bf Performance on high-score objects.} If an inspector votes against the consensus from their peers on high-score objects, their weight is reduced. This favors high-accuracy inspectors.
    \item {\bf Removal of repeat offenders.} Inspectors who consistently mis-identify JF galaxies, voting `no' on objects that received predominately `yes’ votes (down-voters) and vice versa (up-voters), are given a weight of 0, i.e., removed completely. The number of inspectors removed in this manner is 338 (125 down-voters and 213 up-voters). 
\end{itemize}
\footnotetext{Using the adjusted scores, the agreement between the CJF project and the TNG50 pilot project is now \perc{98.9}, and the agreement with the Yun19 sample is \perc{93.7} (\cref{sec:visClassCompare50,sec:visClassCompare100}).}

The inspector weighting algorithm is detailed in \cref{sec:app_InspectorWeightScheme}. With this approach, some objects receive scores based on less or more than 20 votes. However, the final scores of all objects are determined by the votes of at least 13 inspectors, and in \perc{95} of cases the score is determined by 18 or more inspectors.\footnote{The details of the weighting scheme are chosen to enhance the identification of JF galaxies to enable an analysis of the demographics of the JF galaxy population. Future research questions may require a different approach, and a different weighting scheme altogether, and we encourage future users of these data-sets to consider whether and how they wish to formulate a score that best fits their research question.}

In \cref{fig:scoreHistogram_weighted} we show the histogram of the adjusted jellyfish scores (a revised version of \cref{fig:scoreHistogram}). In addition, \cref{fig:scoreComp_raw_weight} shows a comparison between the raw and adjusted scores for the entire inspected sample. As can be seen, at the low-score end the adjusted scores are lower than the original i.e.\@ raw scores, and conversely, at the high-score end the adjusted scores are higher. Due to the score adjustment, an additional \perc{\sim 1} of the entire population is identified as a JF galaxy, as defined in \cref{sec:jfThreshold}. This results in an increase of \perc{37.5} in the number of TNG100 galaxies identified as JF, and an increase of \perc{11.8} for TNG50 galaxies (an increase of \perc{16.6} overall). Namely, adopting the adjusted scores instead of the raw ones returns a total population of IllustrisTNG JF comprising of 5307 objects instead of $4,553$.

\begin{figure*}
  \centering
   \includegraphics[width=18cm,keepaspectratio]{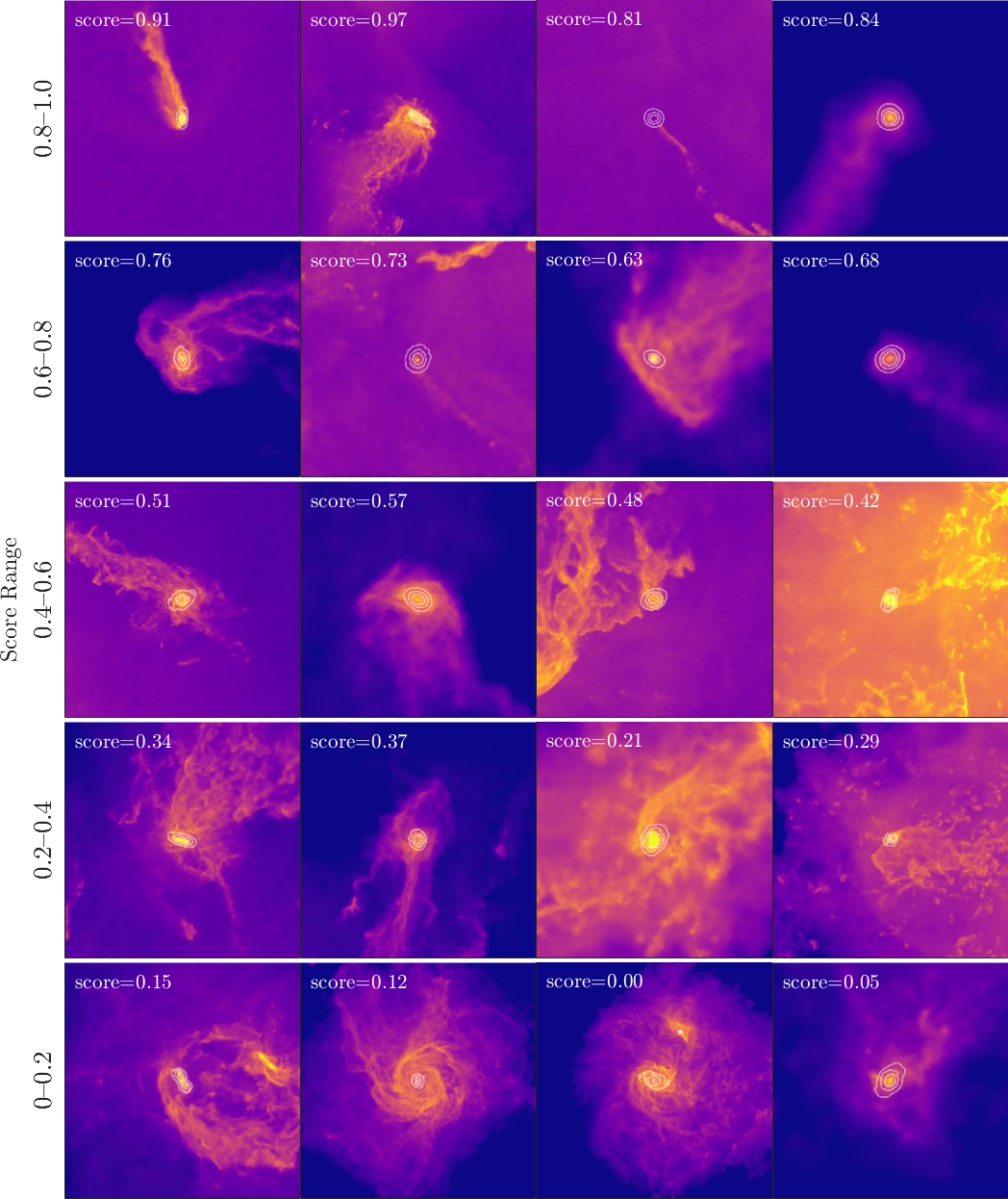}
   \caption{Examples of IllustrisTNG galaxy images of varying jellyfish scores according to the adjusted results from the CJF Zooniverse project. The top row shows objects of the highest scores, $0.8\--1.0$, with the lower rows showing objects of progressively lower scores. The score each image received appears in the top left of each image. The image specifications are the same as in \cref{fig:jf_mosaic}. More examples can be found in \cref{sec:app_imageSample}.}
  \label{fig:score_mosaic}
\end{figure*}

\begin{figure}
  \centering
   \includegraphics[width=9cm,keepaspectratio]{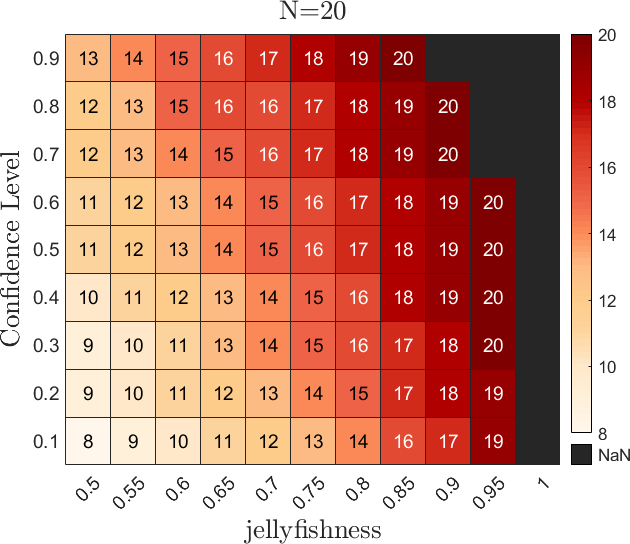}     
   \caption{The relationship between raw jellyfish scores from the CJF project (numbers in the cells), the classification probability of being a jellyfish ($\jness$ or ``jellyfishness'') and the confidence level for the probability values. The confidence level refers to how well a classification probability of $\jness$ or above can be ensured given a score from the $N=20$ inspectors. The black region corresponds to a region which is not obtainable with only 20 inspectors.}
  \label{fig:confLevel}
\end{figure}

\section{Guidelines to decide jellyfish status}
\label{sec:jf_guidlines}

One of the ways in which we envision the use of this inspected sample of galaxies, each with their jellyfish score (adjusted or not), is by establishing an appropriate threshold above which galaxies are considered jellyfish.

In \cref{fig:score_mosaic} we present a sample of 20 images of galaxies that are organized into 5 equally spaced score ranges, as assigned by the classification scheme and adjusted as described above. In the top row we show the high-score objects ($0.8\--1.0$), whereas lower rows show progressively lower score ranges, with the bottom-most showing objects of score $0\--0.2$. Additional, randomly selected examples of images in these score ranges can be seen in \cref{sec:app_imageSample}.

Based on these images we find that, as expected, objects in the highest score bin all appear to be JF galaxies (see also \cref{fig:jf_mosaic}), but some objects in the next score bin ($0.6\--0.8$) could also be considered as JF, as described by the guidelines presented in the classification project (\cref{sec:visClassProcess}). Objects in the low ranges indeed do not resemble JF galaxies. Our fiducial threshold value throughout the paper is 0.8, as we expand upon below.

\subsection{Jellyfish threshold in this study}\label{sec:jfThreshold}

In this work we adopt and hence suggest a fiducial (adjusted, inspector-weighted) threshold score of 0.8 or better to consider a galaxy as a jellyfish. This choice is driven by our requirement for a pure sample population, i.e., one that we could be sure to contain, to a good degree of confidence, very few non-JF galaxies, but that would also be large enough (several thousand objects) to constitute a representative sample of JF across different properties and environments. This requires a relatively high threshold, that may exclude some JF galaxies from our samples (e.g., second row of \cref{fig:score_mosaic}). Visual inspection of the fiducial IllustrisTNG jellyfish sample convinces that the final sample is indeed comprised almost entirely of JF galaxies.

Our choice of jellyfish score threshold is based on the experience gained from the previous pilot projects (\cref{sec:visClassCompare})  and on the considerations above. Furthermore, it is made to suit the needs of our research questions: we encourage future researchers to formulate their own threshold value according to the requirements of the scientific goal in question. Below, we give additional motivations and guidelines on how to determine an optimal threshold.


\subsection{Determining the jellyfish score threshold}\label{sec:jfDef}

We now suggest a statistical framework to assign a confidence level for the choice of a threshold score, and to compare against different threshold choices. We make the following two assumptions in order to define a relationship between an image of a galaxy and its score in the project: 
\begin{itemize}
\item For each galaxy \emph{image} there exists a probability $\jness$ that an inspector will classify the image as a jellyfish galaxy.
\item The probability $\jness$ is defined as the fraction of $N$ inspectors who classify an image as a JF galaxy as $N\to \infty$.
\end{itemize}

Based on this definition, for a total number of $N$ inspectors ($N=20$ in our case), the conditional probability that an image with a given $\jness$ will receive a score $k$ is given by the binomial distribution
\begin{equation}\label{eq:binom}
    P\left(k,N\big|\jness\right)={N \choose k}\jness^k\left(1-\jness\right)^{N-k}
\end{equation}
Using Bayes Theorem we can state the more interesting probability that an image which received a score $k$ has an inherent classification probability of $\jness$
\begin{equation}\label{eq:binom2}
    P\left(\jness\big|k,N\right)=\frac{ P\left(k,N\big|\jness\right)P\left(\jness\right)}{P\left(k,N\right)}
\end{equation}
Since we have no prior knowledge, we choose a flat prior for $\jness$ ($P\left(\jness\right)=\mathrm{const.}$) and the evidence $P\left(k,N\right)=\mathrm{const.}$, thus we find 
\begin{equation}\label{eq:binom3}
 P\left(\jness\big|k,N\right)=\mathcal{A}{N \choose k}\jness^k\left(1-\jness\right)^{N-k}
\end{equation}
where $\mathcal{A}$ is a normalization which ensures \mbox{$\int_0^1 P\left(j\big|k,N\right)\diff{j}=1$}. This defines the Probability Density Function (PDF) for the values of $\jness$ as captured by the score from $N=20$ inspectors.\footnote{The PDF of $\jness$ can be used to explore how a larger or smaller number of inspectors affects our ability to reconstruct $\jness$.}

From this PDF we define the Cumulative Density Function (CDF) which gives the probability that $\jness$ has a given value or less for a galaxy image, given a score of $k$ out of $N$
\begin{equation}\label{eq:cdf}
    P\left(\leq\jness\big|k,N\right)=\int_0^{\jness} P\left(j\big|k,N\right)\diff{j}. 
\end{equation}
Thus, the value $1-P\left(\leq\jness\big|k,N\right)$ gives the probability that, given a score $k$, an image is characterized by a value of $\jness$ \emph{or above}, which is what we look for in defining a threshold for defining a JF galaxy. Furthermore, this value constitutes the \emph{confidence level} associated with a value of $\jness$ or above of the galaxy image. 

In summary, for an image that receives a score $k$, we can not only attach one or more $\jness$ values, but also assess the confidence level of said values. We demonstrate this in \cref{fig:confLevel}, where for each value of $\jness$ (on the x-axis) and desired confidence level (on the y-axis), the number in the intersecting square shows the minimal score in the CJF project that ensures these values. For example, to guarantee a value of $\jness\geq0.8$ at a confidence level of \perc{80} one must choose galaxies with a score of 18 or above, but for a confidence level of \perc{40}, a score of 16 will suffice. In addition, \cref{fig:confLevel} can also be used to find the different $\jness$ values associated with a given score, as well as the confidence level of these values. 

We note that the probability $\jness$ is associated with an \emph{image} of a galaxy within the context of a given project, i.e., based on the image generation technique, instructions for inspectors, and so on. In addition, this model assumes equally weighted inspectors, but can be extended to address an inspector-weighting approach, as in \cref{sec:InspectorWeighting} \citep[e.g.][]{benneyan_useful_2004}. However, this would only increase the confidence level relating a score and $\jness$ value. 

In the case of our fiducial approach, for 20 inspectors, the advocated threshold of 0.8 (16 `yes' votes) ensures that all our images are within the top quartile of ``jellyfishness'' ($\jness\geq0.75$) with a high level of confidence (larger than \perc{60}, thanks to the inspector-weighting).

\begin{table}
\begin{tabular}{@{}rrrrrr@{}}
   \multicolumn{1}{l}{}  & \multicolumn{2}{c}{TNG50}     & \multicolumn{2}{c}{TNG100}    \\ 
 \cmidrule(lr){2-3} \cmidrule(l){4-5} 
 \multicolumn{1}{c}{Redshift} & JF galaxies & JF fraction  &  JF galaxies & JF fraction  \\ 
 \cmidrule(l){1-1}  \cmidrule(l){2-3} \cmidrule(l){4-5} 
 $0\leq z \leq 0.5$     &   3971    & \percSym{8.2}     & 1016 & \percSym{5.7}          \\
 $0.5< z \leq1$         &   138     & \percSym{4.7}     & 23 & \percSym{2.2}          \\
 $1< z \leq2$           &   35     & \percSym{1.6}      & 23 & \percSym{0.7}         \\ 
\cmidrule(l){1-1}  \cmidrule(lr){2-3} \cmidrule(l){4-5} 
\multicolumn{1}{r}{Total}  & 4144 & \percSym{7.7}   & 1163 & \percSym{4.3}       \\ 
 \end{tabular}
\caption{Number of visually-identified JF galaxies, and the resultant JF fractions, in the TNG50 and TNG100 simulations, in 3 redshift bins and according to the adjusted scores from our CJF Zooniverse project. The JF fractions (shown in percentage) are obtained by dividing the JF number by the total number of \emph{inspected} satellites for a given redshift range (see \cref{tab:sampleNumbers}).} 
\label{tab:jfNumbers}
\end{table}


\section{IllustrisTNG jellyfish: basic demographics}
\label{sec:results}

Based on the score distribution of the galaxies in the inspected sample as shown in \cref{fig:scoreHistogram_weighted}, we see that JF galaxies comprise a small percentage of the population of inspected satellites (and an even smaller one of the total satellite population) simulated within IllustrisTNG. For most galaxies there is little doubt that they are \emph{not} JF galaxies: \perc{\sim 44} (\perc{42} in TNG50 and \perc{51} in TNG100) have a score of 0.05 or less, and \perc{\sim 93} have a score of 0.25 or less (\perc{92} in TNG50 and \perc{96} in TNG100).

The total number of JF galaxies identified in the TNG50 and TNG100 simulations via our CJF Zooniverse project is given in \cref{tab:jfNumbers}, where we also separate into redshift bins: 5307 jellyfish in total and available for scientific studies. The total JF fraction out of the inspected satellite sample (score of 0.8 or above) is \perc{6.6} (\perc{7.7} in TNG50 and \perc{4.3} in TNG100). Most of the JF galaxies are found at low redshifts, and the JF fractions are also highest at those times. However, interestingly and as we expand upon in the next Sections, even at redshifts between \zeq{1} and \zeq{2} JF galaxies comprise \perc{\sim 1} of the inspected satellite population.\footnote{These JF fractions are lower than those found in Yun19. While here JF fractions are of order a few percent, in Yun19 the total JF fraction over the entire inspected sample was \perc{13}. However, there is a large difference in the two samples, in terms of satellite stellar mass, host mass and redshift. When comparing the JF fractions under the same selection restrictions for the CJF inspected sample, we find a JF fraction of \perc{\sim 18}.}

In the following Sections we explore the demographics of JF galaxies by examining the number and frequency of IllustrisTNG jellyfish galaxies in relation to satellite stellar mass, host mass, and redshift. 
\begin{figure}
  \centering
  \subfloat[JF fraction vs.\@ host mass in satellite stellar mass bins  ] {\label{fig:jfrac_hmass_smassBin}
  \includegraphics[width=8cm,keepaspectratio]{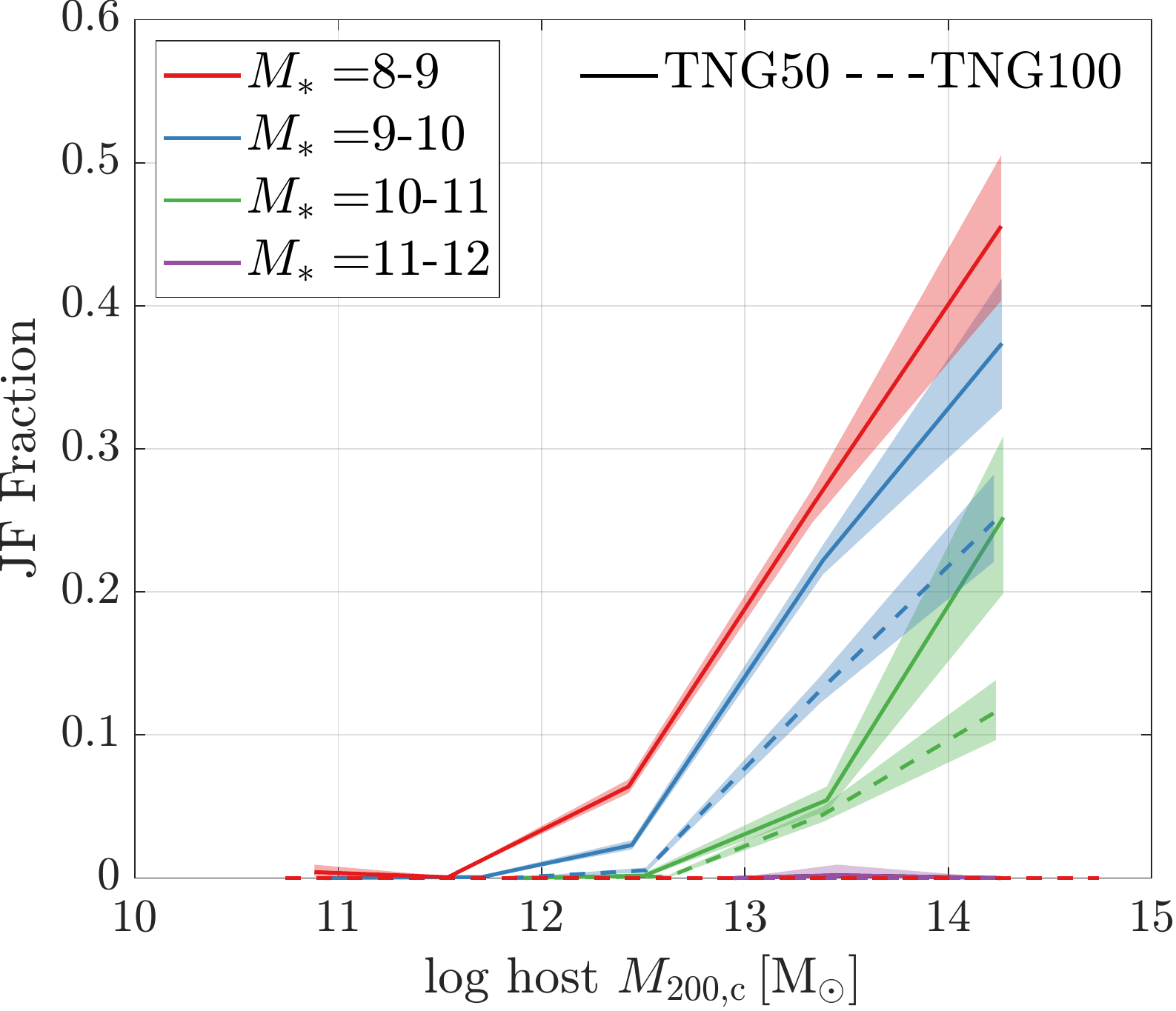}}\\
  \subfloat[JF fraction vs.\@ satellite stellar mass in host mass bins ] {\label{fig:jfrac_smass_hmassBin}
  \includegraphics[width=8cm,keepaspectratio]{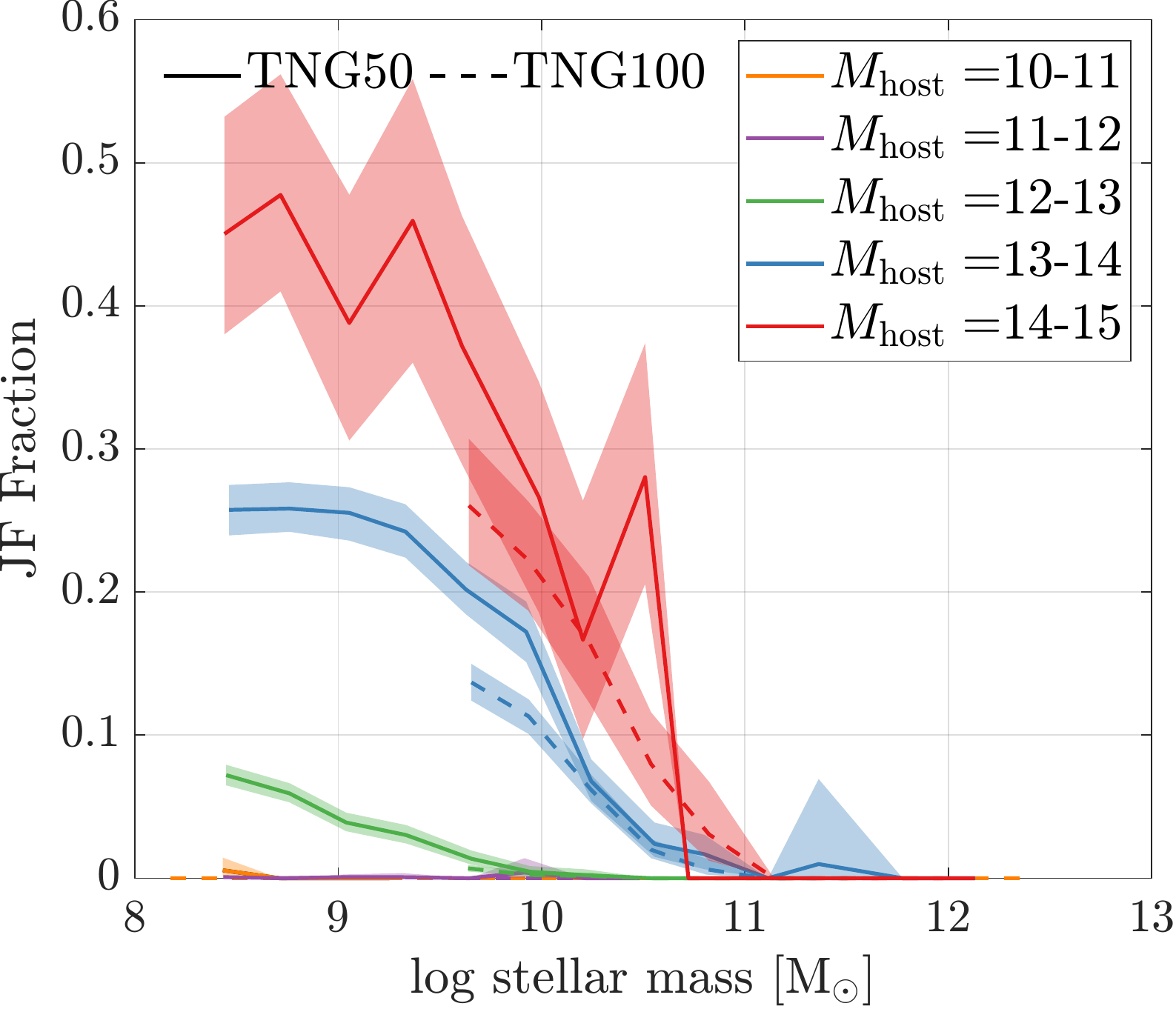}}
  \caption{ The JF fraction according to the IllustrisTNG simulations as a function of satellite stellar mass and host mass bins. The JF fraction is the percentage of JF to the inspected satellite number in each stellar/host mass bin (not all satellites have been visually inspected -- see sample selection criteria in \cref{sec:sample}).  In \subrfig{jfrac_hmass_smassBin} we show the JF fraction vs.\@ the host mass ($\Mv$) in 4 different bins in log stellar mass. The JF fraction vs.\@ the satellite stellar mass is shown in \subrfig{jfrac_smass_hmassBin}, divided into 5 host mass bins. Solid lines show the fractions in the TNG50 sample and dashed lines corresponds to the TNG100 sample. The shaded regions denote the \perc{95} confidence levels intervals based on bootstrapping within each host mass/stellar mass bin. Objects from all redshifts are included here. JF fractions are higher for lower-mass galaxies and in higher-mass hosts.}
  \label{fig:jfrac_smass_hmass}
\end{figure}
\begin{figure}
  \centering
  \subfloat[JF fraction vs.\@ redshift by host mass ] {\label{fig:jfrac_zred_hmassBins}
  \includegraphics[width=8.0cm,keepaspectratio]{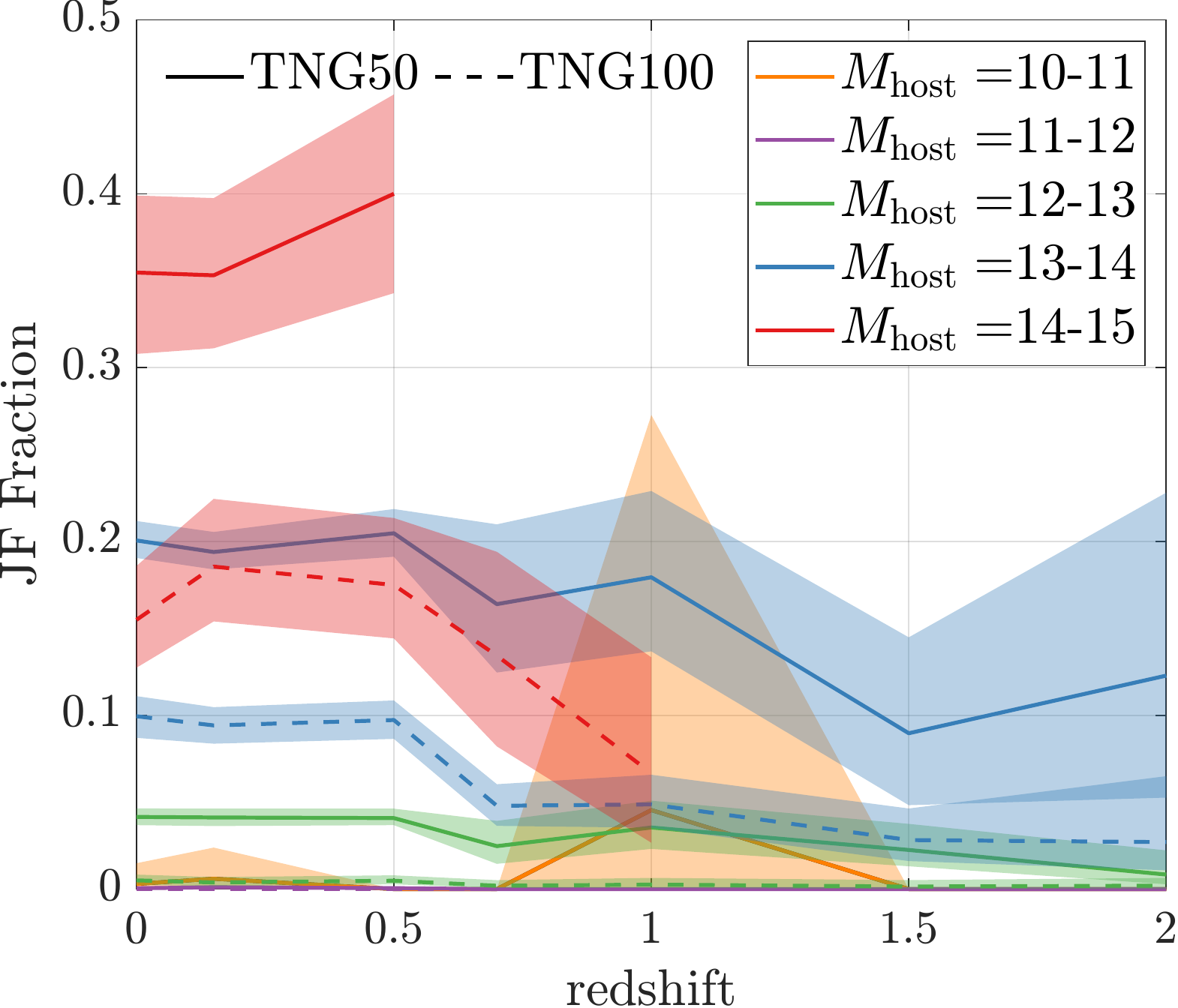}}\\
  \subfloat[JF fraction vs.\@ redshift by satellite stellar mass] {\label{fig:jfrac_zred_smassBins}
  \includegraphics[width=8.0cm,keepaspectratio]{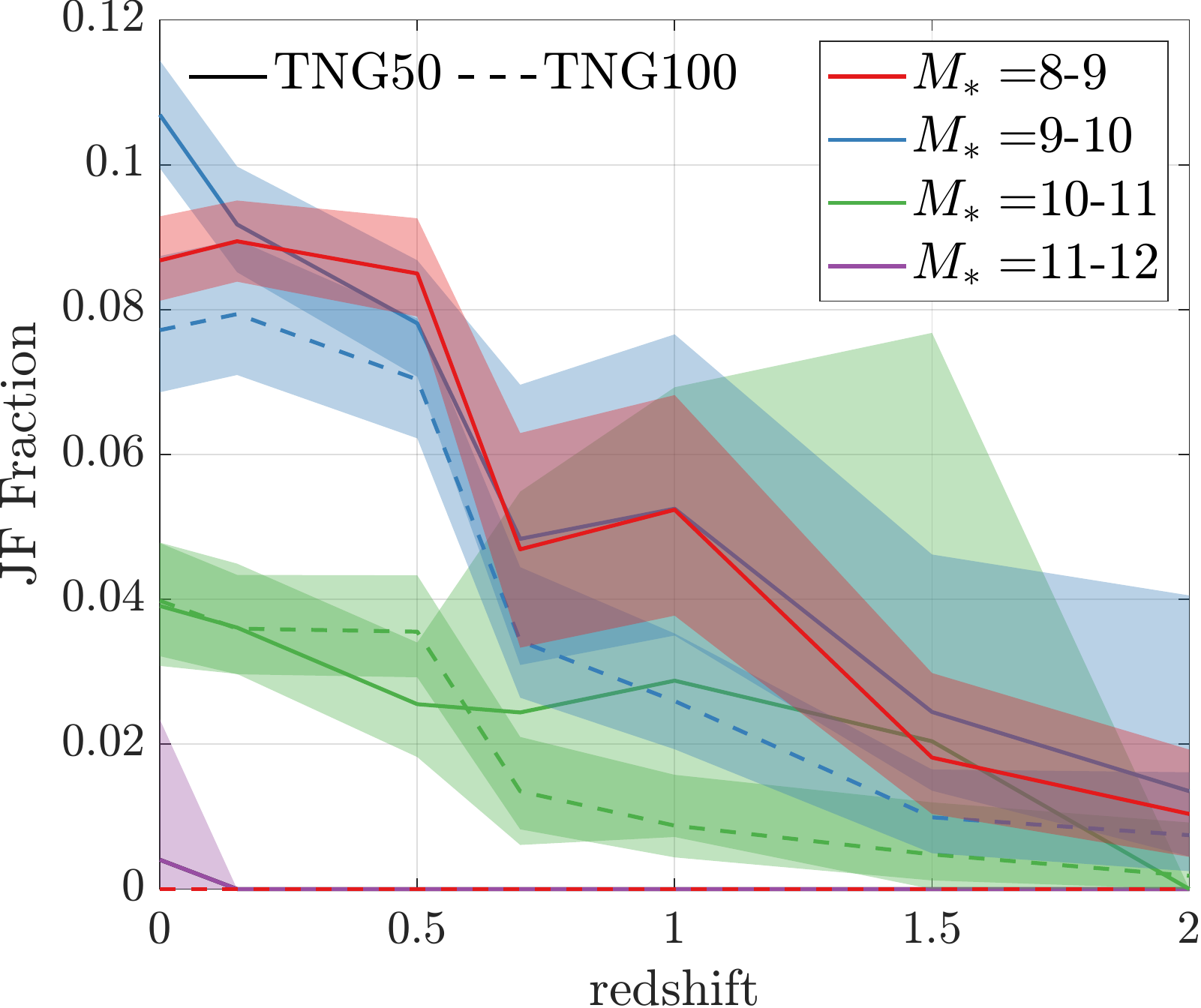}}
  \caption{The redshift evolution of the frequency of IllustrisTNG JF galaxies in host mass bins \subrfig{jfrac_zred_hmassBins} and satellite stellar mass bins \subrfig{jfrac_zred_smassBins}. Annotations are as in \cref{fig:jfrac_smass_hmass}.}
  \label{fig:jfrac_zred_massBins}
\end{figure}

\begin{figure*}
  \centering
  \subfloat[Hosts of TNG50]{\label{fig:hostHistogram_TNG50}
  \includegraphics[width=8.0cm,keepaspectratio]{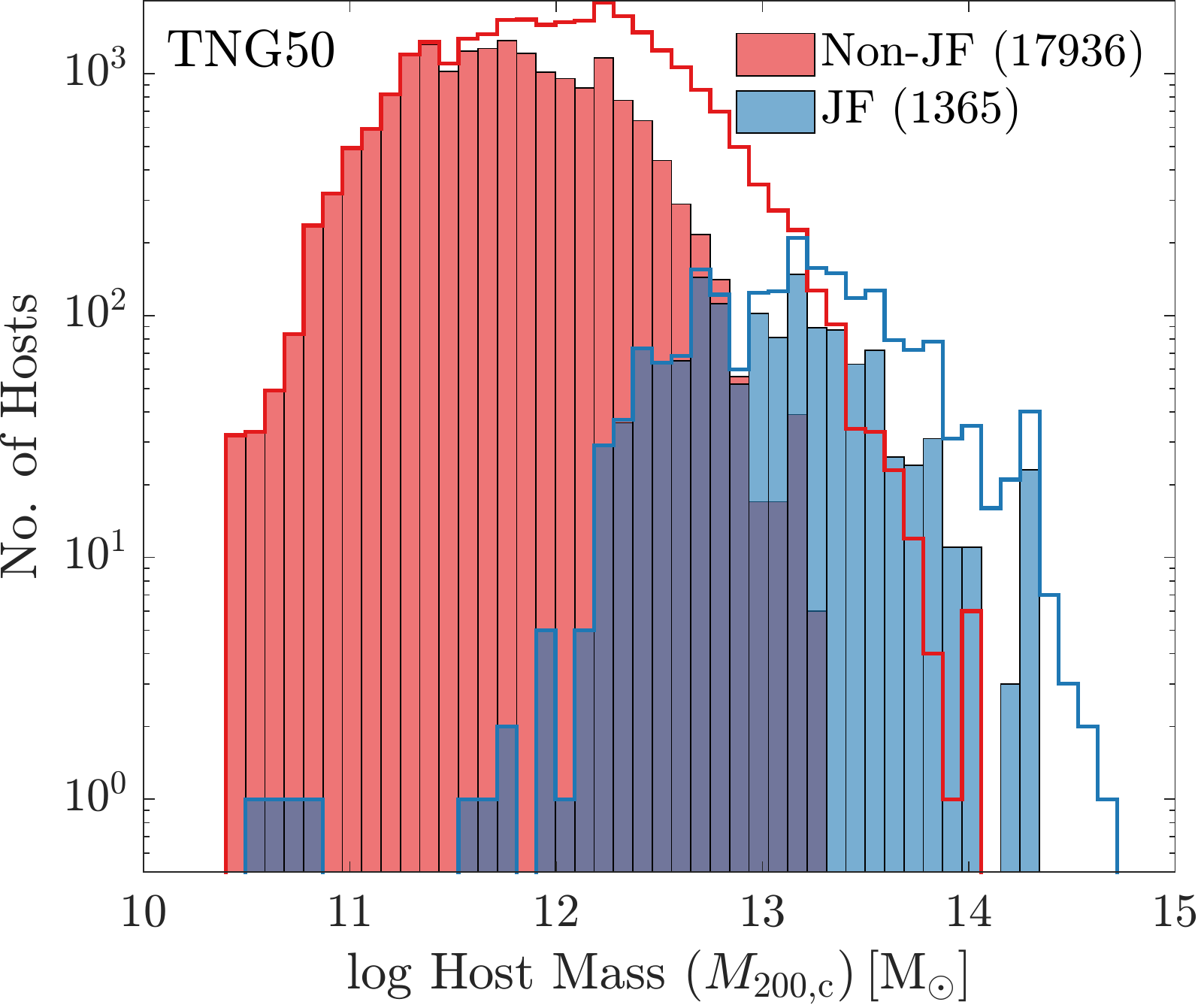}}
   \subfloat[Hosts of TNG100]{\label{fig:hostHistogram_TNG100}
  \includegraphics[width=8.0cm,keepaspectratio]{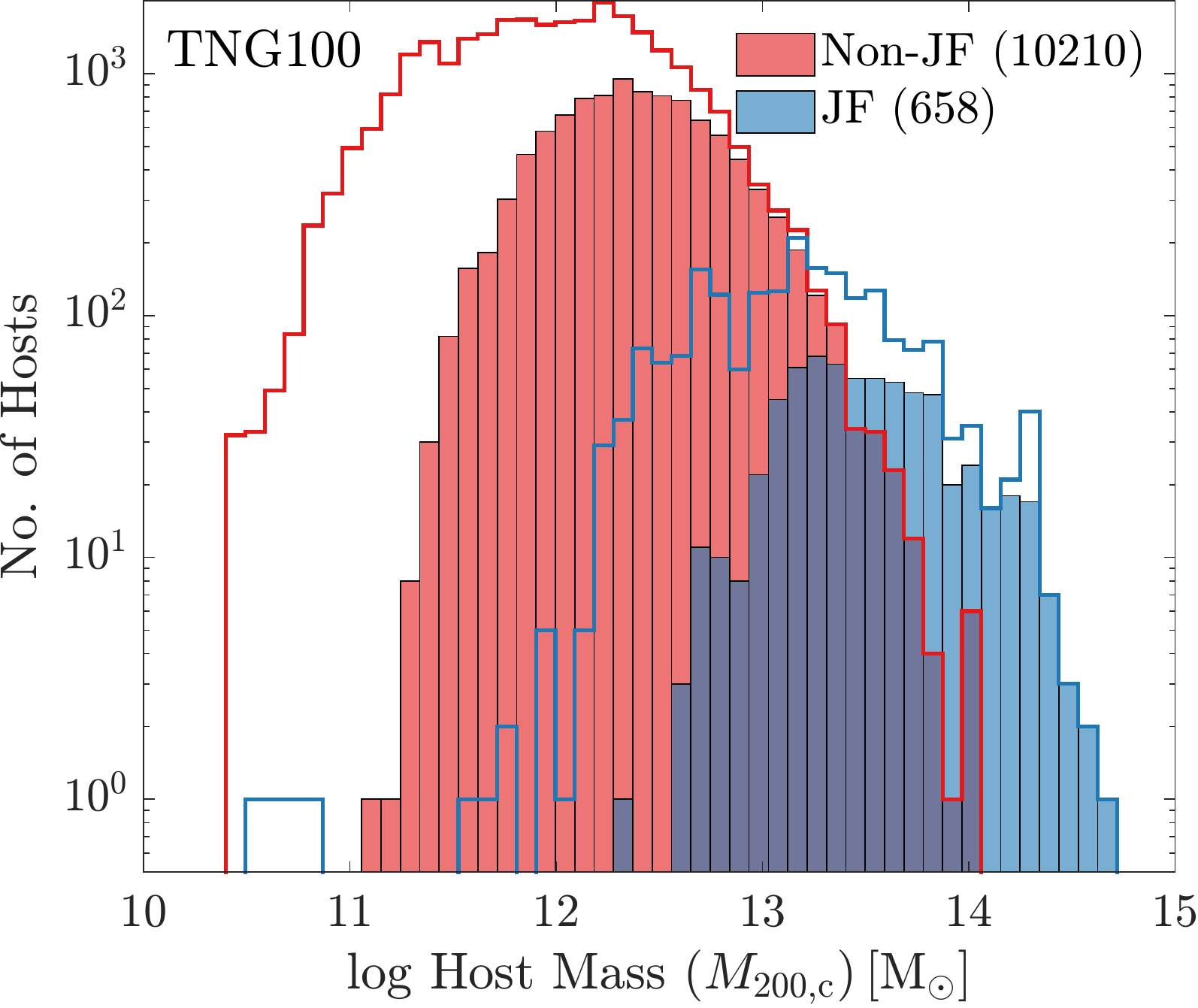}}\\
   \subfloat[JF Fraction of hosts in TNG50 ] {\label{fig:jfFrac_inHost_TNG50}
  \includegraphics[width=8.8cm,keepaspectratio]{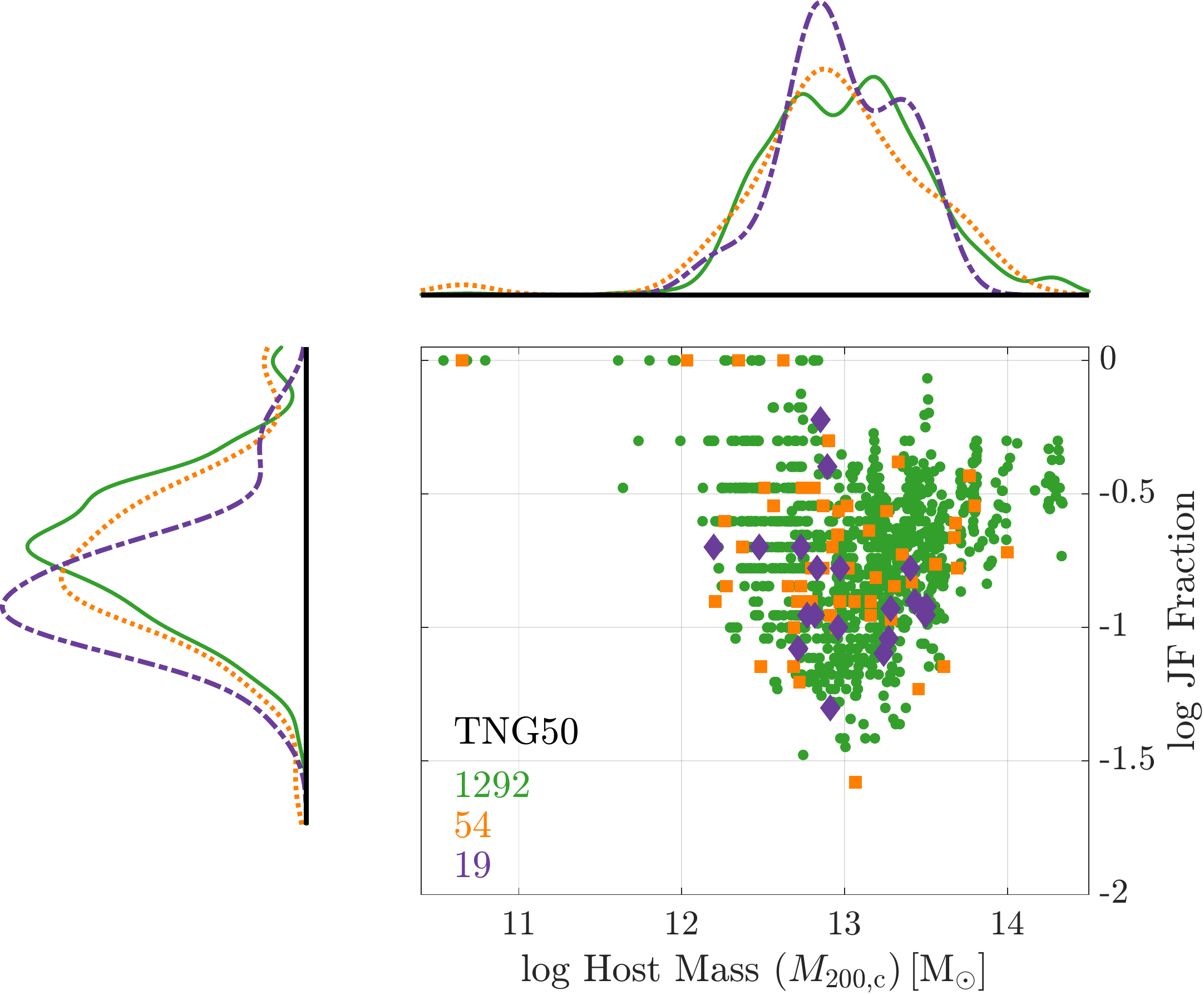}}
  \subfloat[JF Fraction of hosts in TNG100] {\label{fig:jfFrac_inHost_TNG100}
  \includegraphics[width=8.5cm,keepaspectratio]{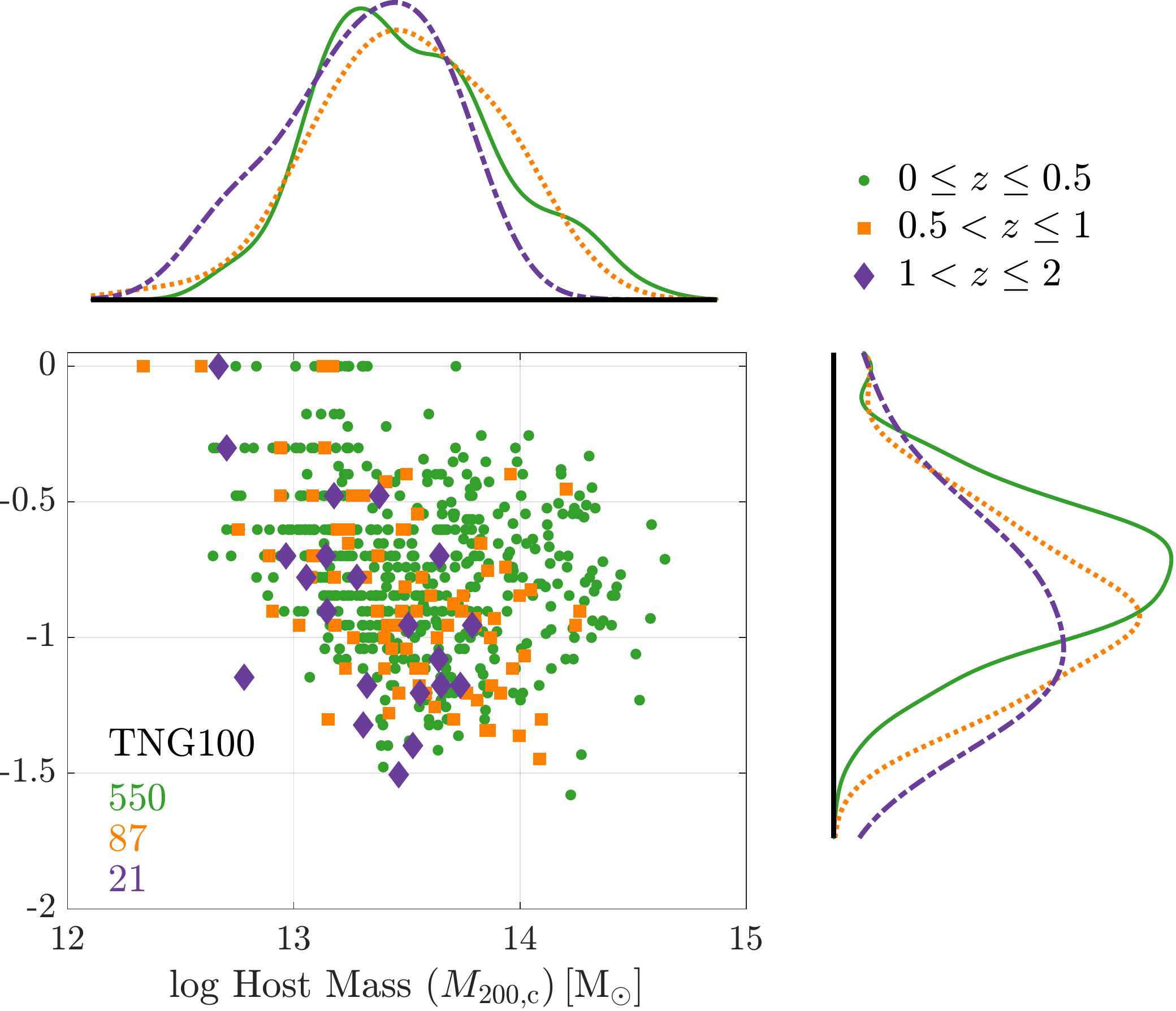}}

  \caption{Demographics of IllustrisTNG hosts that contain JF galaxies. Histograms of the number of hosts of a given $\Mv$ separated into those that contain JF (in red) and those that do not (in blue) are shown for TNG50 and TNG100 in panels \subrfig{hostHistogram_TNG50} \& \subrfig{hostHistogram_TNG100} respectively. The empty histograms show the distribution of the combined inspected sample. For all hosts that contain at least one JF satellite we show the JF fraction vs.\@ the host mass $\Mv$ in the TNG50 and TNG100 samples in the lower panels, \subrfig{jfFrac_inHost_TNG50} \& \subrfig{jfFrac_inHost_TNG100}, respectively. Hosts are color-coded into 3 redshift bins: $z\leq 0.5$ (green), $0.5<z\leq 1$ (orange) and $z>1$ (purple). The total number of hosts in each panel appears in the bottom left corner. Normalized histograms of the projected  distribution along the axes are also shown.}
  \label{fig:hostHistograms_jfFrac}
\end{figure*}

\subsection{Demographics of the JF population and their hosts}\label{sec:jfFrac_demograf}

Since there is a large difference in the TNG50 and TNG100 selected samples in terms of satellite stellar mass, host mass, and redshift ranges (see \cref{fig:sampleNzred,fig:sample_demograf,tab:sampleNumbers}), we present the fraction of JF separately for the two samples. To assess the spread in values of the JF fractions, we use the bootstrap method within a sub-sample of galaxies (defined by a combination of satellite stellar mass, host mass, redshift, etc.\@) and show the resulting \perc{95} confidence level. We note that these fractions relate the number of JF galaxies compared to the entire inspected satellite sample, and not the entire satellite population found in the simulations. However, the change in fractions is at most a factor of 2 at low redshifts , and less at higher redshifts, based on the total number of satellites in the relevant mass range in TNG50 and TNG100 (see \cref{sec:sample}).

\subsubsection{Demographics of JF galaxies}\label{sec:jfFrac_demograf_gals}

In \cref{fig:jfrac_smass_hmass} we examine the JF fractions in bins of satellite stellar mass and host mass. We bin the galaxies in 4 equally sized bins in log stellar mass: 
$10^{8-9}$, $10^{9-10}$, $10^{10-11}$, and $10^{11-12}\,\msun$,
%
%
and 5 equally sized bins in log host mass ($\Mv$):  $10^{10-11}$, $10^{11-12}$, $10^{12-13}$, $10^{13-14}$, and $10^{14-15}\,\msun$. We refer to the latter bin as cluster-sized hosts and notice that, strictly speaking, this is populated not all the way up to $10^{15}\,\msun$ but up to $10^{14.6}\,\msun$, the most massive cluster in TNG100 at $z=0$.


Due to the resolution-dependent mass threshold, the lowest satellite stellar mass bin is not populated in the TNG100 sample, and as a result, the two lowest host mass bins are empty in this sample as well (\cref{fig:sample_demograf}). The highest host mass bin in TNG50 is populated by only 29 objects, most of which are likely different instances of only 1 or 2 clusters in different snapshots.

In \cref{fig:jfrac_hmass_smassBin} we show the fraction of JF galaxies out of the inspected satellite population, over all the hosts in a given mass bin. This includes \emph{all} hosts, and not only those that actually contain JF, which are a small minority of the host population. We focus instead on the JF populations of individual hosts in \cref{sec:jfFrac_demograf_hosts}.  

We see that the JF fractions grow with increasing host mass. In hosts of masses up to several times $\tenMass{12}$ a few percent of satellites are JF, and all of these are necessarily low-mass satellites. However, this frequency grows to \perc{10 \-- 20} in group-mass hosts ($\sim \tenMass{13}$) and up to \perc{40} for low mass satellites in cluster-mass ($\sim \tenMass{14}$) systems in TNG50. The TNG100 values are lower, but still reach values of \perc{10-20} in clusters. 

These JF fractions are qualitatively consistent with recent observations that probe the group-mass scale \citep{roberts_lotss_2021}. Furthermore, this finding explains why the JF fractions are higher at lower redshifts (\cref{tab:jfNumbers}): the number of high-mass hosts increases towards lower redshift, supplying an environment more conducive to the formation of JF galaxies. Additionally, the longer the time galaxies spend in high-dense environments, the higher are the chances for them to undergo RPS.

On the other hand, while rare, JF galaxies can be found even in hosts of mass $\sim \tenMass{12}$, namely around galaxies of mass similar to our own Milky Way and Andromeda. Additional considerations on the presence of jellyfish galaxies around TNG50 Milky Way and M31-like galaxies can be found in \citet{engler_mw_2022}.

In \cref{fig:jfrac_smass_hmassBin} we show JF fraction vs.\@ satellite stellar mass, separated in halo mass bins\footnotemark.The TNG100 sample, which includes roughly 3 times as many objects, exhibits much smoother values. We see that the JF fraction drops with increasing satellite stellar mass. The more massive the galaxies, the harder it is to remove gas by RPS. For low-mass galaxies($<\tenMass{9}$), JF comprise more than \perc{40} of the inspected satellite population in cluster-sized systems, and nearly \perc{30} in groups.
\footnotetext{The fluctuations in clusters in the TNG50 sample are due to small number statistics.There are only two $\sim \tenMass{14}$ hosts in TNG50 at low redshifts \citep{joshi_disc_2020}.}
These results explain the higher JF fractions found in the TNG50 sample compared to the TNG100 galaxies: due to the higher resolution, we can study and hence have many more low-mass galaxies in the TNG50 sample (see selection criteria in \cref{sec:sample}). 

In \cref{fig:jfrac_zred_massBins} we explore how the JF fractions change with redshift, at fixed host mass and satellite stellar mass bins. In \cref{fig:jfrac_zred_hmassBins} we see that the first high-mass clusters only appear after \zeq{1} in TNG100 and \zeq{0.5} in TNG50. The large difference in the JF fraction between the two samples is due to the different mass-cut employed in the samples: the TNG100 sample does not contain galaxies below $10^{9.5}\msun$, where the JF is much higher (\cref{fig:jfrac_smass_hmassBin}).  

We see that JF galaxies can be found as early as \zeq{2}, mostly in groups and proto-clusters of mass $10^{13}\--10^{14}\msun$, where they account for \perc{\sim 10} of all the satellites in these hosts. Even in hosts with masses of $~10^{12}\msun$ we find JF galaxies at these high redshifts. The JF fraction increases with decreasing redshift, by roughly a factor of 2 between \zeq{2} and \zeq{0}. The JF fractions in low-mass hosts of $\sim 10^{10}\msun$ are tenuous: we find only a single such JF galaxy in the TNG50 sample. In fact, there are only 4 objects identified as JF galaxies found in hosts of mass $\sim 10^{10}\msun$, which we study in detail in \cref{sec:4RogueGals}. The bootstrap uncertainty estimates produce large shaded regions for bins containing few galaxies.

The frequency of JF in satellite stellar mass bins also rises with decreasing redshift, as shown in \cref{fig:jfrac_zred_smassBins}. The evolution of the JF fraction is similar in the two samples. At the high-redshift end, the JF fractions are roughly \perc{1} for all mass bins, except for the most massive galaxies which are rare at these epochs (only 24 galaxies of mass $10^{11}\msun$ and above for $z>1$ in the entire inspected sample, none of them JF). The JF fractions rise with decreasing redshift in a similar manner for the lower-mass satellites ($\leq \tenMass{10}$), reaching values of \perc{8\--11}. 

In the higher-mass satellite mass bin, $10^{10}\--10^{11}\msun$, the increase is milder and only reaches \perc{4} or so. In fact, the most massive satellites (with stellar masses of $10^{11}\--10^{12}\msun$) represent a subset of particular interest and complexity: they are JF galaxies only at very low redshifts. In fact, we find only one JF galaxy of mass exceeding $10^{11}\msun$ (TNG50 at \zeq{0.14}), among 1205 inspected ones.  A number of physical processes are at play in the case of these massive galaxies. On the one hand, higher-mass galaxies exert a stronger gravitational pull and, as such, it stands to reason that there should be fewer JF galaxies at these mass ranges. Additionally, as shown e.g.\@ by \citet{terrazas_relationship_2020,zinger_ejective_2020}, at a stellar mass of $\gtrsim \tenMass{10.5}$, the AGN kinetic feedback in IllustrisTNG becomes important and can evacuate most of the gas from the inner regions, or even halos, of galaxies. This may also contribute to the relatively lower number of JF galaxies in this mass range. On the other hand, we have shown that, according to IllustrisTNG, the AGN feedback in massive satellite galaxies is generally hampered by their high-density environments compared to that in similar-mass field galaxies \citep{joshi_disc_2020}. Furthermore, it is not clear whether AGN-driven outflows hinder the emergence of a JF phase by completely removing gas or whether they could actually promote RPS and hence the JF outlook of galaxies by making the gas less gravitationally bound. We postpone to future work more detailed investigations on this.


\begin{figure}
  \centering
  \includegraphics[width=8.5cm,keepaspectratio]{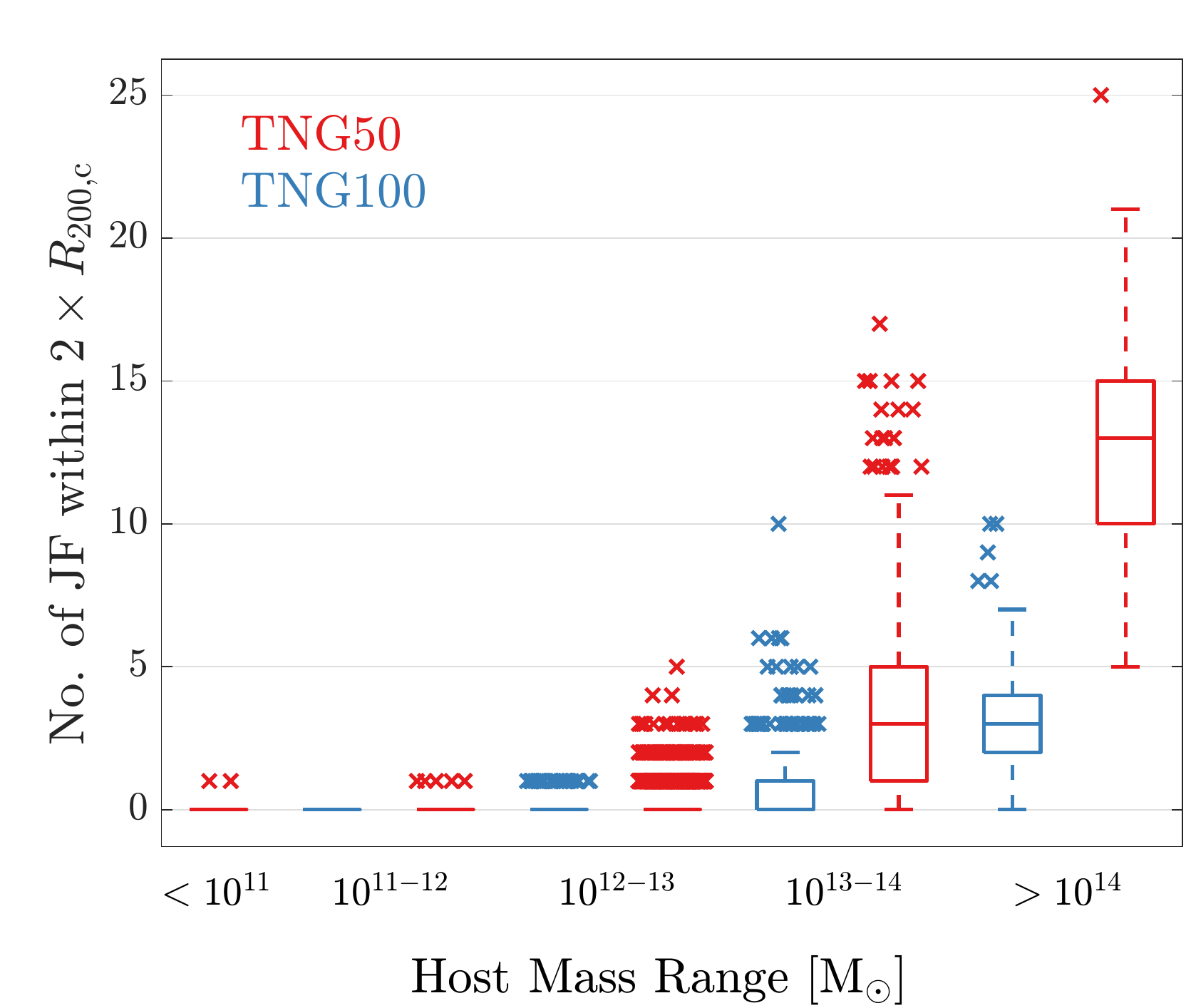}
  \caption{Number of JF galaxies to be expected in individual hosts within $2\Rv$ and across host mass bins, according to the TNG50 and TNG100 simulations (red and blue, respectively). The \perc{25 \-- 75} inter-quartile range is shown by the box, with the median marked by the horizontal line. Outliers, defined as being found 1.5 times the inter-quartile range above/below the 75/25-quantile, are marked with crosses. The dashed lines extend to the farthest values which is not an outlier. There are no $<\tenMass{11}$ hosts in the TNG100 inspected sample.}
  \label{fig:jfNum_inHost_stats}
\end{figure}

\subsubsection{Demographics of hosts that contain JF} \label{sec:jfFrac_demograf_hosts}

In this Section we focus on the demographics of the host halos that contain JF galaxies. In \cref{fig:hostHistogram_TNG50,fig:hostHistogram_TNG100} we show the distribution of $\Mv$ of all hosts that host the galaxies in the inspected sample, separated by simulation. In each simulation sample we show the distribution of the halos that host JF galaxies in red, and the hosts in which no JF galaxies were found in blue. The empty histogram in each panels show the distribution of the combined TNG50 and TNG100 samples, and is thus identical in both panels. It should be noted that, since the inspected sample spans multiple snapshots, there are different instances of the same objects included in each sample (see \cref{sec:branches} for more details).

In general, higher-mass hosts are more likely to have JF galaxies, and most halos of $\tenMass{13.5}$ and above contain at least one JF satellite. Above a certain mass, \emph{all} the hosts in the two samples contain JF galaxies: in TNG50, all hosts above $\tenMass{13.3}$ have a JF satellite, while in TNG100 all but 3 hosts above $\tenMass{14}$ contain JF galaxies. This difference is due to the difference in the underlying numerical resolution and the correspondingly chosen satellite mass threshold in the two simulations. 

Conversely, lower-mass hosts are less likely to host JF galaxies, though this too is a simulation-dependent statement. Because of the minimim stellar mass threshold, the lowest mass halos to host a JF galaxy in TNG100 are of mass $\sim\tenMass{12.5}$. In TNG50 we find 4 low-mass hosts, $\sim\tenMass{10.5}$, with JF galaxies (one each), with the next massive host with JF satellites is found at $\sim\tenMass{11.5}$. As we discuss in \cref{sec:4RogueGals}, of these four JF galaxies, three may be linked to the wrong host and one may be mis-classified. As such, a more conservative estimate for the halo mass threshold for hosting a JF in the TNG50 sample is $\sim\tenMass{11.5}$.

A more detailed view of the hosts of JF galaxies is shown in \cref{fig:jfFrac_inHost_TNG50,fig:jfFrac_inHost_TNG100} where we plot the JF fractions of individual hosts (as long as that fraction is non-zero), vs.\@ the host mass ($\Mv$). Here the JF fraction is the number of JF galaxies associated with the host divided by the total number of inspected satellites of the host. We separate TNG50 (left panel) and TNG100 (right panel), and split each into three redshift bins (three different symbols and colors).

Most hosts that contain JF galaxies are found at low redshifts: $z\leq 0.5$. However, there are tens of hosts containing JF galaxies at earlier cosmic epochs, up to \zeq{2}. We note that some of these objects may be different instances of the same halos in different snapshots; however, since each high redshift bin includes only 2 snapshots (see \cref{fig:sampleNzred}), the actual number of halos containing JF galaxies can be smaller by a factor of 2 at most. Of the 65 hosts with a JF fraction of unity, 63 have one only inspected sample satellite, with the rest having only 2 or 3 satellites. The horizontal lines at certain JF fraction values are likely due to the quantized nature of the JF scores, combined with a small number of inspected galaxies in a given host, a situation more likely for low mass hosts. Overall, the JF fraction distributions for the three redshifts are similar, between \perc{10\--30}.

Now, the demographics considerations above do not allow us to answer the following question: based on the outcome of the IllustrisTNG simulations and of the CJF Zooniverse visual inspections, how many jellyfish galaxies shall we expect to find in any given observed group or cluster of galaxies? In \cref{fig:jfNum_inHost_stats} we therefore present the statistics for the actual number of JF galaxies residing within $2\Rv$ of individual hosts, across the usual five host mass bins: \mbox{$ < \tenMass{10}$}, \mbox{$\tenMass{11\--12}$}, \mbox{$\tenMass{12\--13}$}, \mbox{$\tenMass{13\--14}$}, and \mbox{$> \tenMass{14}$}. We restrict the satellites to $2\Rv$ to demonstrate the numbers of JF galaxies that may be reasonably expected to be found in observations of such hosts. Since the two simulations have different satellite stellar mass and host mass populations (see \cref{fig:sample_demograf}), we again present the JF statistics separately for TNG50 (red) and TNG100 (blue). The median number of JF satellites per host is indicated by a horizontal line, and the \perc{25\--75} inter-quartile is shown by the box. Low-mass hosts ($\tenMass{13}$) with JF satellites are the rare exception, rather than the rule, but at group- and galaxy cluster-scales one can expect to find several JF galaxies. For example, we should expect to find 3 (0) JF galaxies with stellar mass above $10^{8.3}\msun$ ($10^{9.5}\msun$) in the typical Fornax-like group. However, depending on the state and assembly of the host, there could be systems that host up to 10-15 JF galaxies. The numbers are higher for the TNG50 inspected sample due to the lower satellite masses, which implies that probing even lower masses will yield higher numbers of JF satellites.

We note that these JF numbers per host are higher than the values found in observations of the LoTSS survey \citep{roberts_lotss_2021-2}, but not inconsistent with them, given our much lower satellite stellar mass threshold, $\tenMass{8.3}$ in our sample vs.\@ $\tenMass{9.7}$ in LoTSS.

\begin{figure*}
  \centering
  \subfloat[Satellite stellar mass - Host mass, TNG50] {\label{fig:props_sm_hm_50}
  \includegraphics[width=8.5cm,keepaspectratio]{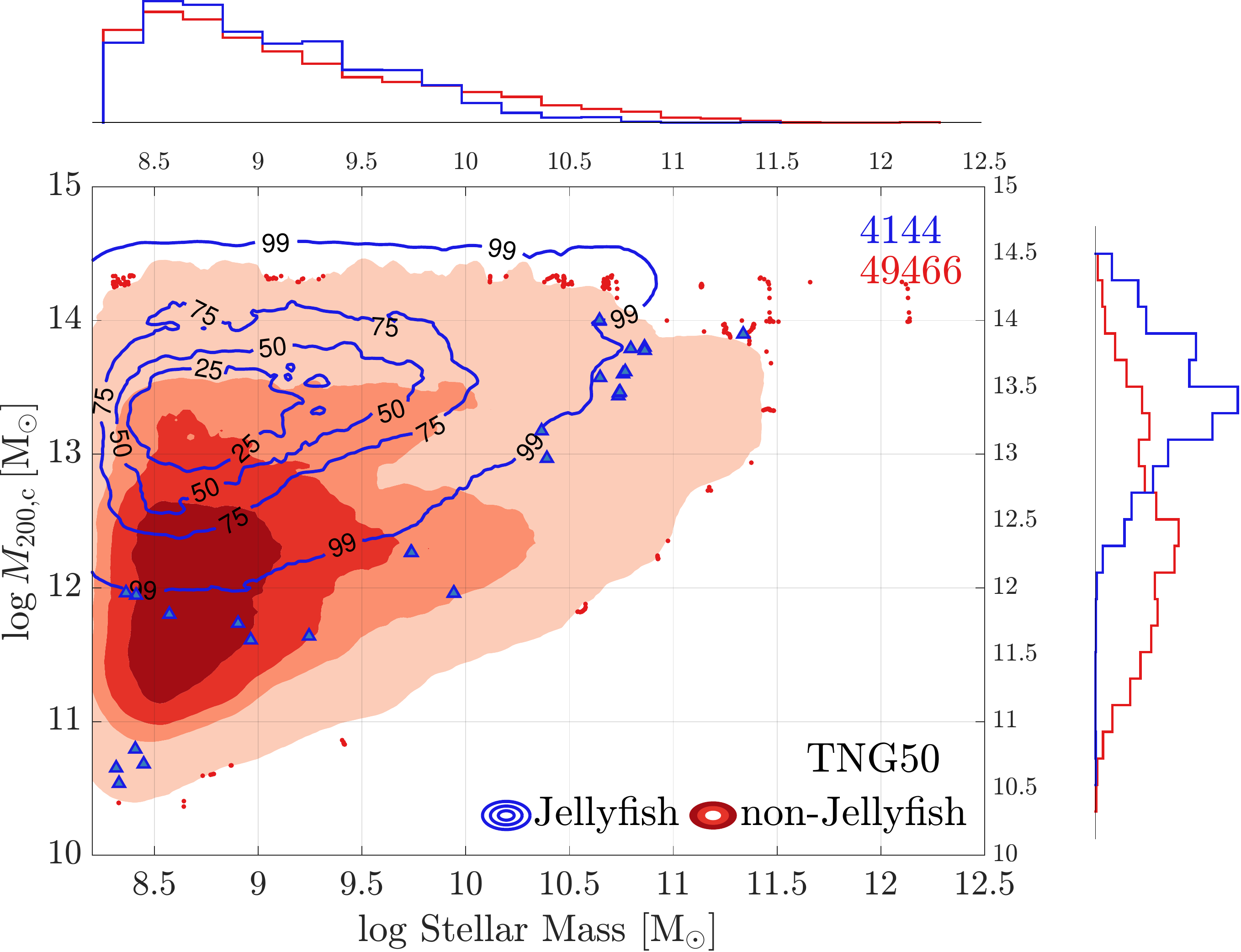}}
  \subfloat[Satellite stellar mass - host mass, TNG100] {\label{fig:props_sm_hm_100}
  \includegraphics[width=8.5cm,keepaspectratio]{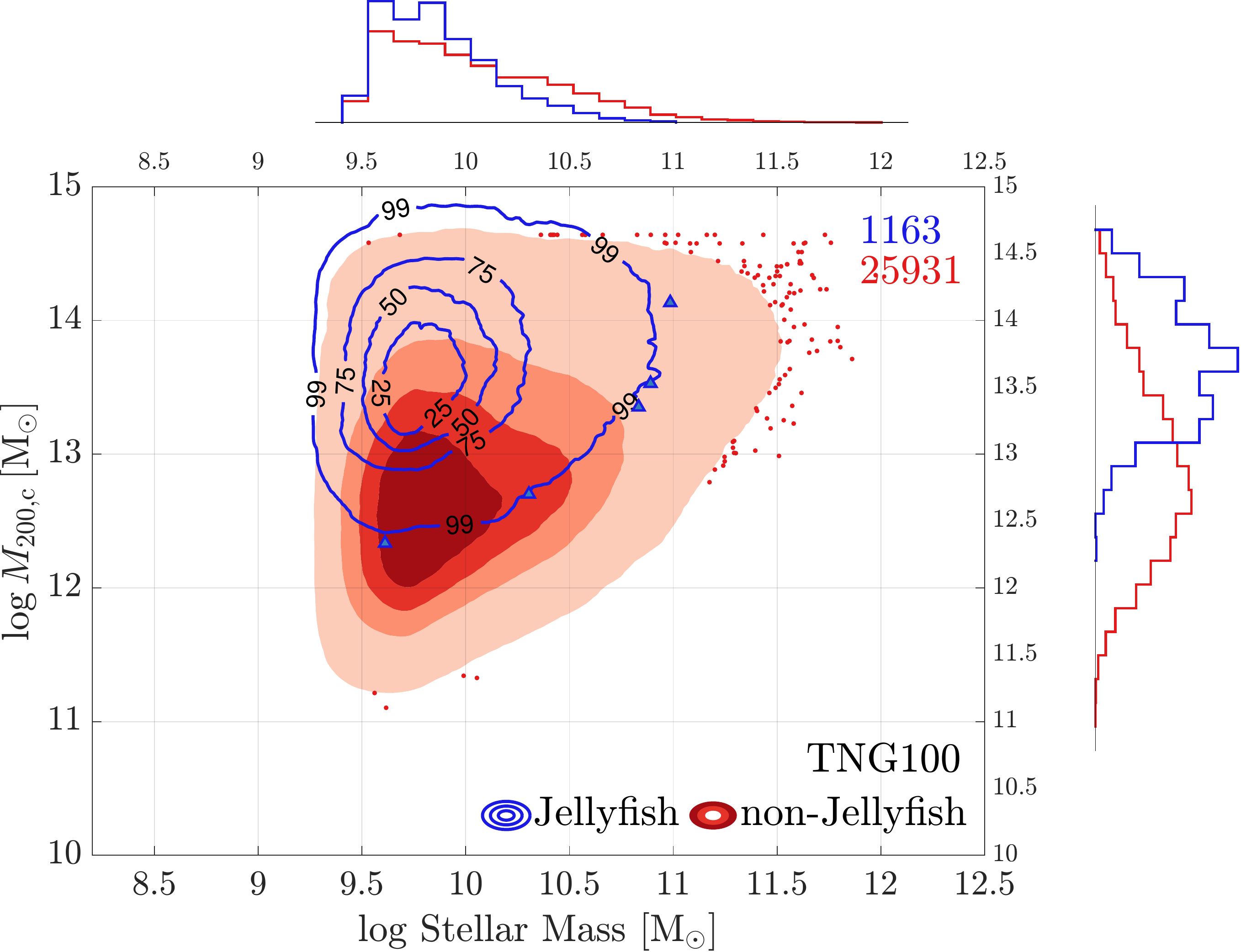}}\\
  \subfloat[Radial position - satellite stellar mass, TNG50 ] {\label{fig:props_rpos_sm_50}
  \includegraphics[width=8.5cm,keepaspectratio]{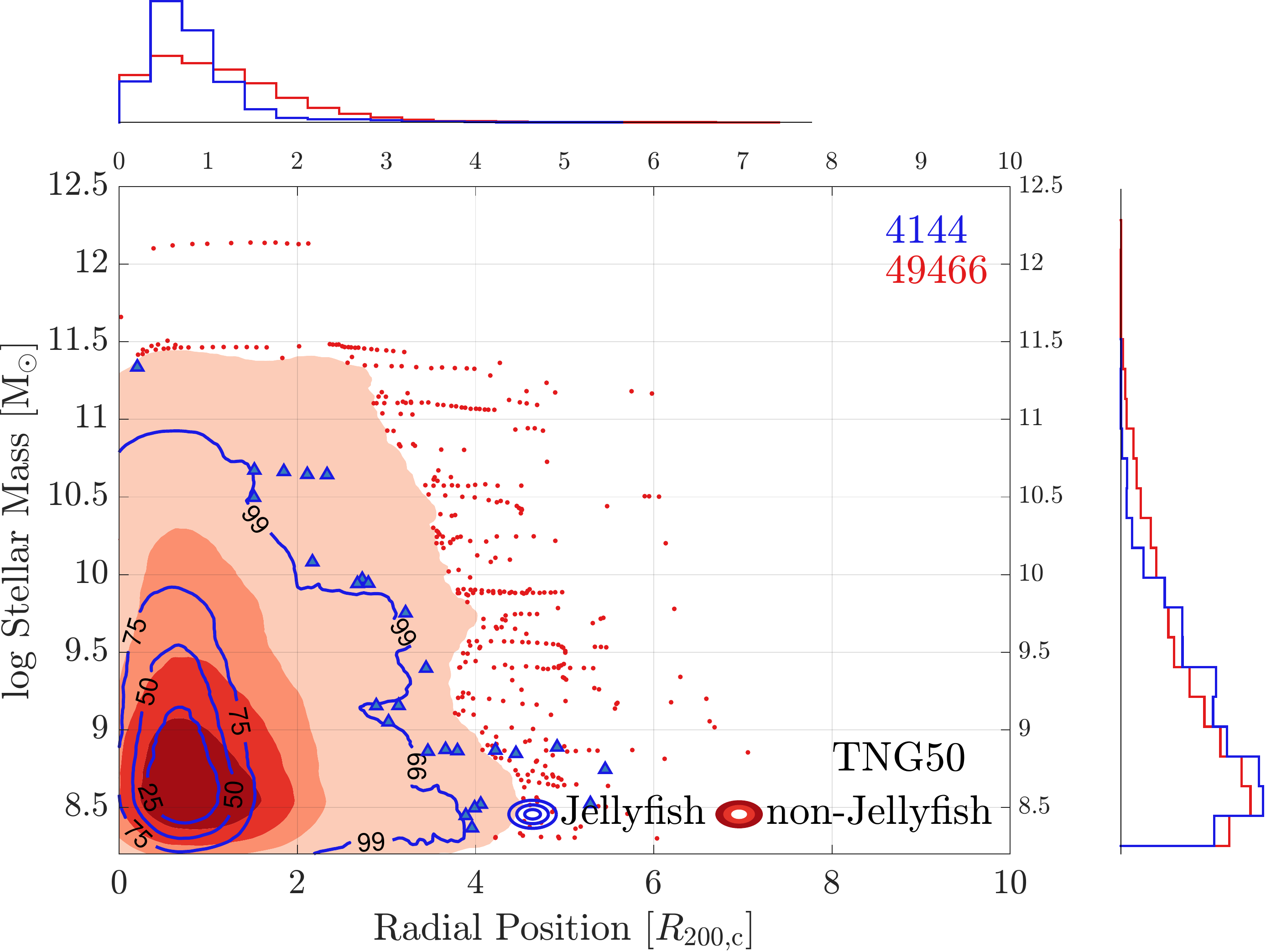}}
  \subfloat[Radial position - satellite stellar mass, TNG100] {\label{fig:props_rpos_sm_100}
  \includegraphics[width=8.5cm,keepaspectratio]{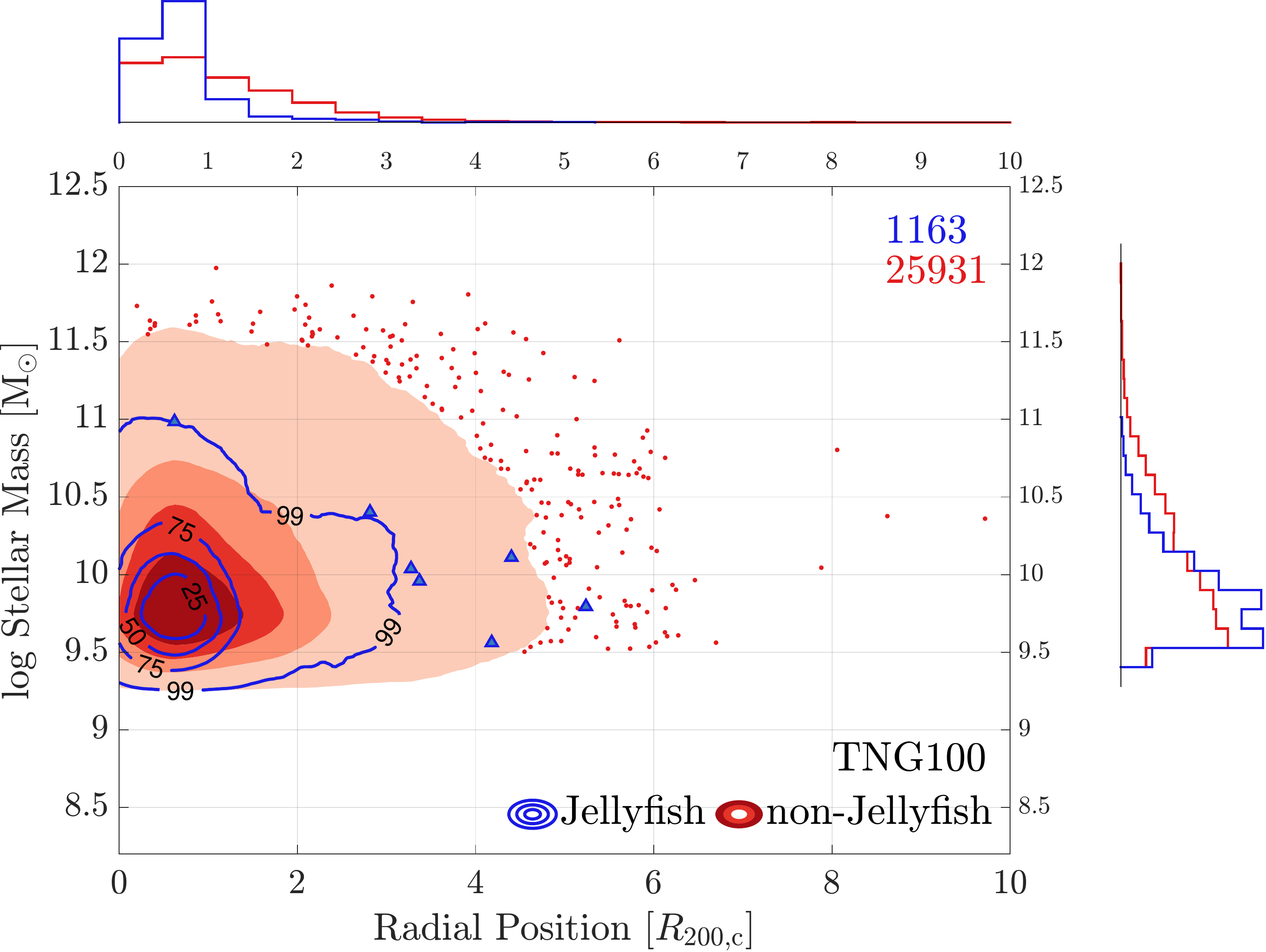}}\\
  \subfloat[Radial position - Host mass, TNG50] {\label{fig:props_rpos_hm_50}
  \includegraphics[width=8.5cm,keepaspectratio]
  {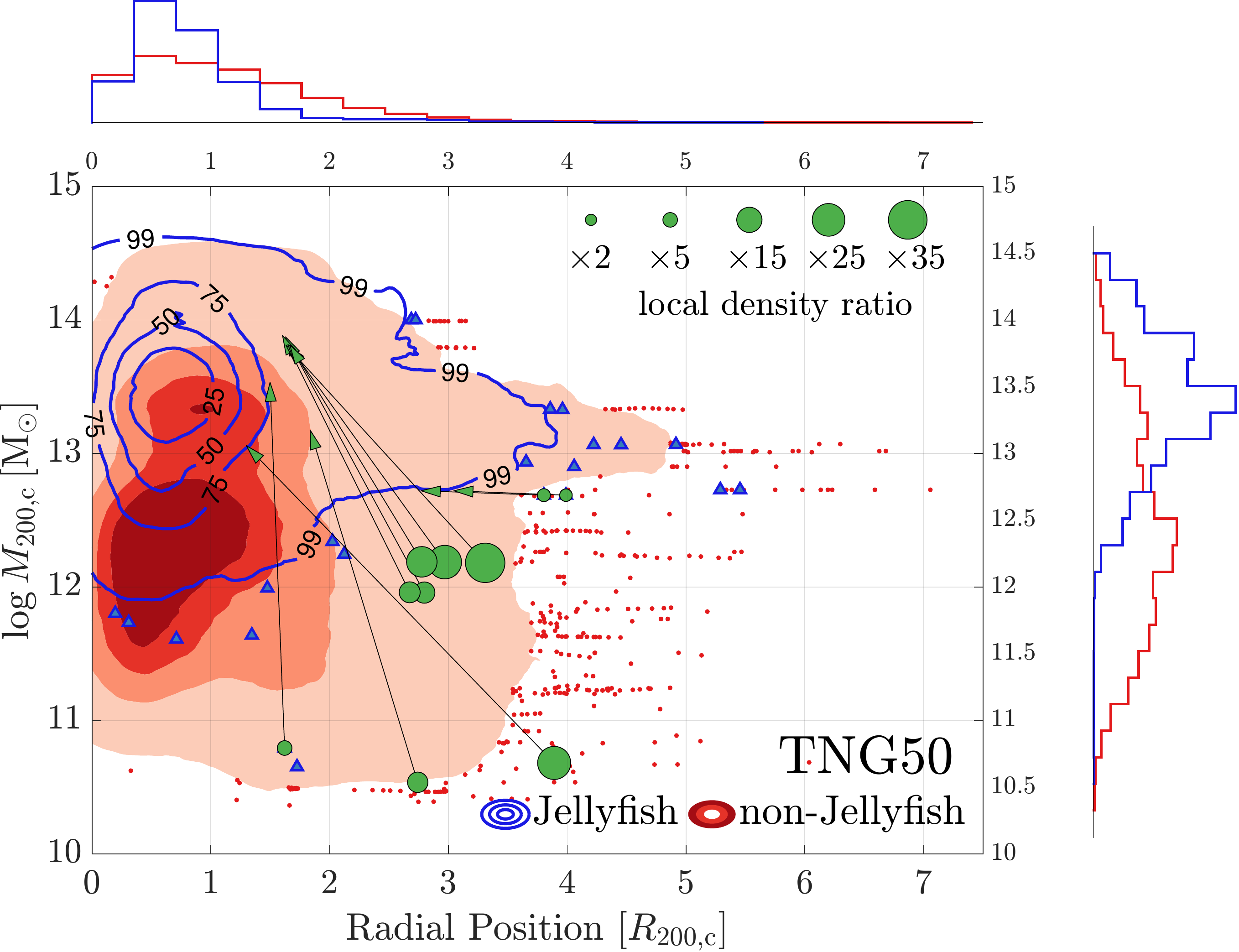}}
  \subfloat[Radial position - Host mass, TNG100] {\label{fig:props_rpos_hm_100}
  \includegraphics[width=8.5cm,keepaspectratio]{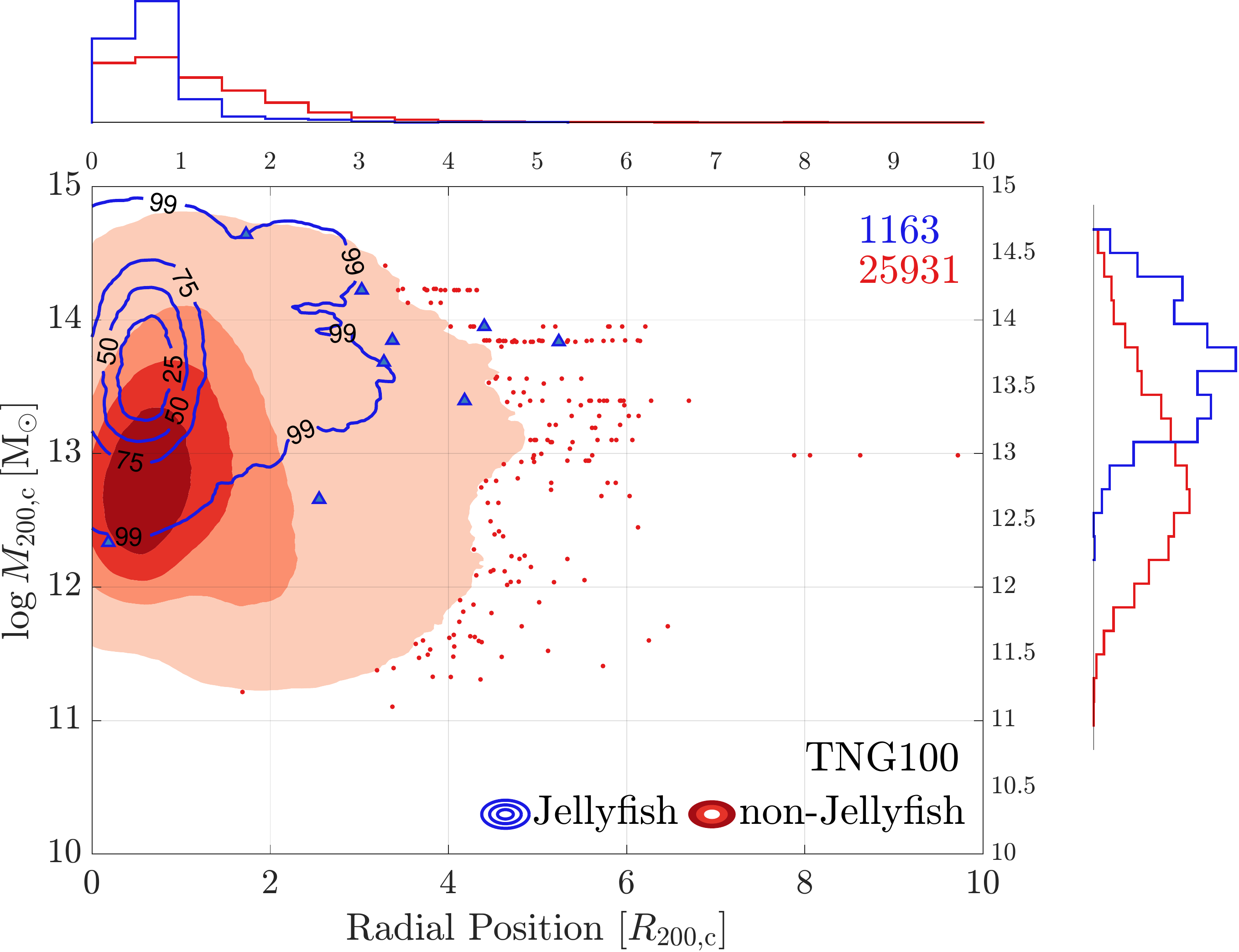}}
  \caption{Distribution of inspected satellites and of JF galaxies on the satellite stellar mass - host mass - radial position planes for TNG50 galaxies (left column) and TNG100 galaxies (right columns).  The radial position is the 3D distance to the center of the host in units of $\Rv$. JF galaxies are shown by blue contours enclosing 25, 50, 75 and 98 percent of the JF population. JF galaxies found beyond the \perc{98} contour are marked with blue triangles. The non-JF population is similarly marked by red contours and dots. Histograms show the normalized distributions along each axis. In \subrfig{props_rpos_hm_50}, ten JF galaxies with possibly mis-attributed hosts are marked in green, as explained in \cref{sec:4RogueGals}. The size of the green marker shows the ratio of highest local density over all hosts and the density of the assigned host. Arrows point to the location of the JF galaxy on the plane if the high-density host is the true one.}
  \label{fig:props_sm_hm_rpos}
\end{figure*}

\subsection{Stellar mass, host mass and radial distance of JF galaxies}

\cref{fig:props_sm_hm_rpos} summarizes in one visualization the richness of phenomenology described so far: there we show the distribution of the TNG100 and TNG50 JF populations in terms of their stellar mass, host mass and radial distance, in units of $\Rv$. The distribution of the non-JF galaxies is also shown for comparison, and populations from each simulation are separated (left vs.\@ right). 

The distributions are shown in the satellite stellar mass - host mass plane (\cref{fig:props_sm_hm_50,fig:props_sm_hm_100}), the 3D radial position - satellite stellar mass plane (\cref{fig:props_rpos_sm_50,fig:props_rpos_sm_100}) and the 3D radial position - host mass plane (\cref{fig:props_rpos_hm_50,fig:props_rpos_hm_100}). The JF population is traced by the blue contours, which enclose the 25, 50 and 75 and 98 percentiles: objects beyond the outermost contour are indicated by blue triangles. The non-JF population is likewise shown by the red contours. The number of JF and non-JF galaxies are indicated by the blue and red numbers in the upper right corner. Histograms of the normalized distribution along each axis are also given.

As already quantified in previous Sections, \cref{fig:props_sm_hm_rpos} confirms that most of the JF galaxies reside in high-mass hosts, with roughly half the population inhabiting group- and cluster-sized hosts of mass $\tenMass{13}$ and above: the peak of the distribution is found at $\sim \tenMass{13.5}$. Two opposite trends lead to this configuration: conditions for forming JF galaxies become more favorable with increasing mass (\cref{fig:jfrac_smass_hmass}) while the abundance of hosts decreases with mass (\cref{fig:hostMass_hist}). This is in stark contrast to the non-JF population where most satellites are found in hosts less massive than $\tenMass{13}$, since there are many more of these objects in the inspected sample (\cref{fig:hostMass_hist,fig:satNum_hist}). 

In terms of satellite stellar mass the distributions of the two populations are very similar. However, there is a relative deficit in JF at mass of $\tenMass{10}$ and above, which is seen most clearly in the TNG100 sample -- see considerations discussed in \cref{sec:jfFrac_demograf_gals}.

When examining the radial positions of JF galaxies in \cref{fig:props_sm_hm_rpos} we find that the majority of JF galaxies are found within $\Rv$ of their host halo\footnotemark, with the distribution peaking at $\gtrsim0.5\Rv$ and dropping to low numbers beyond $\Rv$, in contrast to the non-JF distribution that is flatter and declines more gradually. There is a drop in JF numbers in the innermost regions, similar to what was found in Yun19: this is due to the smaller volumes these region represents (there is a similar drop in the non-JF distribution, though not as sharp), but may also be due to the proximity to the central galaxy and other satellites in these regions: inspectors were explicitly instructed not to classify galaxies as JF if there were other galaxies nearby in the image (\cref{sec:visClassProcess}). Finally, as shown by Yun19, the lack of JF in the cores of groups and clusters is due to the fact that, satellites who reside mostly in the innermost regions of their host, have typically already lost the majority if not all their gas: this both acts against their selection for inspection as well as against the possibility of exhibiting gaseous tails.
 \footnotetext{The association of a satellite galaxy with a host and its position with respect to it are determined by the host halo and sub-structure identification methods used in the simulations, namely Friends-of-Friends for the host halos and \textsc{subfind} for the satellites.} 
\begin{figure}
  \centering
  \includegraphics[width=8.5cm,keepaspectratio]{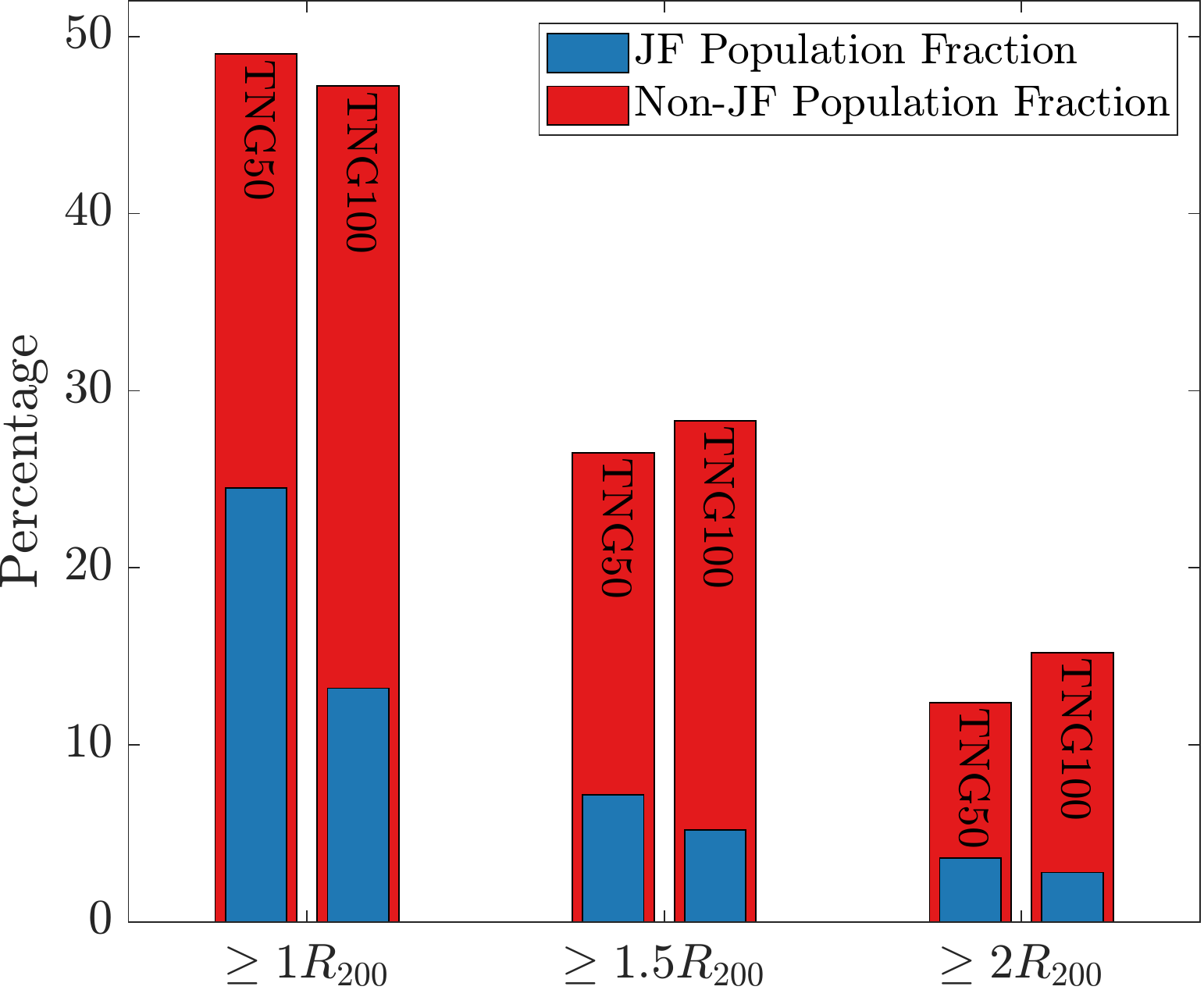}
  \caption{The fraction of the JF and non-JF  population found beyond $1\Rv$, $1.5\Rv$ and $2\Rv$, for the TNG50 sample and the TNG100 separately (right and left columns, respectively). JF data is shown in blue while the non-JF is in red. For the non-JF population, almost half of the satellites are found beyond $\Rv$ (in both samples). only a quarter of the JF are found beyond $\Rv$ in TNG50, while the fraction of TNG100 JF in that region is only \perc{13}.}
  \label{fig:jfFrac_dist}
\end{figure}

\subsubsection{JF galaxies beyond $\Rv$}\label{sec:beyond_rv}

In the TNG100 sample the fraction of JF found beyond $\Rv$ is small, but in TNG50 sample there is a substantial number of JF galaxies beyond $\Rv$ and even beyond $2\Rv$. This is in contrast to the non-JF population where the decline with host-centric distance is more gradual, and a non-negligible fraction of satellites can be found even up to $3\Rv$. 

We explore this further in \cref{fig:jfFrac_dist} where we show the fraction of both the JF and non-JF populations found beyond $1\Rv$, $1.5\Rv$, and $2\Rv$, for each of the inspected samples (TNG50 and TNG100). The fractions for the non-JF population are similar, with almost half of all satellites found beyond $\Rv$, and \perc{12} (\perc{15}) beyond $2\Rv$ in the TNG50 (TNG100) sample. 

For the TNG50 JF populations we see that a \emph{quarter} of all JF reside beyond $\Rv$, while only \perc{13} of TNG100 are found in these regions. Comparing \cref{fig:props_rpos_sm_50,fig:props_rpos_hm_100} we see that this difference is driven by the differences in satellite stellar masses between the samples: low-mass galaxies are more susceptible to RPS and the chances that they will be JF galaxies in the outer regions of their host is higher. Likewise, in higher mass hosts JF galaxies are more common even in the outer regions. Even beyond $2\Rv$ one can find JF galaxies that comprise several percent of the JF population, suggesting the presence of ambient gas and environmental effects even at these large distances. 

\subsubsection{JF galaxies in low-mass hosts}\label{sec:low_mass_host}

The most favorable conditions for the formation of JF galaxies are found in high-mass cluster-sized host, where the higher ambient density and large infall velocities create a strong RPS force. This is clearly evident in our inspected sample: nearly all hosts of mass $\tenMass{14}$ and above host JF galaxies (\cref{fig:hostHistograms_jfFrac}), and the highest JF fractions are found in these objects (\cref{fig:jfrac_smass_hmass}). Indeed, most observational surveys for JF galaxies have focused on these objects. 

However, although most JF galaxies are found in group-sized hosts of masses $10^{13}\--10^{14}\msun$ (see \cref{fig:props_sm_hm_rpos}), there is a sizeable population of halos below the group scale that host JF galaxies. Within our inspected sample there are  $\sim 700$ hosts of mass below $\tenMass{13}$ which host roughly \perc{17} of all JF galaxies. Nearly \perc{4} of JF galaxies are found in hosts of mass  $\tenMass{12.5}$ and below. There are even 9 hosts of masses of $10^{11.6}\--10^{12}\msun$ with identified JF galaxies.

\subsubsection{The case of four JF galaxies found in $\sim 10^{10}\msun$ hosts}\label{sec:4RogueGals}

In previous figures we noted 4 JF galaxies found in hosts of very low mass, $\sim \tenMass{10.5}$. However, the next most massive hosts that contain JF galaxies are a full order of magnitude more massive. In addition, as seen in \cref{fig:props_rpos_hm_50}, all four satellites are found relatively far from their hosts, with distance ranging between $\sim 1.5\Rv$ to $\sim 4\Rv$, and all are the only satellite galaxy found in the host within the inspected sample (\cref{fig:jfFrac_inHost_TNG50}). While these may be true JF galaxies within these hosts, other explanations exist. 

First, a classification error may falsely report these as JF galaxies. The scores assigned to these objects range between 0.8 and 0.88. However, a visual inspection by the team experts confirmed that three of the four are clearly JF galaxies, with the fourth also exhibiting some features of JF galaxies. Alternatively, these satellites, while bound to their low-mass hosts, may actually be in the sphere of influence of a more massive host that is responsible for the JF status. We measure the local ambient gas density of the direct hosts of these satellites and compare it to the gas density profiles of other hosts. We examine our entire inspected sample in this manner.

To do so, for each galaxy in the inspected sample, we find its position with respect to \emph{all} halos within the simulation and then estimate the resulting gas density from \emph{each halo separately} by assuming an NFW profile for the halo density distribution. We also assume the gas density follows the dark matter density, as $\rho_\mathrm{gas}=f_\mathrm{gas}\rho_\mathrm{NFW}\left(r;\Mv,\cvir\right)$, where $r$ is the distance between the galaxy and the halo center. The NFW density profile for each host is set by its virial mass parameter, $\Mv$ in this case, and the concentration parameter $\cvir$. The concentration parameter is randomly selected from the $\cvir\--\Mv$ relation of \citet{dutton_cold_2014}. The gas fraction $f_\mathrm{gas}$ is set as the ratio of the total gas mass and the total dark-matter mass in the halo. For each galaxy we evaluate all these gas density values and identify the halo with the maximal gas density at the location of the galaxy. We compare it to the gas density of the host halo assigned by the halo-finder. We define the ratio of these two densities as $\eta=\rho_\mathrm{gas,max} / \rho_\mathrm{gas,host}$.

For the entire inspected sample we find that less than \perc{1} of all galaxies have $\eta > 1.01$ (a \perc{1} excess). None of the galaxies with $\eta>1$ are found within $\Rv$ of their assigned host - the closest case is for a satellite found at $\sim 1.4\Rv$. These rare cases of competing influence are only relevant in the outskirts of the hosts. There are only ten JF galaxies with values of $\eta>1$ in the entire inspected sample, all from TNG50.

In \cref{fig:props_rpos_hm_50} the location of these ten JF galaxies is marked with green circles. The size of the circle corresponds to the ratio $\eta$. The arrows in the figure point to the location on the plane set by the mass of the host exerting the dominant influence. We see that all but one of these ten galaxies are found at large distances ($>2.5\Rv$) from their assigned hosts. In nearly all cases, the host exerting the stronger influence is much more massive, and the JF galaxies are actually relatively closer to the host (in terms of the new host $\Rv$). 
In particular, three of the four JF found in hosts of $\sim\tenMass{10}$ are indeed affected by a substantially more massive host ($\gtrsim\tenMass{13}$).  

\begin{table*}
\centering
\begin{tabular}{@{}lcccccc@{}}
   \multicolumn{1}{l}{}  & \multicolumn{3}{c}{TNG50}     & \multicolumn{3}{c}{TNG100}    \\ 
 \cmidrule(lr){2-4} \cmidrule(l){5-7} 
 \multicolumn{1}{c}{Redshift} & Total & JF Random & JF Opt.  & Total  &  JF Random & JF Opt.\\ 
 \cmidrule(l){1-1}  \cmidrule(l){2-4} \cmidrule(l){5-7} 
 \zeq{0.5}     &   1501    & 94 (\percSym{6})  &  130 (\percSym{9}) & 3064 & 110 (\percSym{4}) &   177 (\percSym{6})  \\
 \zeq{0}         &   1417     &  118 (\percSym{8}) & 161 (\percSym{11})  & 2780 & 210 (\percSym{8})  &  203 (\percSym{7})   \\
 \cmidrule(l){1-1}  \cmidrule(lr){2-4} \cmidrule(l){5-7} 
\multicolumn{1}{r}{Total}  & 2918 & 212 (\percSym{7}) & 291 (\percSym{10}) & 5844 & 320 (\percSym{5}) &  380 (\percSym{7})    \\ 
 \end{tabular}
\caption{Selection and results for the test on the viewing angle of the gas maps for the visual inspection of JF galaxies. A subset of all inspected galaxies from both TNG50 and TNG100 at two different redshifts have been used for this comparison: we give here the total numbers of inspected objects as well as the number of JF galaxies identified in the random vs. optimal orientations. JF fractions to the inspected samples are given in parenthesis.} \label{tab:viewAngleSample}
\end{table*}

\section{Discussion}
\label{sec:discuss}

\subsection{Jellyfish galaxies across their evolutionary pathways}
\label{sec:branches}
In the previous Sections, we have focused on {\it populations} of galaxies selected at various cosmic epochs from the IllustrisTNG simulations. This is formally akin to what is typically possible with observations, with the difference that, within the simulated volumes, galaxies of given epochs are the progenitors or descendants of galaxies at other epochs. Hence, not all among our visually-identified JF are unique.

In fact, the sample we identified for visual inspection for the CJF Zooniverse project includes all relevant satellite galaxies from the snapshots of the TNG50 and TNG100 simulations, as described in \cref{sec:sample}. In many cases, our inspected sample includes several instances of the same galaxy across a number of different snapshots, which is thus inspected multiple times along its evolutionary path. The sample selection process, and of course the visual classification, are totally agnostic to this. However, scientific results based on our jellyfish scores and analysis need to be interpreted accordingly. In fact, it is also possible to exploit all this by following individual galaxies, and hence their jellyfish score, across their evolutionary pathways.

With the available simulation data, we can link galaxies that have been inspected at multiple cosmic epochs using the {\sc sublink\_gal} merger trees \citep{rodriguez-gomez_merger_2015}.  Whereas the total sample of inspected satellites amounts to more than $80,000$ objects, of these, $5,023$ and $9,052$ represent unique galaxy evolutionary tracks in TNG50 and TNG100, respectively. Therefore, on average, each TNG50 (TNG100) galaxy meeting the selection criteria have been inspected $\sim 11$ ($\sim3$) times. TNG50 galaxy tracks have been inspected more frequently because every snapshot of this simulation since \zeq{0.5} is included in our inspected sample. We expand on this complementary way to analyse and look at the visually-inspected JF in IllustrisTNG in two companion papers (see \textcolor{blue}{Rohr et al 2023}, Section 2.3, and \textcolor{blue}{Goeller et al 2023}, Section 3.5), where we follow satellites across times to quantify the modalities of RPS and their star formation histories. Here we note that, of the $\sim 14,000$ unique branches in total inspected in the CJF Zooniverse project, 935 and 922 from TNG50 and TNG100, respectively, are classified as JF at least at one inspected snapshot across their lifetime (adjusted scores).

\begin{figure*}
  \centering
  \subfloat[] {\label{fig:angleComp_optJF_1}
  \includegraphics[width=9cm,keepaspectratio]{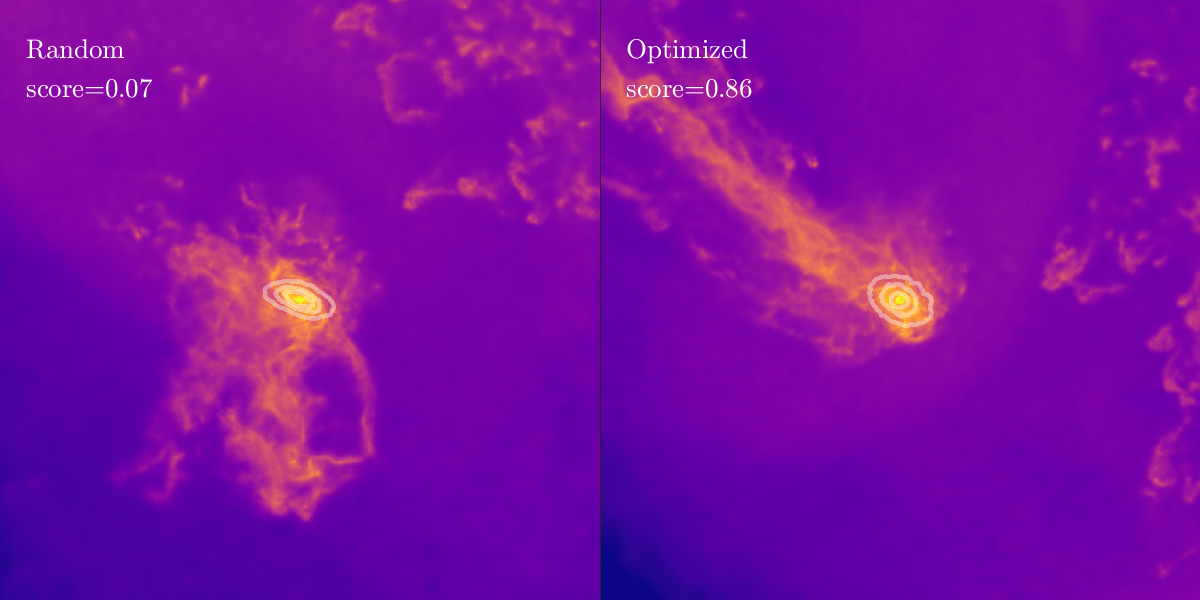}}
  \subfloat[] {\label{fig:angleComp_optJF_2}
  \includegraphics[width=9cm,keepaspectratio]{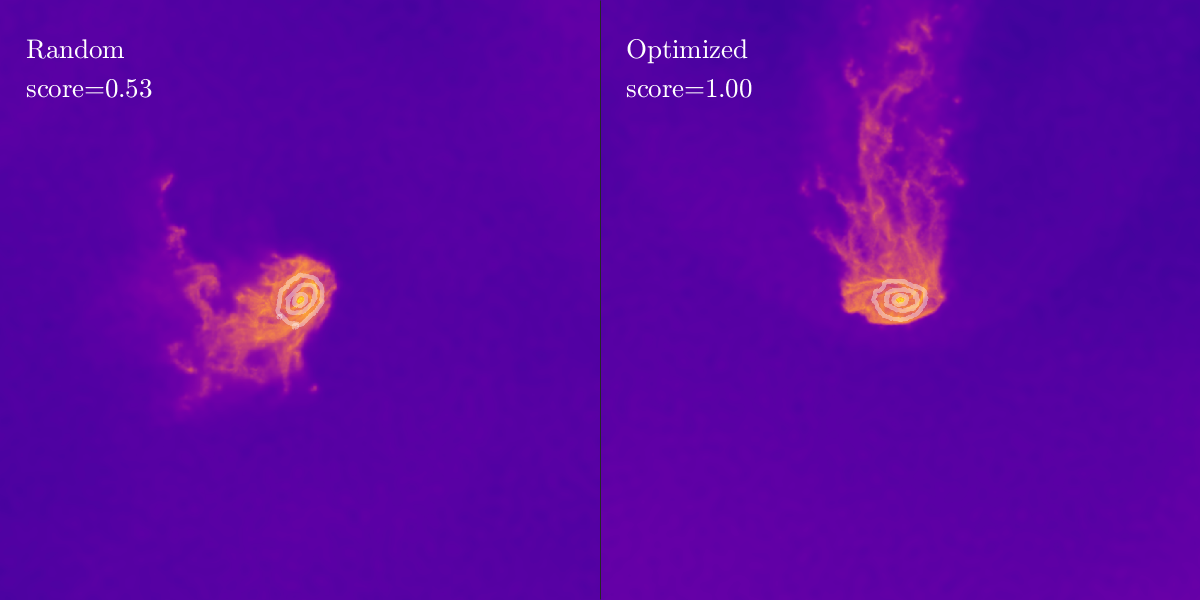}}\\
  \subfloat[] {\label{fig:angleComp_optJF_3}
  \includegraphics[width=9cm,keepaspectratio]{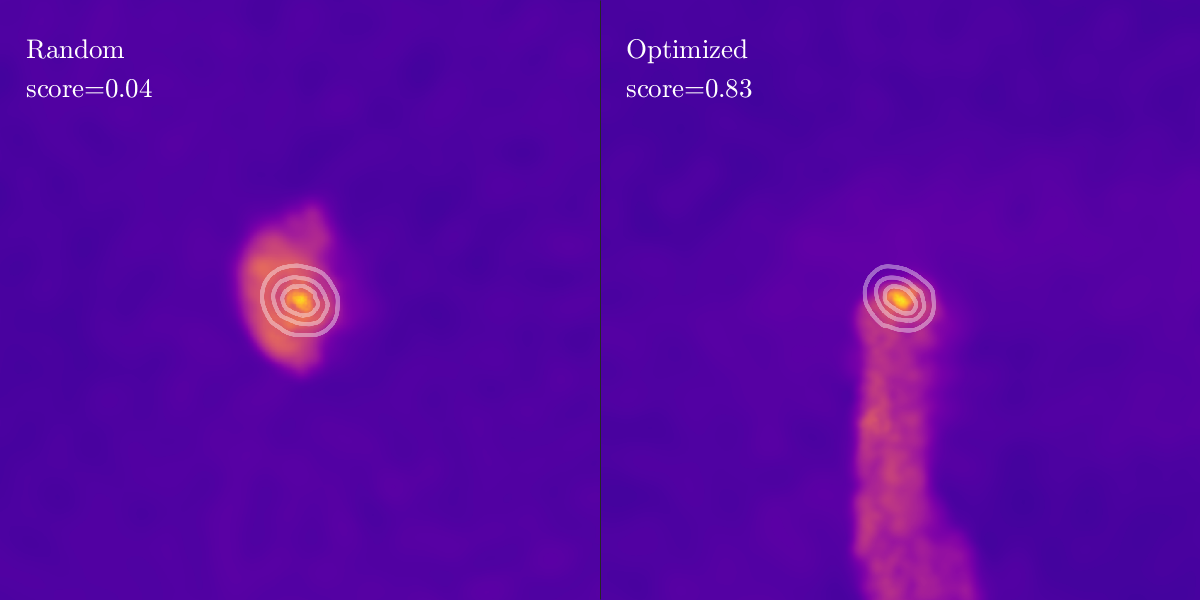}}
  \subfloat[] {\label{fig:angleComp_optJF_4}
  \includegraphics[width=9cm,keepaspectratio]{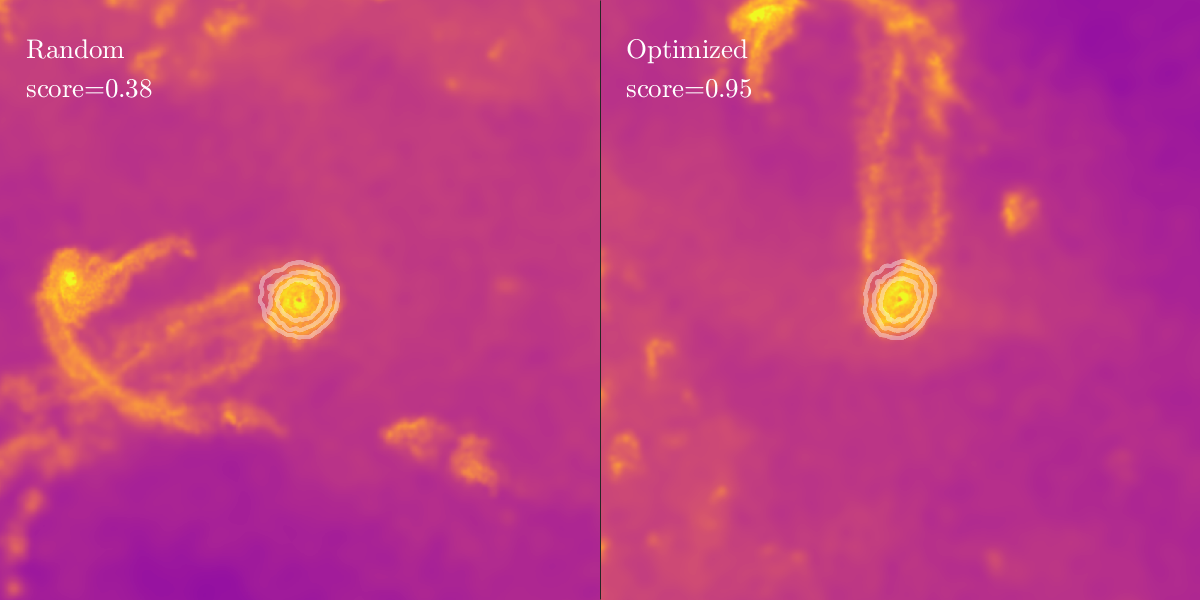}}\\ 
  \subfloat[] {\label{fig:angleComp_randJF_1}
  \includegraphics[width=9cm,keepaspectratio]{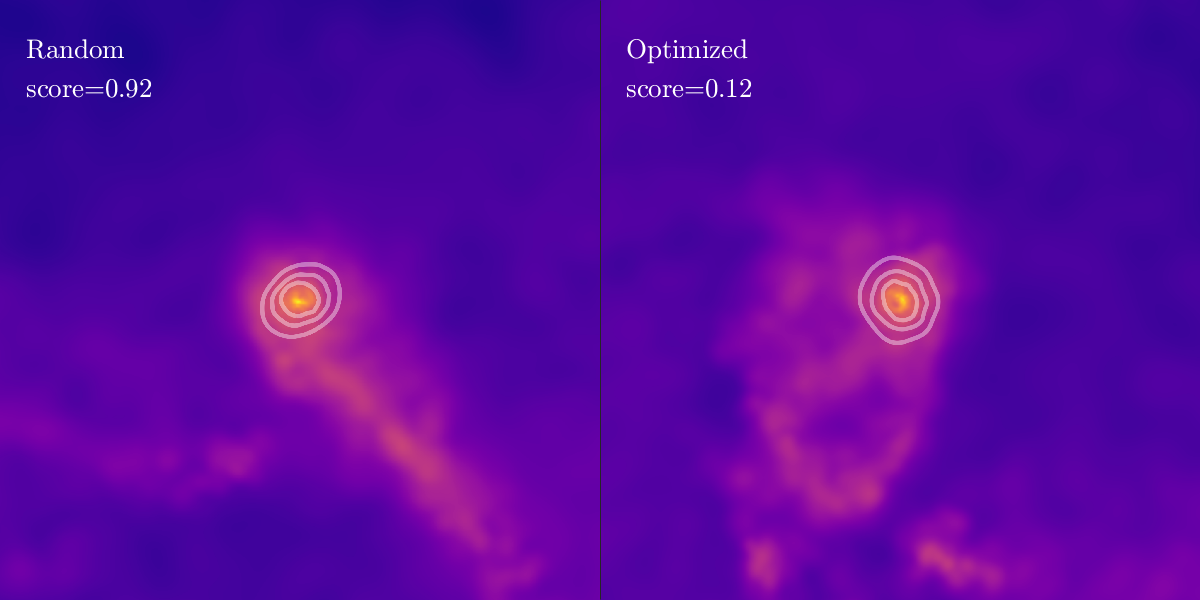}}
  \subfloat[] {\label{fig:angleComp_randJF_2}
  \includegraphics[width=9cm,keepaspectratio]{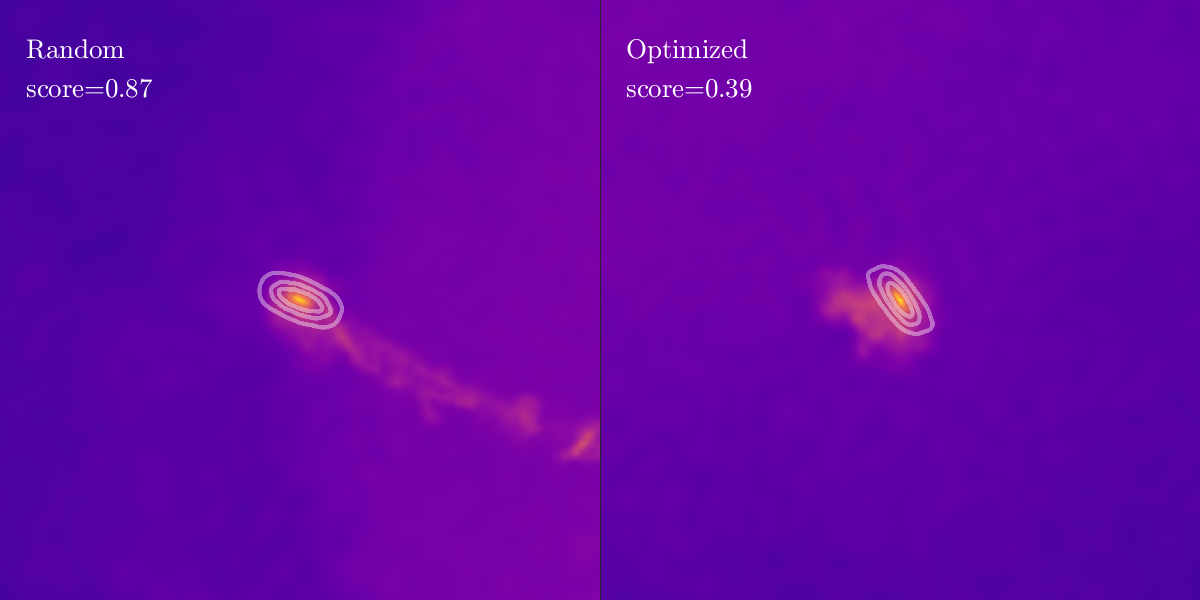}}\\
  \subfloat[] {\label{fig:angleComp_randJF_3}
  \includegraphics[width=9cm,keepaspectratio]{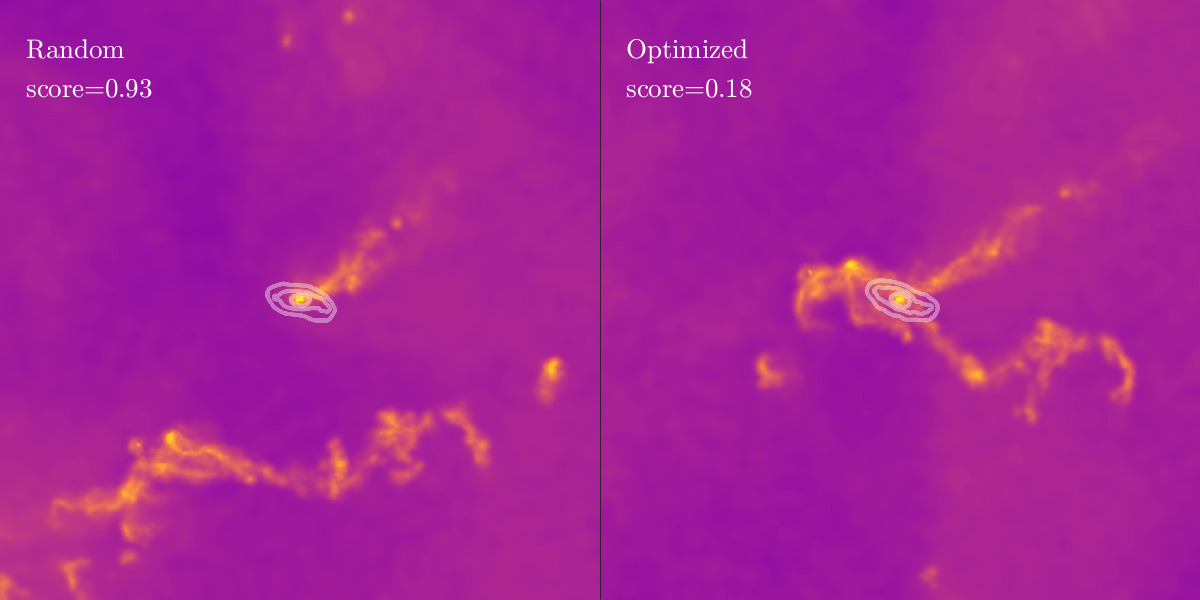}}
  \subfloat[] {\label{fig:angleComp_randJF_4}
  \includegraphics[width=9cm,keepaspectratio]{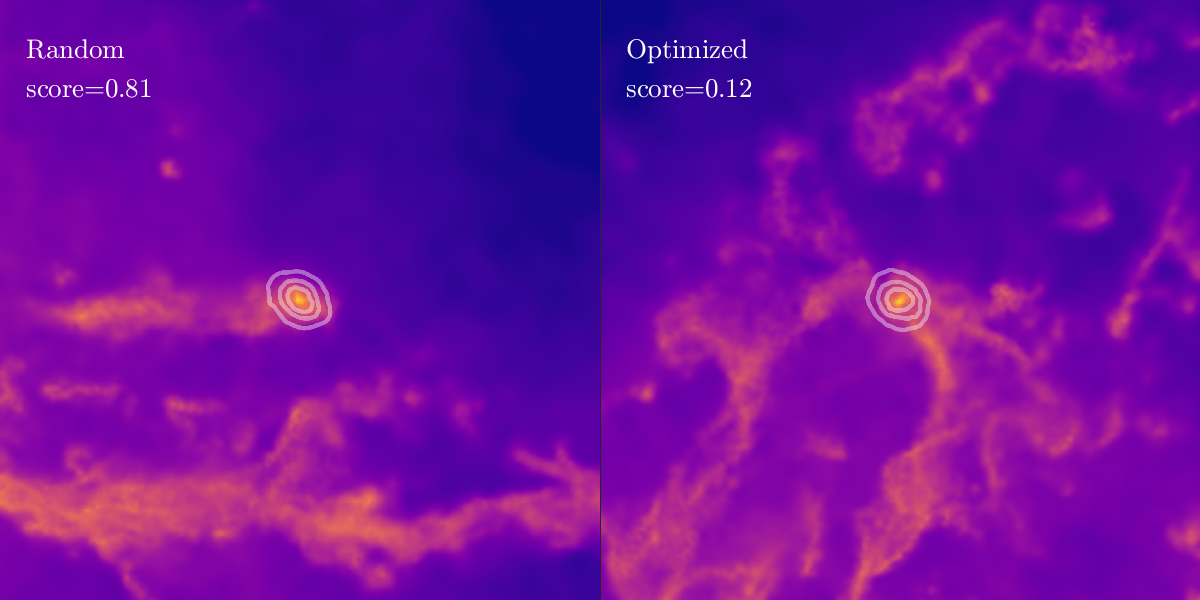}}  
  \caption{Effect of viewing angle i.e.\@ quantification of projection effects for the identification of jellyfish galaxies in gas maps. We show pairs of images for the same IllustrisTNG objects where the optimized viewing angle resulted in a higher score and JF classification (\cref{fig:angleComp_optJF_1,fig:angleComp_optJF_2,fig:angleComp_optJF_3,fig:angleComp_optJF_4}), and examples of cases where the \emph{random} viewing angle resulted in a higher score and JF classification (\cref{fig:angleComp_randJF_1,fig:angleComp_randJF_2,fig:angleComp_randJF_3,fig:angleComp_randJF_4}). The latter cases are rarer than the former ones.}
  \label{fig:viewAngleComp_mosiac}
\end{figure*}

\begin{figure*}
    \subfloat[Score distribution for the two orientations] {\label{fig:viewAngleComp_scoreHist}
  \includegraphics[width=8.5cm,keepaspectratio]{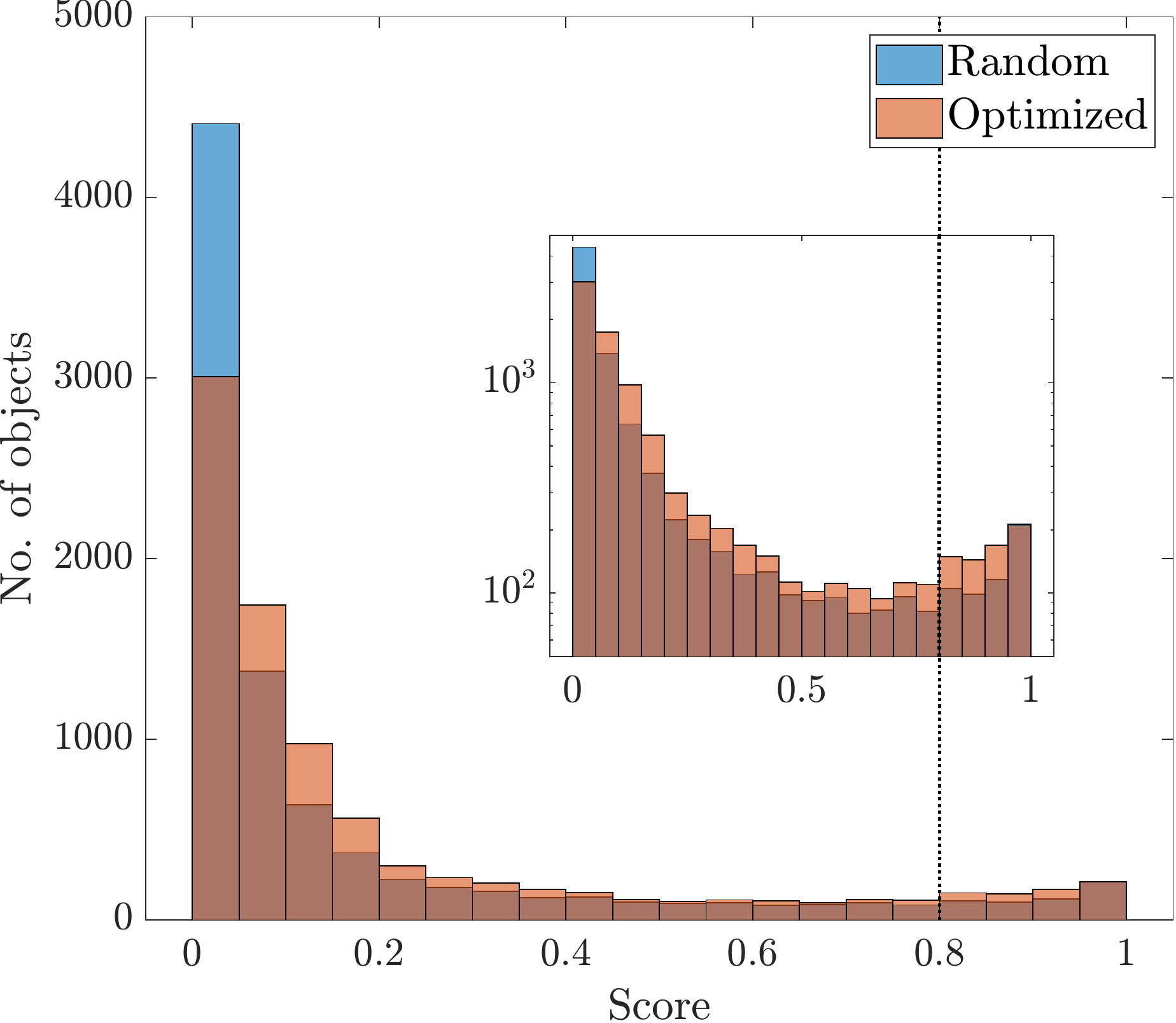}}
   \subfloat[2D histogram of score comparison] {\label{fig:viewAngleComp_heatmap}
  \includegraphics[width=8.5cm,keepaspectratio]{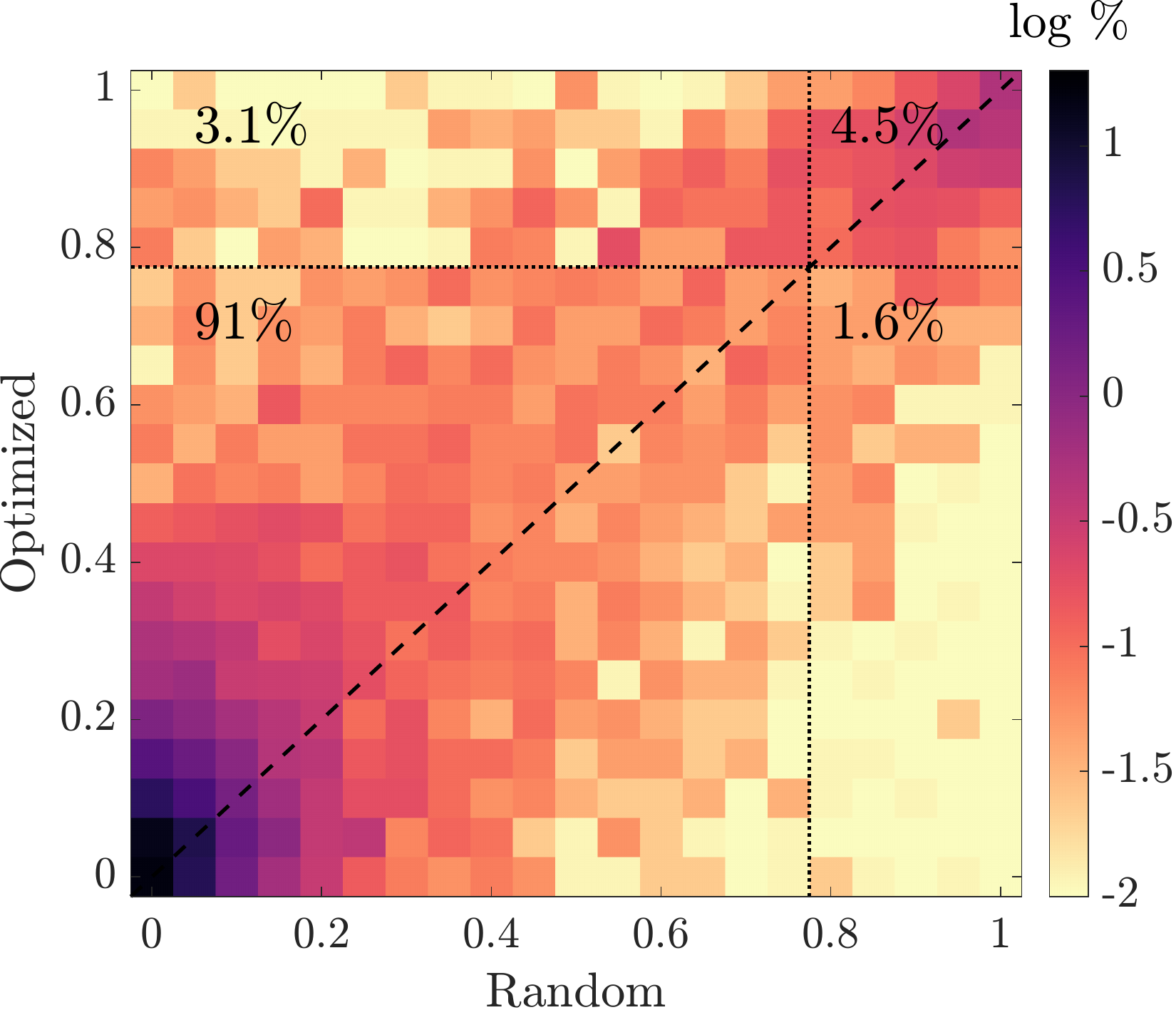}}
    \caption{Comparison of jellyfish scores (adjusted) between images shown in random vs.\@ optimized orientations. The distributions of scores for the random (blue histogram) and optimized (red histogram) projections are shown in panel \subrfig{viewAngleComp_scoreHist}, with the inset showing the histograms on a log scale. The vertical dotted lines denote our fiducial JF threshold score of 0.8. Panel \subrfig{viewAngleComp_heatmap} shows the 2D histogram comparing the classifications with different viewing angles, with the axes reporting the scores based on the randomly-oriented (x-axis) and optimized (y-axis) images. The colors are the percentage of the population in each pixel. The horizontal and vertical dotted lines mark the JF threshold score, and the percentage of the population in each quadrant is also provided.}
  \label{fig:viewAngleComp_scoreHist}
\end{figure*}

\begin{figure*}
  \centering
  \subfloat[] {\label{fig:angleComp_same_1}
  \includegraphics[width=9cm,keepaspectratio]{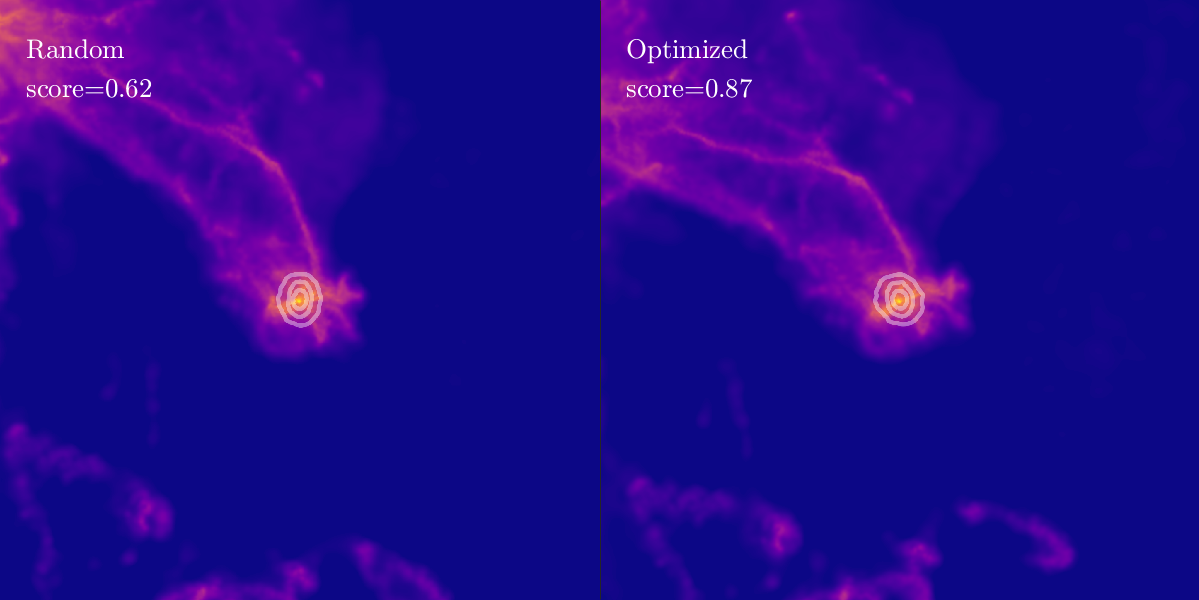}}
  \subfloat[] {\label{fig:angleComp_same_2}
  \includegraphics[width=9cm,keepaspectratio]{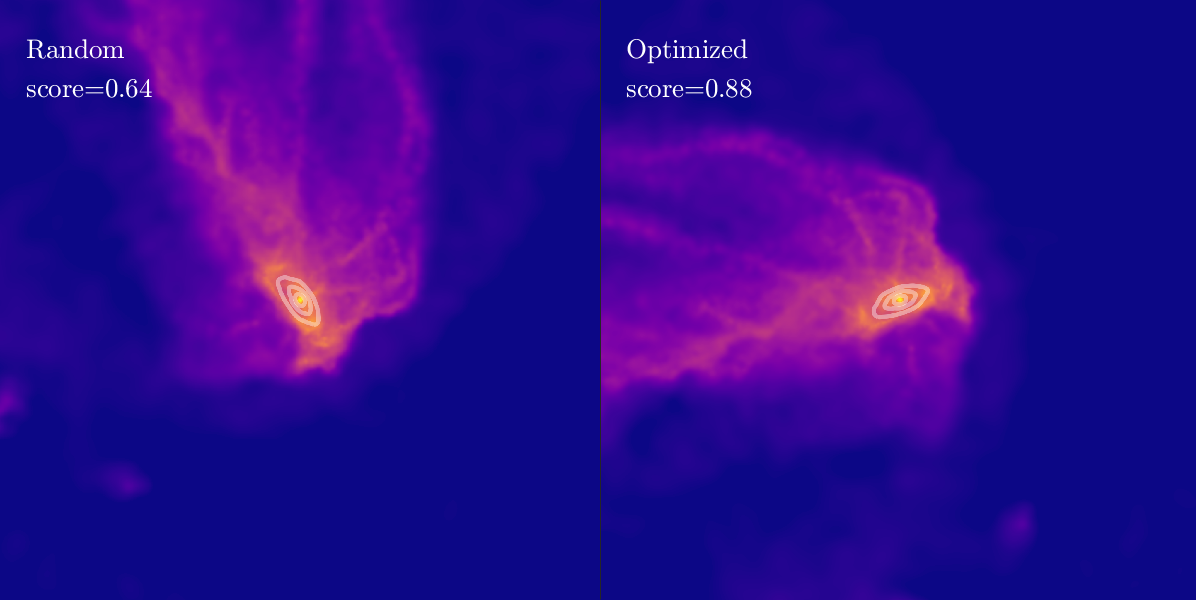}}\\
  \subfloat[] {\label{fig:angleComp_same_3}
  \includegraphics[width=9cm,keepaspectratio]{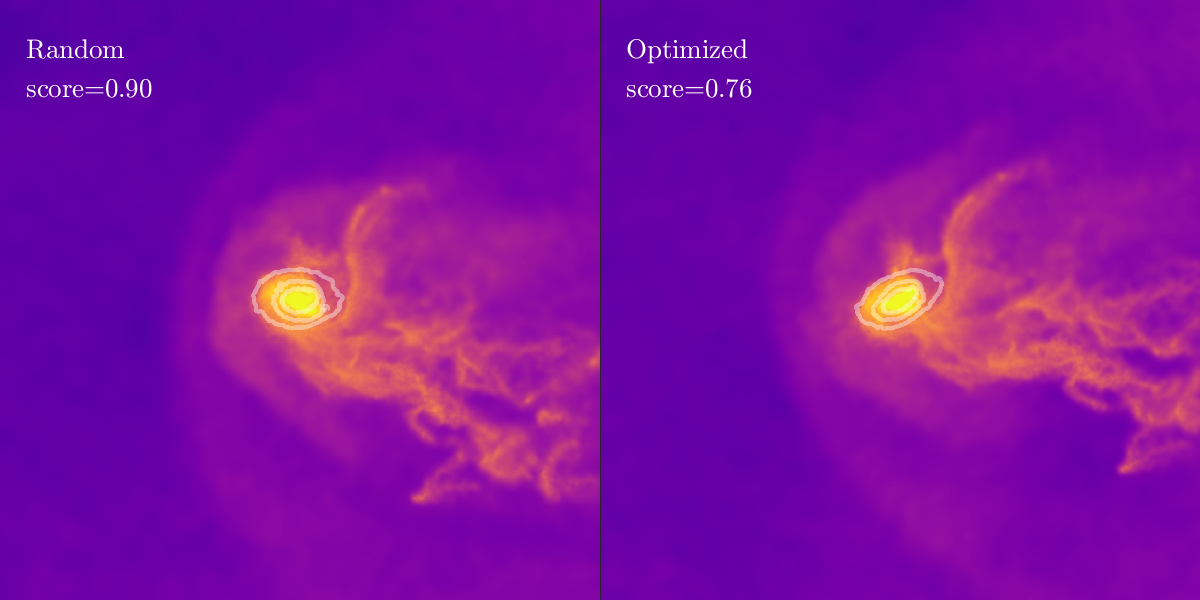}}
  \subfloat[] {\label{fig:angleComp_same_4}
  \includegraphics[width=9cm,keepaspectratio]{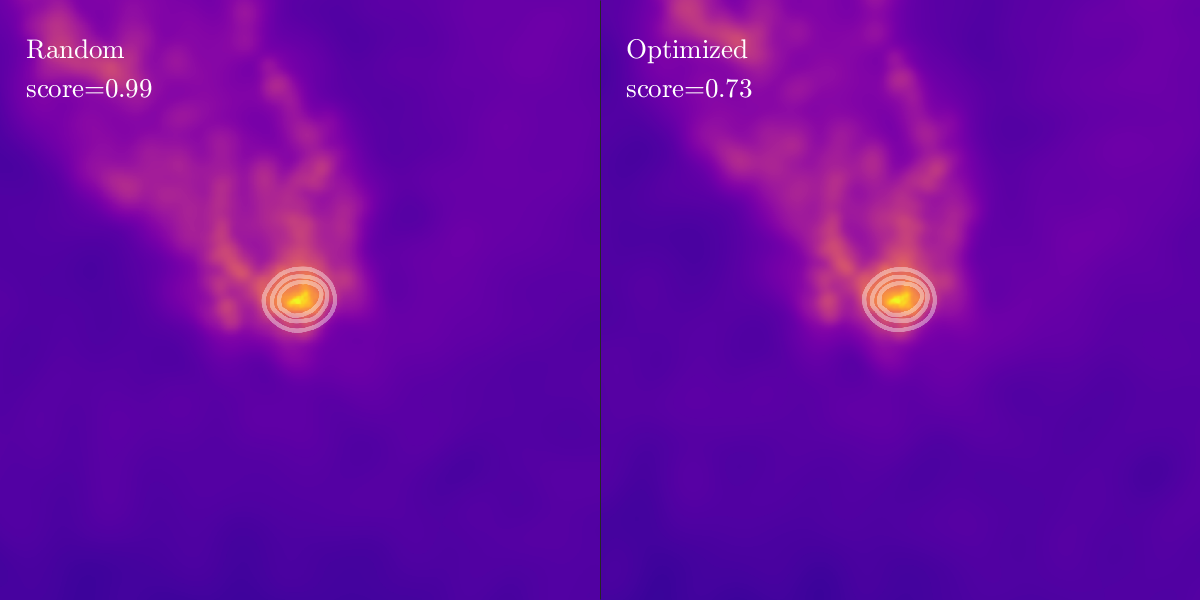}} \\ 
  \subfloat[] {\label{fig:angleComp_same_5}
  \includegraphics[width=9cm,keepaspectratio]{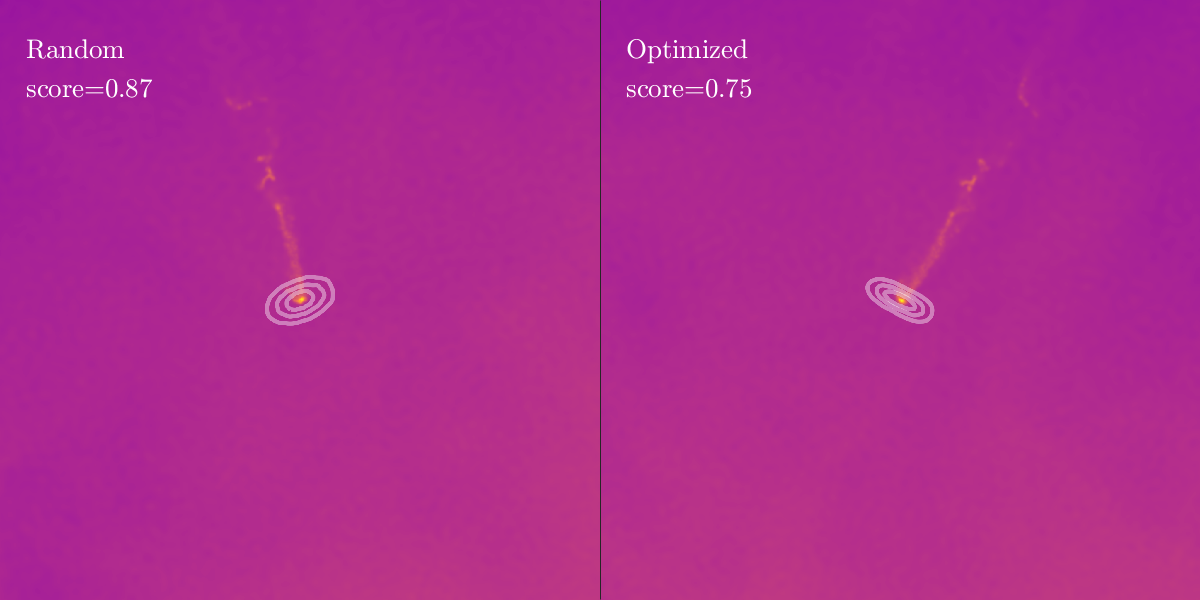}}
  \subfloat[] {\label{fig:angleComp_same_6}
  \includegraphics[width=9cm,keepaspectratio]{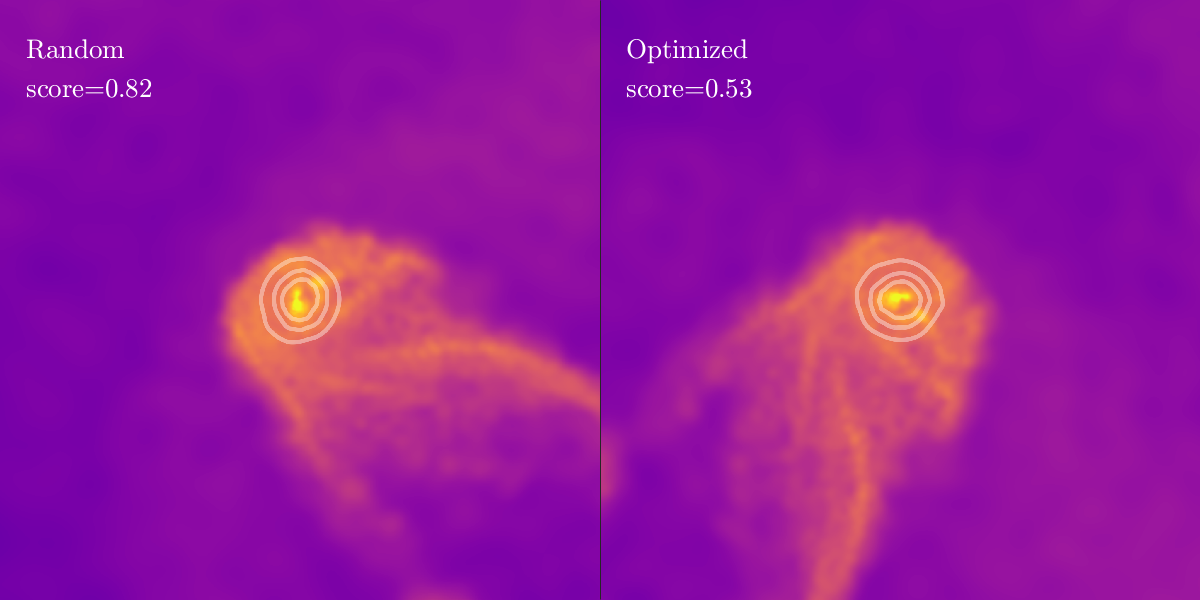}} 
  \caption{Examples of IllustrisTNG galaxies where the optimized viewing angle resulted in images that are very similar to those in random orientation, but with a difference in score (see text for details).}
  \label{fig:viewAngleComp_same}
\end{figure*}

\subsection{Random versus optimized image orientation}
\label{sec:viewAngleComp}

The signature feature of a JF galaxy is the asymmetric gas `tails' that trail the main stellar body. As such, identifying a JF galaxy depends on the direction in which one views a JF galaxy. For example, a head-on viewing angle may partially or completely obscure the tails. 

To assess this impact, i.e.\@ the projection effects that would also affect any observational survey, we select a sub-sample of objects to show in two orientations: once with a random orientation (the default for the entire inspected sample), and again with an image generated with a viewing angle optimized for identifying tails generated by RPS. For the images in preferred orientations, the gas cell and star particle positions are first rotated about the centre of the galaxy, such that the velocity vector of motion of the galaxy is within the plane of the image, but in a random orientation within the plane. The velocity vector is measured as the bulk peculiar velocity of all particles/cells belonging to the subhalo\footnotemark. These images allow us to examine the impact of image orientation on the classifications, by placing any potential gas tails parallel to the image plane, under the assumption (verified in Yun19) that the gas tails are formed in the direction opposite to the direction of motion of the galaxy.
\footnotetext{We confirmed that using the velocity of the subhalo relative to its host FoF group did not produce a significantly different image and therefore, chose to use the subhalo velocity for simplicity.}

The rest of the classification procedure is the same: the optimally aligned images are shown to inspectors without any special distinction to avoid bias in the classification. A similar comparison in the Yun19 pilot project concludes that as many as \perc{30} of JF galaxies may be missed due to the viewing angle. 

For this comparison study, we use all inspected galaxies from both TNG50 and TNG100 in two snapshots: \zeq{0,0.5}. This test sample consists of $8,762$ objects, of which $532$ (\perc{6}) are JF based on the randomly oriented images and $671$ are JF based on the optimized orientation. The composition of this test sample in terms of simulation and snapshot and the results of the visual classification are shown in \cref{tab:viewAngleSample}, as well as the number and fraction of JF galaxies based on the random and optimized orientations. Similarly to Yun19, we quantify that as many as $20-30$ per cent of JF galaxies may be missed because of an unlucky projection.

In \cref{fig:viewAngleComp_mosiac} we show several examples of galaxies that were not identified as JF in one orientation but that, upon changing the viewing angle, received a higher score above the JF threshold. In \cref{fig:angleComp_optJF_1,fig:angleComp_optJF_2,fig:angleComp_optJF_3} the randomly oriented image obscures the true extent of the gas tails resulting in images that are rightfully given low scores. The re-orientation of the image brings the gas tails into full view as well as evidence of bow shocks, which further cement a JF identification. In \cref{fig:angleComp_optJF_4} a galaxy with prominent tails is seen with an additional companion, leading to a low score (as requested in the classification instructions). On the other hand, the optimally-oriented image shows that the proximity to the other object is due to projection effects and the galaxy is given a much higher score. 

However, the re-orientation of the objects does not always lead to better classification. In \cref{fig:angleComp_randJF_1,fig:angleComp_randJF_2,fig:angleComp_randJF_3,fig:angleComp_randJF_4} we showcase 4 examples of objects classified as JF in the random orientation that received scores \emph{below} the JF threshold in the optimal orientation (out of 136 such cases). In \cref{fig:angleComp_randJF_1,fig:angleComp_randJF_2} we see that the optimal viewing angle actually makes the tails \emph{less} visible, while in \cref{fig:angleComp_randJF_3,fig:angleComp_randJF_4} the re-orientation leads to projection effects which mask the appearance of the gas tails. These cases are however twice as rare (within this testing sample) than the ones whereby the optimal projection returns higher jellyfish scores, as we quantify next. 

\cref{fig:viewAngleComp_scoreHist} quantifies the distribution of scores these objects received, both in the random and optimized orientations. As a result of the improved viewing angle, many ($\gtrsim 1000$) objects of score zeros now have slightly higher scores. In the JF region ($\geq 0.8$), there is an increase of several hundred objects. However the overall shape of the distribution does not change dramatically. 

A more detailed view can be found in \cref{fig:viewAngleComp_heatmap}, where we show a 2D histogram contrasting the scores based on the random and optimized orientations. Horizontal and vertical lines mark the JF threshold score and divide the plane into quadrants: two quadrants (bottom-left and top-right) in which the objects do not change their JF classification despite a change in score and two quadrants (top-left and bottom-right) in which the change in score led to a different classification - non-JF to JF and vice versa. The percentage of the sample in each quadrant is shown as well. 

Over half of the objects in the sample increase in score (\perc{55}), about \perc{30} decrease in score and \perc{15} do not change at all. There is a net increase of \perc{26} in the number of JF galaxies identified by the optimized orientation. Within the inspected test sample, there are 136 objects that are identified as JF galaxies in the random orientation but not in the optimized orientation. Most of these objects have scores that are close to the JF threshold in the random orientation.   

If we consider the optimized classifications as the true JF population and the random classifications as a representation of what an observational survey may produce, we see that the viewing angle may lead to missing \perc{40} of the true JF galaxies (objects in the top-left quadrant). In addition, roughly \perc{26} of the objects thought to be JF are in fact not (bottom-right quadrant), resulting in the \emph{net} of \perc{26}. However, upon inspection of the objects in this quadrant, a large number of them are indeed JF, with the optimized projection either reducing the visibility of the tails (as shown in \cref{fig:viewAngleComp_mosiac}), or presenting a very similar image as the random projections but with a reduced score. If the objects in the lower-right quadrant are still considered JF, then adding the JF objects from the new projection results in a \perc{50} increase in JF numbers.

To summarise, roughly a third of all jellyfish galaxies may go undetected because of viewing angle related issues, in agreement with Yun19. 

\subsubsection{Different scores for (nearly) identical images}\label{sec:sameImage}

Inspection of the objects in the top-left and bottom-right quadrants of \cref{fig:viewAngleComp_heatmap} yield a set of $\sim 35$ galaxies (out of 329) whose images in the two orientations are so similar as to be practically identical (though some are rotated in the image plane), but which received significantly different scores. We show 6 examples of these objects in \cref{fig:viewAngleComp_same}. The score difference for all such objects we found is on average $\sim 0.2$ in the $0-1$ score range. We discuss the implication of this finding in \cref{sec:disc_jf_identification}.

\subsection{Overall validity of the jellyfish identification method} \label{sec:disc_jf_identification}

The results in this paper and and future ones that will be based on the CJF Zooniverse project rely on the classification method used to identify JF galaxies in our inspected sample. In this Section, we discuss the advantages and drawbacks of our method and suggest improvements for the future.

Our classification method has its drawbacks, some common to all citizen-science projects and some particular to the field of JF galaxies. The primary issue is that the majority of the  classifications are carried out by volunteers who have little to no background knowledge of what a galaxy is, let alone a jellyfish galaxy. We argue that the nature of the task makes this inexperience a minor handicap: a galaxy that is clearly not a JF and one that clearly is, is relatively easy to identify. But in cases with some ambiguity, such as a messy background, multiple tails, or extended gas halos surrounding the stellar body, the choice is not always clear even to experts. In such cases, prior experience and expert knowledge is important to find the secondary features that help classifications: the existence of bow-shocks, signatures of outflows, tail continuity, etc. 

The tutorial and training module provided to volunteers partially address this lack of knowledge and experience, and provide a common level of basic knowledge required for the task. The tutorial includes several examples of the galaxies of interest, and vice versa. However, the diversity and richness in the images makes it impossible to show examples of all the different features volunteers may encounter while performing the task. 

An additional challenge is the subjectivity and bias inherent in any human-based classification. A citizen-science task must be simple and straightforward. Our task-question: ``Do you think that the galaxy looks like a jellyfish?'' is by its nature asking the inspector for a subjective answer. As the examples shown in \cref{sec:sameImage} exemplify, two nearly identical images may garner different scores from different inspectors. Even when galaxies appear identical, a built-in bias for tail direction may increase or decrease the likelihood of a `yes' answer to our question. The effect of human bias in citizen-science has also been observed when selecting chirality and spiral structure in the Galaxy Zoo project \citep{land_galaxy_2008,hayes_nature_2017}.

One way to alleviate these issues is to expand the number of inspectors per image who determine the final score. If enough people cast their votes, personal inexperience, errors in judgement and biases can be compensated for. The common wisdom is that the larger the number of votes, the more dependable the result is (as demonstrated in \cref{sec:jfDef}). However, this choice must also take into account practical considerations: requiring too many votes per image may extend the duration of the classification phase. In addition, if a project appears to be advancing very slowly, volunteers may lose interest and abandon the project altogether. This choice must be made while carefully balancing scientific necessity and resource-management.\footnote{The main resource is the volunteer pool, but maintaining a citizen-science project also requires significant time and effort from the research teams in preparing the project, interacting with volunteers via message boards, volunteer recruitment efforts, and so on.} While our choice of 20 independent votes per image may have led to some cases of mis-classification, we believe it achieves the dual goals of providing a mostly reliable classification within a reasonable time-frame. 

As a result, we have carried out post-processing steps to increase our confidence in the results. First, the classification comparison between the CJF project and the pilot projects (\cref{sec:visClassCompare}) yields good agreement between volunteers and expert classifications. Second, the inspector-weighting scheme marginalizes over inspectors with little experience or bad track-records, and adds weight to experts (\cref{sec:InspectorWeighting,sec:app_InspectorWeightScheme}). Third, the high threshold for identifying JF galaxies was chosen to achieve a very pure JF sample, even at the expense of leaving some JF galaxies out. 

Thanks to these steps we are confident that our final sample is indeed comprised of JF galaxies and can be relied upon to produce accurate information about the JF population. 

Additional measures may further improve the quality of citizen-science classifications in the future. For example, carrying out expert classifications on a random sampling of objects. This allows the team to gauge how well the classification aligns with an expert opinion while adding the expert vote to final tally. This may be particularly useful for borderline scores just below the threshold, where there are surely JF galaxies that have been missed, as is evident in \cref{fig:score_mosaic,fig:scoreBin4_mosaic}. Studies into the bias and overall quality of classifications may also be useful, e.g., presenting the same galaxy from various viewing angles to further explore the dependence of classification on viewing angle, exploring image-enhancing techniques such as background subtraction and showing the same \emph{image} multiple times rotated in different directions to detect possible bias. 


\section{Summary and conclusions}\label{sec:summary}

In this work we have presented the results of the Cosmological Jellyfish (CJF) Zooniverse project: a citizen science effort to identify and classify jellyfish (JF) satellite galaxies in the IllustrisTNG simulations. Over $6,000$ volunteers classified nearly $90,000$ images of over $80,000$ satellite galaxies.

The inspected galaxy sample is unprecedented in terms of its size, the satellite stellar mass and host mass ranges, and the redshift coverage: it contains all satellite galaxies in the TNG50 and TNG100 simulations above a satellite stellar mass of $\tenMass{8.3}$ and $\tenMass{9.5}$, respectively, and with some gas. There is no explicit minimal or maximal mass cut for the hosts, which range from $\tenMass{10.4}$ up to $\tenMass{14.6}$ and include over $2,000$ group-sized and over $100$ cluster-sized systems (TNG100). 10 snapshots of the TNG100 box were included between \zeq{2} and \zeq{0} (intervals of $\sim 1 \units{Gyr}$). For TNG50, 4 snapshots between \zeq{2} and \zeq{0.5} were examined, identical in redshifts to the TNG100 snapshots, and in addition, \emph{all} available snapshots between \zeq{0.5} and \zeq{0} were inspected allowing us to examine the state of a galaxy (JF or not) over time intervals of $\sim 150\units{Myr}$. Some of the images hence include the same galaxies across several snapshots (\cref{sec:sample} and \cref{sec:branches}).

The aim of our classification process was a sample of simulated galaxies that could be confidently called ``jellyfish'' galaxies. We preferred a pure sample rather than a complete one and this choice set the guidance we provided the volunteers with, the algorithm chosen to assess the inspectors, and the threshold chosen to identify JF galaxies. 

Images of the gas density field for all galaxies in the inspected sample were created along a random orientation (\cref{sec:visClassImage}), and presented on the Zooniverse platform to at least 20 volunteers with a simple yes/no question: ``Does this galaxy look like a jellyfish?'' (see \cref{sec:visClassProcess} for details), resulting in a set of raw scores ranging between 0 and 20. To validate the process, a comparison against the classifications by a team of experts on two previous smaller visual-inspection campaigns was carried out: the \cite{yun_jellyfish_2019} sample (Yun19, based on TNG100) and a pilot project of TNG50 images. The agreement with the TNG50 pilot is excellent, whereas  the agreement with the Yun19 sample is slightly lower, likely due to differences in the types of images used for the classification (\cref{sec:visClassCompare}). In any case, both comparisons lend support to the usage of the classification by non-experts, with a high enough number of classifications per object, for the purpose of identifying jellyfish galaxies in gas maps.

Furthermore, we have implemented an inspector-weighting algorithm to minimize the impact of inexperienced volunteers and those that consistently voted against the consensus, and added weight to experts inspectors, i.e., research team members (\cref{sec:InspectorWeighting,sec:app_InspectorWeightScheme}). This results in a set of \emph{adjusted} scores that we recommend for usage and that range between 0 (definitely \emph{not} a JF) and 1 (most definitely a JF), with the threshold value for a JF galaxy set to be 0.8 (\cref{sec:jfThreshold}). 

An additional set of images, produced along an optimized orientation for tail identification for a subset of the TNG50 and TNG100 galaxies (at \zeq{0.5} and \zeq{0}) allowed us to carry out a preliminary analysis of the effect of viewing angle i.e.\@ projection effects. By a conservative estimation, a third of all JF galaxies are mis-classified (i.e.\@ missed) due to the viewing angle (\cref{sec:viewAngleComp}). A set of galaxies in which the two orientations give practically-identical images, but which receive different scores (by $\sim 0.2$ on average in the $0-1$ score range), showed the inherent bias issues in citizen-science projects.  

\subsection{Guidelines for the use of the CJF dataset}\label{sec:conc_guidelines}

With this publication we make the full dataset of jellyfish scores generated by the CJF project publicly available via the \href{https://www.tng-project.org/data}{IllustrisTNG website}, for future usage and reference. The dataset includes separate galaxy catalogs for each snapshot in the TNG50 and in the TNG100 simulations. Each catalog lists the raw scores for each galaxy, as well as the adjusted score (after implementing the inspector weighting), both based on the map inspected with random orientations. The total weight of the combined inspectors and the number of experts that viewed each galaxy image is also included in the dataset. In addition, the scores for the optimized orientation images are also available. In this paper, we have motivated our recommended score threshold of 0.8 (in the $0-1$ range of adjusted scores) above which a galaxy can be called ``jellyfish'' and we have inspected the basic demographics of so-defined JF galaxies from the IllustrisTNG simulations. In \cref{sec:jfDef} we suggest a statistical framework that can be used to gauge the confidence level associated with this or an alternative choice of threshold score.

\subsection{Demographics of the JF population}\label{sec:conc_results}

Our JF sample contains $5,307$ galaxies from the TNG50 and TNG100 simulations, comprising \perc{\sim 7} of the \emph{inspected} satellite population. As a reminder, the latter does not include the entire populations of simulated satellites above a certain stellar mass but includes only those with at least some gravitationally-bound gas -- a fundamental requirement to witness ram-pressure stripping in action.

{\bf General demographics: } Over the entire population of inspected satellites, jellyfish galaxies are more frequent in higher-mass hosts and for smaller satellite stellar masses. Among the lower-mass inspected satellites ($\lesssim \tenMass{10}$) in cluster-sized systems, about \perc{40} are JF. Within he inspected satellite population in group-sized system JF galaxies comprise \perc{\sim 25}.
The JF galaxies are more abundant at low masses than their non-JF counterparts, with very few JF of stellar mass above $\tenMass{10}$  - This is due in part to the increased gravitational binding force, but also may be affected by the onset of the kinetic feedback mode in the IllustrisTNG physical model (at $\sim \tenMass{10.5}$), that may deplete the gas from the satellite. 

The distribution of host masses for JF galaxies is skewed to higher masses compared to general satellite population. Most halos above $\tenMass{13.5}$ contain JF galaxies and nearly every cluster sized system, $\tenMass{14}$ and above, has a JF population, with JF fractions of \perc{10\--40}. (\cref{sec:jfFrac_demograf,fig:jfrac_smass_hmass,fig:hostHistograms_jfFrac,fig:props_sm_hm_rpos})

{\bf JF in low mass hosts: }
Over half of all JF are found in groups and clusters. While JF fractions increase with halo mass, the peak of the distribution of JF in halos mass is at the group scale, $\tenMass{13.5}$, due to the much higher number of hosts in lower masses. However, nearly $1,000$ JF galaxies reside in hosts of mass less than $\tenMass{13}$ of which 200 are in hosts of mass $\tenMass{12.5}$ and below. The least massive hosts that host JF galaxies are $\tenMass{11.5}$ in TNG50 and $\tenMass{12.5}$ in TNG100. (\cref{sec:low_mass_host,fig:hostHistograms_jfFrac,fig:props_sm_hm_rpos})

{\bf JF beyond $\Rv$:}
While most JF are concentrated within $\Rv$ (compared to the general satellite population), a full \emph{quarter} of all JF may be found beyond it. A few percent may even be found beyond $2\Rv$. These JF in the outskirts are more likely to be of lower masses and to reside around higher mass hosts. (\cref{sec:beyond_rv,fig:props_sm_hm_rpos,fig:jfFrac_dist})

{\bf JF at high redshifts:} Our inspected sample extends up to \zeq{2}, and we find JF galaxies at all redshifts. At redshifts of \zeq{1.5\--2},  they are found in small groups and proto-clusters ($\tenMass{12\--14}$) where they comprise \perc{\sim 10} of the satellite population in these systems (there are $\sim 40$ such objects that represent \perc{3} of all halos in this epoch and in this mass range) .  
JF fractions as a function of host mass show little redshift dependence, while the JF fractions in satellite stellar mass bins grow with decreasing redshift. (\cref{sec:jfFrac_demograf,fig:jfrac_zred_massBins,fig:jfFrac_inHost_TNG50,fig:jfFrac_inHost_TNG100} )

The results presented above showcase the enormous potential this dataset represents, thanks to the combination of the large volumes simulated within the IllustrisTNG project with the capacity of citizen-science projects such as those configurable on Zooniverse . The ability to study such a large sample of JF galaxies, which span a large range of properties, environments and cosmic epochs, will enable us to address many open questions about galaxy evolution in groups and clusters and in particular about this important phase in the life-cycle of satellite galaxies. In two companion papers, we study JF galaxies along their evolutionary paths and learn when and where they lose their cold gas (\textcolor{blue}{Rohr et al 2023} and quantify their star-formation properties and histories (see \textcolor{blue}{G{\"o}ller 2023,}).

 The sheer size of the inspected and JF samples provided here will also allow us to expand the identification to even larger datasets of simulated and observed galaxies (e.g. to the TNG300 simulation and others), via machine learning. 
The dataset presented in this work can constitute a valuable resource both for training  convolutional neural networks -- preliminary tests in this direction with the Yun19 data showed much promise -- as well as for comparison to the results of unsupervised or self-supervised classification schemes.

Finally, whereas the primary goal of the CJF Zooniverse project was a scientific one, offering a (surprisingly) fast way of carrying out visual classification of a large number of images, an additional, and in our opinion highly valuable, goal was achieved:  the exposure of the general public to the research of galaxy evolution in general and the study of JF galaxies in particular. This was done through the educational material on the website, interaction on the message boards, and volunteer recruitment efforts (press releases, public talks and others). We hope this may lead, in some small way, to increasing science literacy and to inspiring scientific study, especially in younger audience.

\section*{Acknowledgements}

This publication uses data generated via the \href{https://www.zooniverse.org/}{Zooniverse.org} platform, development of which is funded by generous support, including a Global Impact Award from Google, and by a grant from the Alfred P. Sloan Foundation. We wish to extend our thanks to the team at Zooniverse with their advice and assistance in building and running this project. We also thank the thousands of volunteers who invested their time and effort to assist us in this project. ER is a fellow of the International Max Planck Research School for Astronomy and Cosmic Physics at the University of Heidelberg (IMPRS-HD). DN acknowledges funding from the Deutsche Forschungsgemeinschaft (DFG) through an Emmy Noether Research Group (grant number NE 2441/1-1). GJ acknowledges funding from the European Union’s Horizon 2020 research and innovation programme under grant agreement No. 818085 GMGalaxies. The authors also thank Martina Donnari and Allison Merritt for their participation in the pilot classification project that we performed to verify the function of the ``Cosmological Jellyfish'' Zooniverse website and that we have used to assess its consistency. EZ thanks Yehuda Hoffman for fruitful advice and discussions. 


\section*{Data Availability}

All the simulation data of the IllustrisTNG simulations, and hence of TNG100 and TNG50 used here, are publicly available and accessible at \url{https://www.tng-project.org/data} \citep[][]{nelson_illustristng_2019}. With this paper, we release the scores of the fiducial jellyfish classification justified, produced and analyzed here, so that interested researchers and citizen scientists can make further usage of our simulation data and of the visual classification.

The Scripts used for the data analysis and figures  can be found here: \href{https://github.com/eladzing/matlab_scripts.git}{https://github.com/eladzing/matlab\_scripts.git}. Please contact the corresponding author for assistance at \href{mailto:elad.zinger@mail.huji.ac.il}{elad.zinger@mail.huji.ac.il}. 

\bibliographystyle{mnras}
\bibliography{refs}


\appendix

\begin{figure}
  \centering
  \subfloat[Classification Histogram] {\label{fig:demograf_num_mstar_50}
     \includegraphics[width=9cm,keepaspectratio]{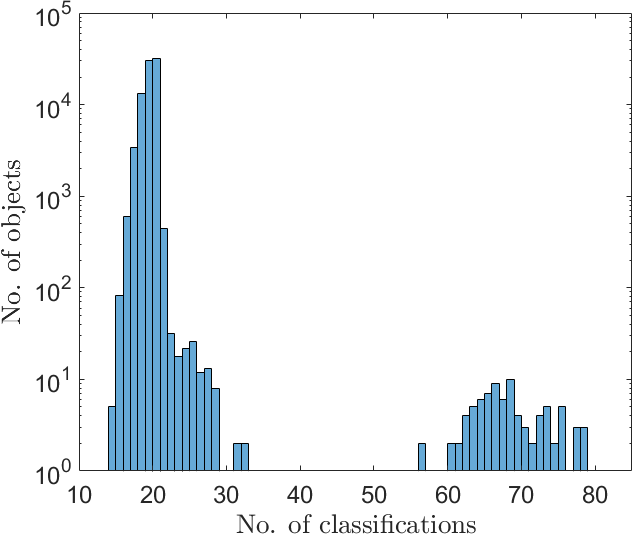}}\\
     \subfloat[Weight Histogram] {\label{fig:demograf_frac_mstar_50}
       \includegraphics[width=9cm,keepaspectratio]{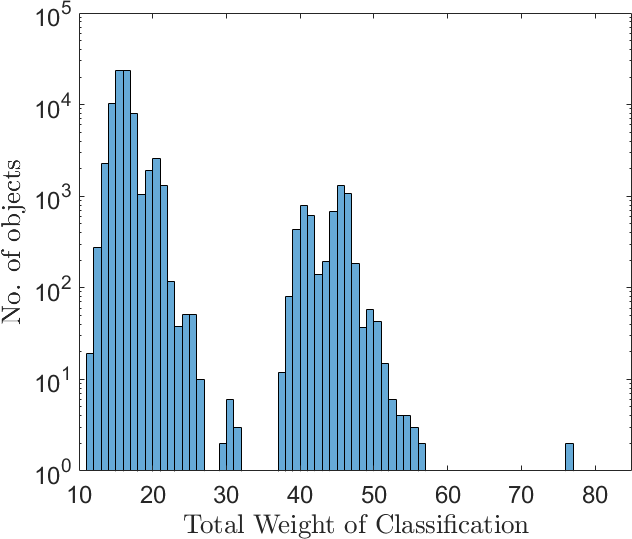}}
              \caption{The number of classifications for the objects in the inspected sample and the total weight of the inspectors. }
  \label{fig:clsNum_weight_hist}
\end{figure}

\section{Inspector ranking}\label{sec:app_InspectorWeightScheme}

The following describes the procedure for generating weights for all inspectors in the CJF project. Inspector identification is based on the user-name assigned to inspectors who are logged into the platform. If an inspector performs a classifications while not logged on, a random identifier is assigned that signifies that the user is not logged on. 

For each inspector we assign a set of weights based on different criteria, with the final weight determined by a combination of these as detailed below.

\begin{itemize}
    \item{\bf Inexperienced inspectors - } Inspectors that have classified fewer than 10 images are all given a weight of 0.5. As can be seen in \cref{fig:classificationHistogram}, this applies to several hundreds of inspectors.  
    \item{\bf Expert inspectors - } Members of our research group are all deemed experts and given a weight of 5. Classifications from the previous classification projects, the TNG50 Pilot and the Yun19 projects, are also treated as being carried out by experts. 
    
    \item{\bf Performance on high score objects - } We assess how each inspector performs compared to their peers when classifying high-score objects. In this way we are giving more weight to inspectors who have shown they can correctly identify a JF when presented with one. The score of an object is the fraction of `yes' votes an object receives in the initial tally. High-score objects are defined as objects that received a score greater than 0.5, and the higher the score of the object, the more important it is in setting the inspector ranking. 
    
    The inspector weight is defined as the weighted mean of all the `yes' (1) votes of a given inspector over all high-score objects
    \begin{equation}\label{eq:weightHi}
    W_\mathrm{high}=\frac{\sum_{\mathrm{score\geq0.5}} w_\mathrm{o} v}{\sum_{\mathrm{score\geq0.5}} w_\mathrm{o} }, 
    \end{equation}
    With $w_\mathrm{o}=2\times\mathrm{score}-1$  being the object weight and $v$ being the vote given for each object (either 0 or 1).
    
    \item {\bf Identifying up/down-voters - } We identify up-voters as inspectors who consistently mis-identify JF galaxies, voting `no' on objects that received predominately 'yes' votes (down-voters) and vice versa (up-voters). We assign a rank to each inspector to appraise whether or not they are an up-/down-voter by finding the weighted mean of votes cast against the consensus on objects in the low-/high-end of the initial tally. The higher the rank, the more likely an inspector \emph{is} a an up-/down-voter. 
    
    The ranking of down-voters is calculated on objects of initial score 0.75 and above: 
    \begin{equation}\label{eq:downVote}
    R_\mathrm{down}=\frac{\sum_{\mathrm{score\geq0.75}} w_\mathrm{d} \left(1-v\right)}{\sum_{\mathrm{score\geq0.75}} w_\mathrm{d} }, 
    \end{equation}
    With the object-weighting set by \cref{tab:updownWeight}. Similarly, the up-vote ranking is set by 
    \begin{equation}\label{eq:upVote}
    R_\mathrm{up}=\frac{\sum_{\mathrm{score\leq 0.25}} w_\mathrm{d} v}{\sum_{\mathrm{score\leq0.25}} w_\mathrm{d} }, 
    \end{equation}

    An inspector that has voted against the consensus at least 3 times on objects of scores $0.75$ and above, and has a down-vote ranking of at least $0.5$ is designated a down-voter and given an inspector weight of $0$, i.e., all of their classifications are ignored. In a similar fashion, an inspector that has voted against the consensus at least 3 times on objects of scores 0.25 and below, and has a up-vote ranking of at least 0.5 is designated an up-voter, and given an inspector weight of 0. 
    
    In total we identified 125 down-voters and 213 up-voters.
    
\end{itemize}

The overall inspector weight is set to be $W_\mathrm{high}$, unless the inspector is either inexperienced (weight of 0.5) or an expert (weight of 5), or if they have been identified as and up-voter or down-voter and are removed completely from tally (weight of 0). 

In \cref{fig:clsNum_weight_hist} we show the histogram for the number of classifications for the objects in the inspected sample as well as the histogram of the sum of the inspector weights. Although there are now objects that were classified by fewer than 20 inspectors, all objects were classified by at least 13 inspectors. 

\begin{table}
\begin{tabular}{llc}
Object Score (low-end) & Object Score (high-end) &Object Weight \\ \hline
0.05 and below  & 0.95 and above   & 5             \\
0.05-0.1 &  0.9-0.95     & 4             \\
0.1-0.15 & 0.85-0.9   & 3             \\
0.15-0.2 & 0.8-0.85     & 2             \\
0.2-0.25 & 0.75-0.8   & 1             \\
otherwise   & & 0            
\end{tabular}
\caption{Object weighting for the up-/down-voter ranking.} 
\label{tab:updownWeight}
\end{table}

\section{Sample of Galaxy Images in score bins}\label{sec:app_imageSample}

\begin{figure*}
  \centering
       \includegraphics[width=18cm,keepaspectratio]{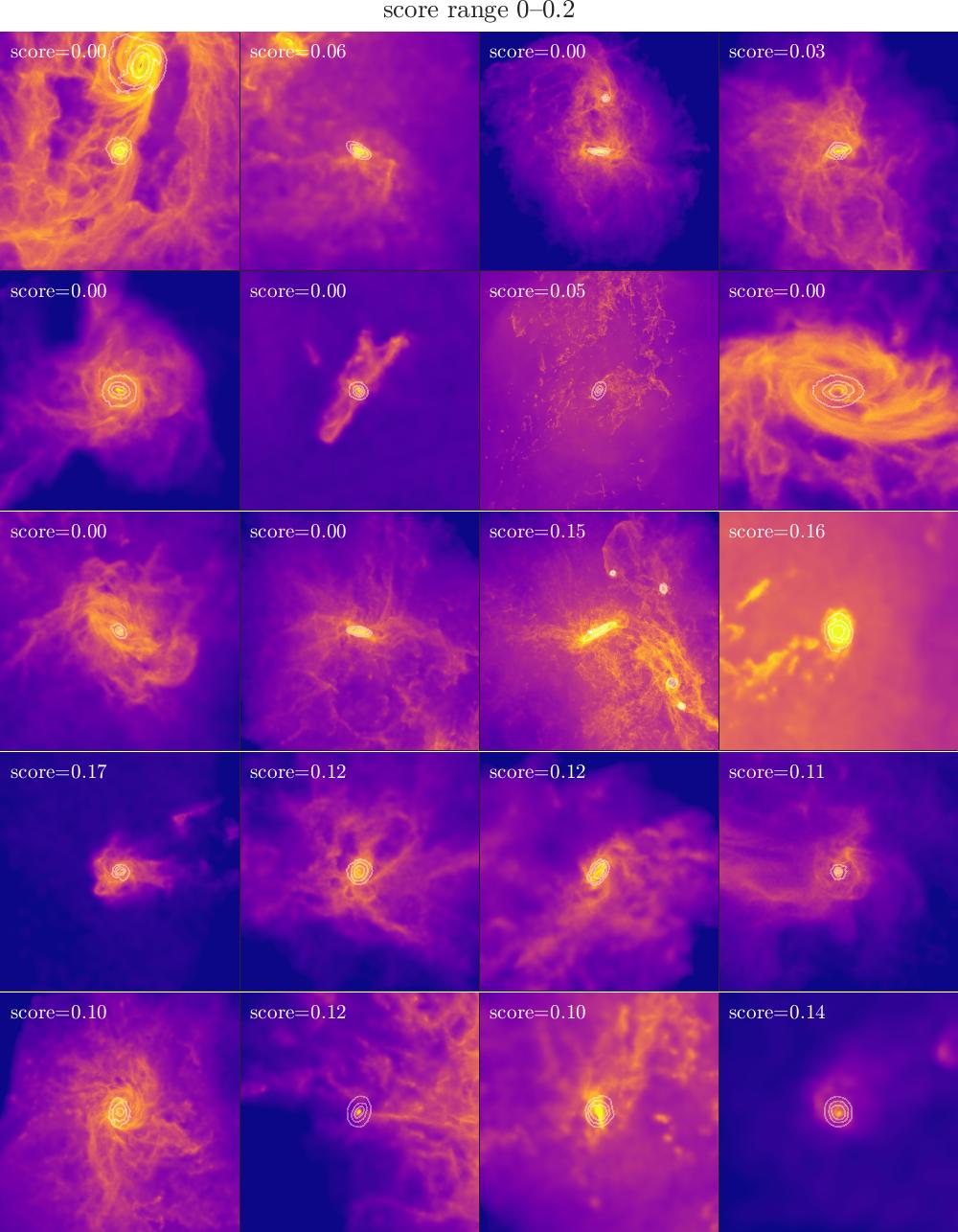}
       \caption{Sample of galaxy images of scores 0--0.2}
  \label{fig:scoreBin1_mosaic}
\end{figure*}

\begin{figure*}
  \centering
       \includegraphics[width=18cm,keepaspectratio]{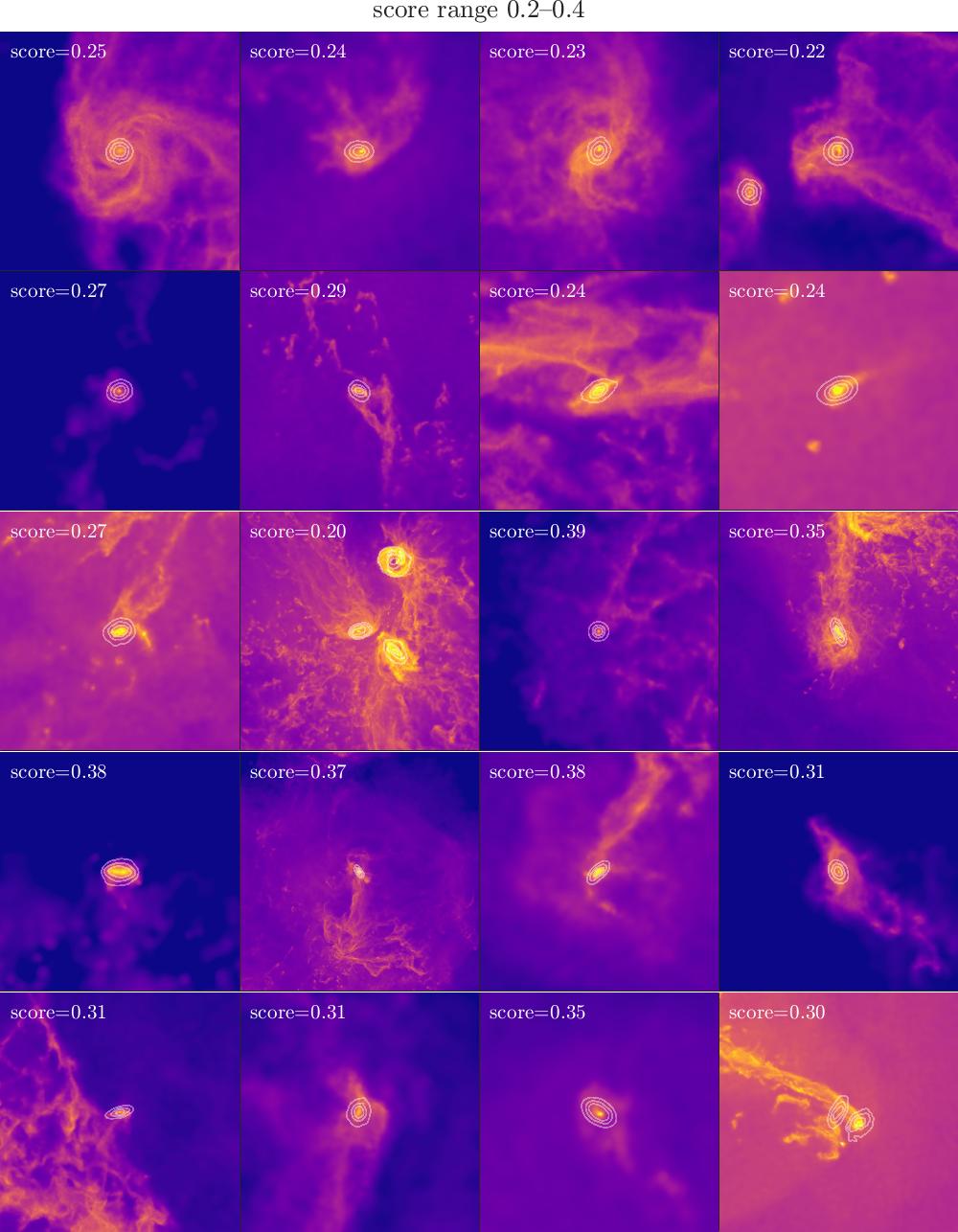}
       \caption{Sample of galaxy images of scores 0.2--0.4}
  \label{fig:scoreBin2_mosaic}
\end{figure*}

\begin{figure*}
  \centering
       \includegraphics[width=18cm,keepaspectratio]{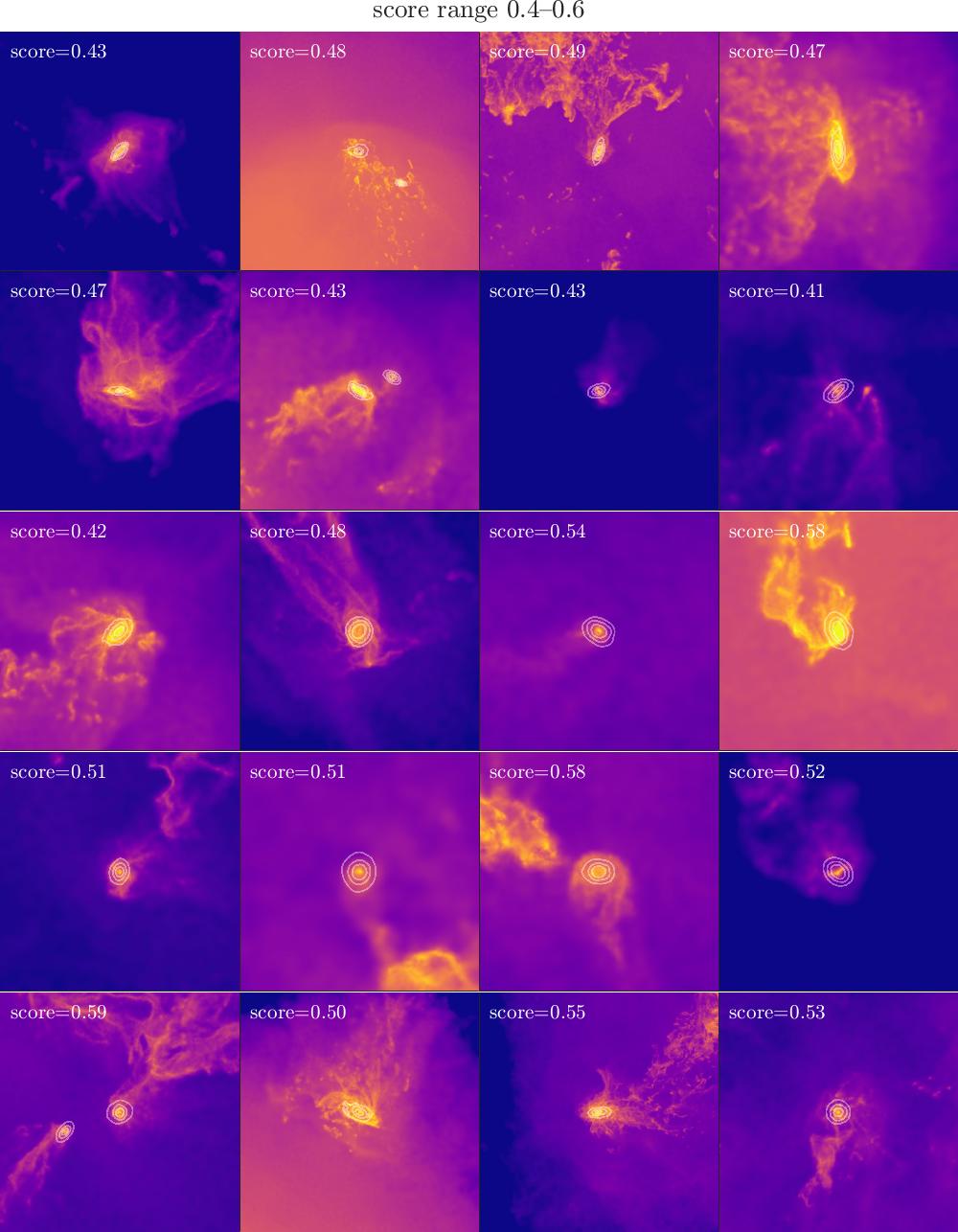}
       \caption{Sample of galaxy images of scores 0.4--0.6}
  \label{fig:scoreBin3_mosaic}
\end{figure*}

\begin{figure*}
  \centering
       \includegraphics[width=18cm,keepaspectratio]{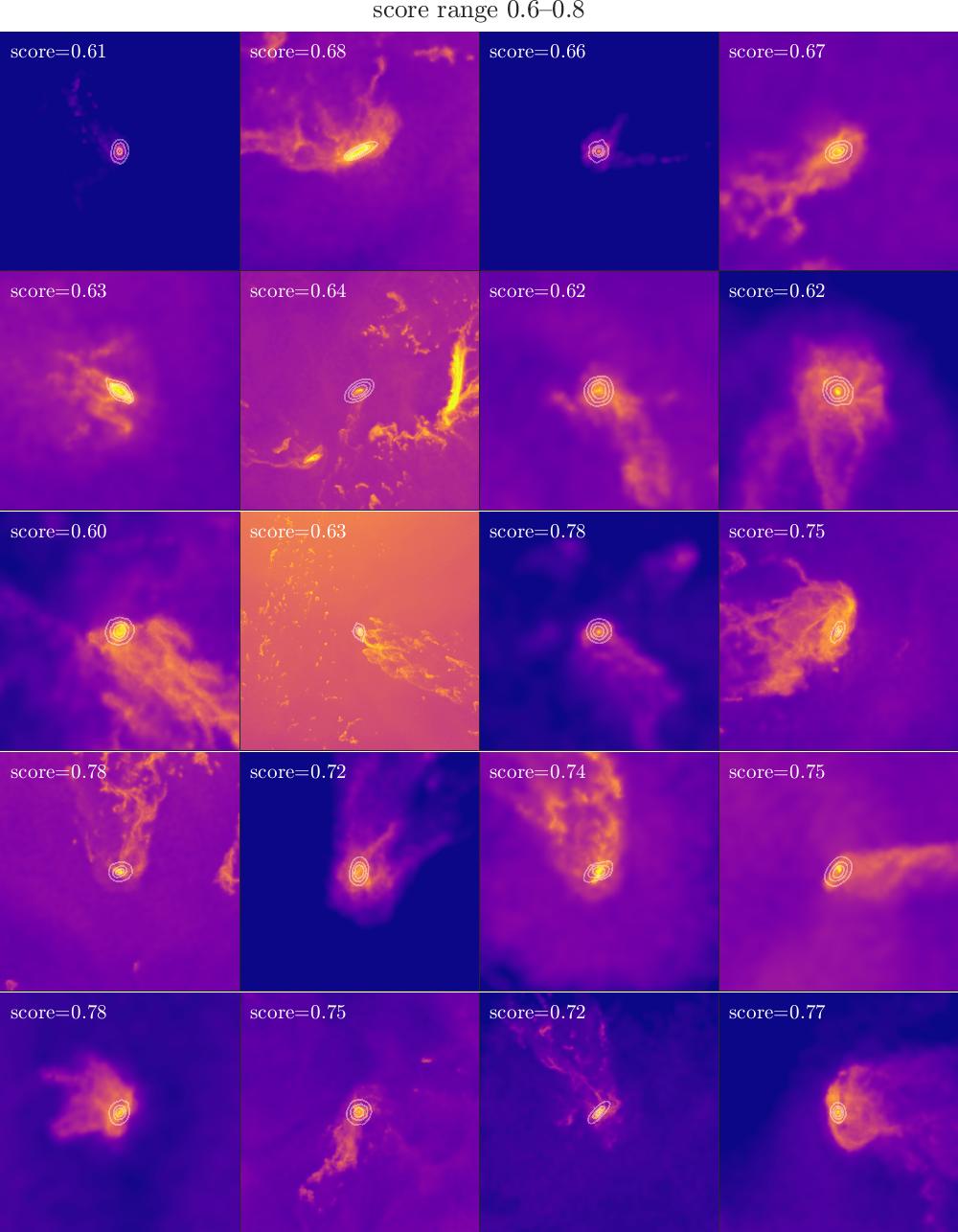}
       \caption{Sample of galaxy images of scores 0.6--0.8}
  \label{fig:scoreBin4_mosaic}
\end{figure*}

\begin{figure*}
  \centering
       \includegraphics[width=18cm,keepaspectratio]{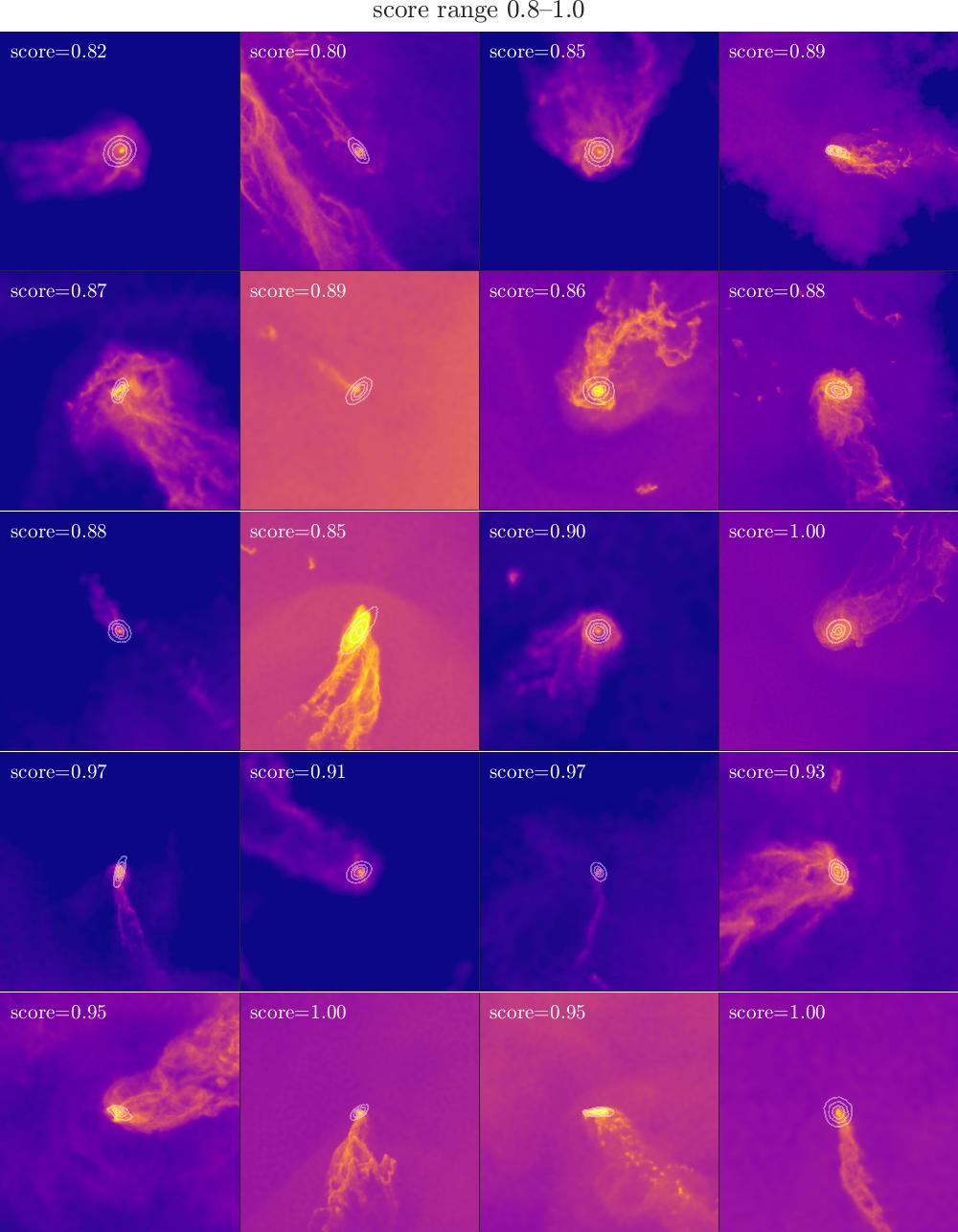}
       \caption{Sample of galaxy images of scores 0.8--1.0}
  \label{fig:scoreBin5_mosaic}
\end{figure*}

In \cref{fig:scoreBin1_mosaic,fig:scoreBin2_mosaic,fig:scoreBin3_mosaic,fig:scoreBin4_mosaic,fig:scoreBin5_mosaic} we show several examples of galaxies in five bins of JF scores ranging from 0-0.2 up to 0.8-1.

 \label{lastpage}
 \end{document}